\documentclass[iti,ngerman,masterarbeit,pdf,report]{unitext}
\usepackage{hyperref}
\pdfoutput=1

\usepackage[utf8]{inputenc}

\usepackage{enumitem}
\makeatletter
\let\note\@undefined          
\makeatother

\usepackage{calligra}
\usepackage{amsmath}
\usepackage{amsthm}
\usepackage{stmaryrd}
\usepackage{colonequals}
\usepackage{float}
\usepackage{xspace}
\usepackage{subfig}
\usepackage{framed}
\usepackage{listings}
\usepackage{bussproofs}
\usepackage{xargs}
\usepackage{xifthen}

\makeatletter
\let\c@table\c@figure
\makeatother

\usepackage{catchfilebetweentags}
\usepackage{etoolbox}
\makeatletter
\patchcmd{\CatchFBT@Fin@l}{\endlinechar\m@ne}{}
  {}{\typeout{Unsuccessful patch!}}
\makeatother

\usepackage{algorithmic}
\usepackage{algorithm}
\floatname{algorithm}{\algorithmName}
\numberwithin{algorithm}{chapter}
\algsetup{linenosize=\normalsize, linenodelimiter=.}

\makeatletter
\let\c@algorithm\c@figure
\makeatother

\usepackage{tikz}
\usetikzlibrary{positioning}
\usepackage{environ}
\makeatletter
\newsavebox{\measure@tikzpicture}
\NewEnviron{scaletikzpicturetowidth}[1]{%
  \def\tikz@width{#1}%
  \begin{lrbox}{\measure@tikzpicture}%
  \BODY
  \end{lrbox}%
  \pgfmathparse{#1/\wd\measure@tikzpicture}%
  \BODY
}
\makeatother

\lstdefinelanguage{lama}{
    morekeywords={ and, bool, definition, enum, input, int, let, match,
        mod, not, real, select, tel, typedef, xor, assertion, 
        constants, div, false, invariant, local, node, or, record, 
        sint, transition, uint, automaton, constr, edge, initial, default,
        ite, location, nodes, project, returns, state, true, use
    },
    otherkeywords={', =>, \#},
    sensitive=true,
    morecomment=[l]{--}
}

\lstdefinelanguage{scade}{
    morekeywords=[1]{%
      type, else, if, node, returns, then
      , returns, var, let, tel, package, public
      , private, imported, open, group
      , const, sensor, function, when, match, not
      , or, and, xor, mod, end, pre, last
      , default, sig, automaton, state, restart
      , resume, every, times, do, emit
      , case, of, fby, merge, flatten
      , true, false, activate, where, numeric
      , specialize, guarantee
    },
    morekeywords=[2]{bool, char, int, real, clock, probe},
    morekeywords=[3]{initial, final, unless, until, synchro},
    otherkeywords={', ->},
    sensitive=true,
    morecomment=[s]{/*}{*/}
}

\ExecuteMetaData[content/Grammar.tex]{grammarhead}

\newfloat{listing}{h}{lol}[chapter]
\floatname{listing}{\listingname}
\makeatletter
\let\c@listing\c@figure
\makeatother

\usepackage[draft]{pdfpages}

\newtheoremstyle{note} 
                {10pt} 
                {10pt} 
                {\itshape} 
                {} 
                {\bfseries} 
                {} 
                {.5em} 
                {} 

\newtheoremstyle{definition} 
                {10pt} 
                {10pt} 
                {} 
                {} 
                {\bfseries} 
                {} 
                {.5em} 
                {} 

\newtheoremstyle{theorem} 
                {10pt} 
                {10pt} 
                {\itshape} 
                {} 
                {\bfseries} 
                {} 
                {.5em} 
                {} 

\newtheoremstyle{example} 
                {10pt} 
                {10pt} 
                {} 
                {} 
                {\bfseries} 
                {} 
                {\newline} 
                {} 

\theoremstyle{definition}
\newtheorem{definition}{Definition}[chapter]

\newtheorem{translation}[definition]{Übersetzung}

\theoremstyle{theorem}

\theoremstyle{example}
\newtheorem{example}[definition]{\exampleName}

\theoremstyle{note}
\newtheorem{remark}[definition]{\remarkName}

\newcommand{\chapRef}[1]{\chaptername \xspace \ref{chap:#1}}
\newcommand{\secRef}[1]{\sectionName \xspace \ref{sec:#1}}

\newcommand{\transRef}[1]{Übersetzung \xspace \ref{trans:#1}}
\newcommand{\figRef}[1]{\figureName \xspace \ref{fig:#1}}
\newcommand{\algRef}[1]{\algorithmName \xspace \ref{alg:#1}}
\newcommand{\lstRef}[1]{\listingname \xspace \ref{lst:#1}}
\newcommand{\tabRef}[1]{Tabelle \xspace \ref{tab:#1}}

\newcommand{\eqRef}[1]{(\ref{eq:#1})}
\newcommand{\seeR}[1]{\seename \xspace \ref{#1}}
\newcommand{\exRef}[1]{Beispiel \xspace \ref{ex:#1}}

\newcommand{\secRefO}[1]{\sectionName \xspace #1}

\newcommandx{\term}[2][1]{\emph{#2}\ifthenelse{\isempty{#1}}{\index{#2}}{\index{#1}}}

\let\oldcdot=\cdot
\def\cdot{\negthinspace\oldcdot\negthinspace}

\mathchardef\breakingcomma\mathcode`\,
{\catcode`,=\active
  \gdef,{\breakingcomma\discretionary{}{}{}}
}
\newcommand{\mathlist}[1]{$\mathcode`\,=\string"8000 #1$}

\newcommand{\B}{\mathbb{B}}
\newcommand{\N}{\mathbb{N}}
\newcommand{\Z}{\mathbb{Z}}

\newcommand{\R}{\mathbb{R}}

\newcommand{\func}[3]{\ensuremath{{#1} \colon {#2} \to {#3}}}
\newcommand{\deriv}[2]{{#1}^{(#2)}} 
\newcommand{\vect}[1]{\mathbf{#1}}

\newcommand{\powSet}[1]{\mathcal{P}(#1)}

\newcommand{\setDef}[2]{\left\{{#1} \:|\: {#2}\right\}}

\newcommand{\stream}[1]{\left[\N, #1 \right]}

\newcommand{\lang}[1]{\textsc{#1}\xspace}
\newcommand{\Lama}{\lang{Lama}}
\newcommand{\Scade}{\lang{Scade}}

\DeclareMathOperator{\states}{states}
\newcommand{\Usage}{U\!sage}
\newcommand{\Mode}{M\!ode}
\newcommand{\ite}{ite}
\newcommand{\iteS}{\ite_{\mathrm{strict}}}
\newcommand{\iteL}{\ite_{\mathrm{lazy}}}
\newcommand{\Ident}{\mathrm{Ident}}
\newcommand{\Term}[1]{\mathrm{Term}_{#1}}
\newcommand{\seqTerm}{\Term{\stream{A}}}

\newcommand{\syn}[1]{\texttt{{#1}}}
\newcommand{\ty}[1]{\textbf{#1}}

\newcommand{\tyRule}[1]{\text{\textit{#1}}}

\def \lstLm {\lstinline[language=lama]}

\newcommand{\arr}{\rightarrow}
\DeclareMathOperator{\pre}{pre}
\DeclareMathOperator{\last}{last}
\newcommand{\lastApp}{\last\; ' \!}
\def \lstSc {\lstinline[language=scade]}

\usetikzlibrary{matrix,arrows,decorations.pathmorphing,positioning}

\newdimen{\ArrowLength}

\newcommandx{\doublexrightarrow}[2][1]{%
\setlength{\ArrowLength}{\maxof{\widthof{$#1$}}{\widthof{$#2$}}}%
\tikz[minimum height=0ex,baseline]
  \path[->]
   node (a) {}
   node [right=\ArrowLength of a] (b) {}
   (a.north)  edge node[above]{$#2$} (b.north)
   (a.center)  edge node[below]{$#1$} (b.center);%
}

\newcommandx{\triplexrightarrow}[2][1]{%
\setlength{\ArrowLength}{\maxof{\widthof{$#1$}}{\widthof{$#2$}}}%
\tikz[minimum height=0ex,baseline]
  \path[->]
   node (a) {}
   node [right=\ArrowLength of a] (b) {}
   (a.north)  edge node[above]{$#2$} (b.north)
   (a.center) edge (b.center)
   (a.south)  edge node[below]{$#1$} (b.south);%
}

\graphicspath{{content/images/}}

\title{Transformation von Scade-Modellen zur SMT-basierten Verifikation}
\author{Henning Basold}
\date{12. Oktober 2012}
\dozent{Prof. Dr. Jiří Adámek}
\betreuer{Dr. Stefan Milius}

\makeindex

\begin{document}

\titelblatt             
\erklaerung             
\cleardoublepage

\begin{abstractKeyw}
In dieser Arbeit wird ein Verfahren zur vollautomatischen Verifikation
von Sicherheitseigenschaften von \Scade-Modellen entwickelt. Dazu
transformieren wir jedes dieser Modelle in eine SMT-Instanz
(Satisfiability Modulo Theories) und übergeben dies an einen Solver.

SMT wurde gewählt, da es Logiken umfasst die ausdrucksstärker als
Aussagenlogik sind, während deren Solver sehr gute Geschwindigkeiten
erreichen. Die Ausdrucksstärke dieser Logiken erlaubt es symbolisches
Modelchecking zu implementieren und damit
eine Berechnung des gesamten Zustandsraumes zu vermeiden.

Um die Komplexität zu reduzieren, teilen wir die Tranformation von
\Scade-Modellen in SMT-Instanzen in zwei Schritte auf. Zuerst
werden die \Scade-Modelle auf Programme
einer synchronen Datenflusssprache \Lama reduziert.
Diese Sprache hat eine einfachere Semantik als \Scade, erhält
aber einige Abstraktionen des Programmierers.
Im zweiten Schritt interpretieren wir diese Programme
als Systeme quantorenfreier Formeln erster Ordnung.

Die Abstraktionen,
die in \Lama erhalten bleiben, können in einer weiteren Arbeit
genutzt werden, um diese Systeme zu vereinfachen. Dies wiederum
kann zu einer Beschleunigung des Verifikationsprozesses führen
und mehr Eigenschaften verifizierbar machen.

Die beschriebenen Transformationen wurden erfolgreich in einer Software
umgesetzt. Wir vergleichen diese abschließend mit der existierenden
Verifikations-Software „\Scade Design Verifier“ aus der \Scade Suite.
\end{abstractKeyw}

\selectlanguage{english}
\begin{abstractKeyw}
In this work we develop a fully automatic verification procedure of
safety properties of \Scade models. We transform each such model
into an SMT instance (Satisfiability Modulo Theories) and feed this
to a solver.

The choice of SMT is determined by the fact that it offers more
expressive logics than propositional logic, yet their solvers have
recently been shown to perform very well. The expressiveness of
SMT logics allows us to implement symbolic model checking thus avoiding
the expansion of the complete state space of the models during the
verification.

In order to reduce the complexity we transform the \Scade
models into SMT instances in two steps. First the models are reduced
to programs of a synchronous data flow language \Lama. This language
has simpler semantics than \Scade while still preserving some of the
programmer's abstractions.
Next we interpret such a \Lama program as a system of quantifier
free first-order formulas.

The remaining abstractions in \Lama can be
used to simplify these systems. This in turn could lead to speeding
up the verification process and allowing more properties to be
verifiable.

We implemented these transformations successfully in a software.
This work is concluded by a comparison of this software to the existing
verification software ``\Scade Design Verifier'' which comes
with the \Scade Suite.

\end{abstractKeyw}
\selectlanguage{ngerman}



\cleardoublepage
\tableofcontents        

\starttext
\chapter{Einleitung}
\label{chap:intro}

\Scade ist eine Sprache zur Modellierung
\term[synchrones System]{synchroner Systeme}.
Diese Klasse von Programmiersprachen hat sich als besonders
geeignet zur Beschreibung einzelner reaktiver Systeme herausgestellt.
Dies liegt in der einfachen Semantik begründet. Während allgemeinere
Programmiersprachen Schleifen bzw. Rekursion zur Berechnung zur
Verfügung stellen, arbeiten synchrone Sprachen taktbasiert. Der
Entwickler beschreibt dazu die Berechnungen \emph{eines} Taktes,
wobei er sich auf Berechnungen vergangener Takte beziehen kann.
Ein \term{Lauf}\footnote{Ein Lauf eines Systems ist eine (unendliche)
Folge von Eingaben, Zuständen, Ausgaben und evtl. weiteren
Informationen die nötig sind, um die Reaktion des Systems
nachvollziehen zu können.} eines solchen Systems
ergibt sich nun als Aneinanderreihung dieser Takte.

Aufgrund dieser reduzierten Komplexität der Semantik eignet sich
diese Klasse von Sprachen auch besser zur automatischen Verifikation
gewünschter Eigenschaften eines Programms. In der
\Scade-Suite\footnote{Die \Scade-Suite ist ein Softwarepaket zur
Unterstützung der Entwicklung mit \Scade. Sie enthält unter anderem
ein Programm zur graphischen Modellierung von \Scade-Programmen.}
existiert bereits ein Programm (Design Verifier), mit dem eine solche
Verifikation betrieben werden kann. Allerdings hat sich gezeigt
(s. z.B. \cite{Huhn12}), dass dieses Programm sehr schnell
an seine Grenzen stößt. Leider ist dies ein kommerzielles Produkt,
dessen Quelltext nicht verfügbar ist. Daher ist es auch nicht möglich
herauszufinden, an welchen Stellen die Komplexitätsprobleme auftreten
und diese zu beheben. Dies wollen wir in dieser Arbeit änderen,
indem wir ein Quelloffenes System entwickeln.


Dieses System soll auf \term{SMT}
(\term{Satisfiability Modulo Theories}) aufbauen. In letzter Zeit hat
sich gezeigt, dass SMT-Solver durch große Fortschritte in
der Technologie der SAT-Solver sehr gute Performance-Eigenschaften
besitzen. Dies kann man sich gerade in Systemen mit numerischen
Berechnungen (ganzzahlig oder reell) zu Nutze machen. Aber auch
bei Systemen über zweiwertige Logiken bringen diese noch
Vorteile. Dabei bezieht sich dieser Vergleich auf
\term{Modelchecking}, bei dem ein System durch ein
Transitionssystem beschrieben wird. Hier muss der Zustandsraum
vollständig exploriert werden (vgl. \secRef{model_checking}).

\Scade unterstützt aber auch numerische Berechnungen. Wenn nun bspw.
der Typ \lstSc!int! als $\Z$ interpretiert wird, ist eine vollständige
Exploration unmöglich. Bei einer beschränkten Darstellung wird
dagegen der Zustandsraum enorm groß. Daher wollen wir
\term[Modelchecking!symbolisches]{symbolisches Modelchecking}
(\ref{sec:model_checking}) verwenden. Dies ist mit
SMT besonders leicht umzusetzen.

\section{Aufbau der Arbeit}
\label{sec:thesis_structure}

Die Arbeit gliedert sich zwei wesentliche Transformationsschritte.

Zunächst wird \Scade in eine einfachere Sprache \Lama transformiert.
Diese wurde eigens für diese Arbeit aus anderen Sprachen heraus
entwickelt. Syntax und Semantik der Sprache \Lama wird in
\chapRef{lama} beschrieben. In \chapRef{lama_smt} gehen wir auf die
Transformation in SMT ein.

Die Transformation von \Scade in \Lama wird in \chapRef{scade2lama}
beschrieben.

Die Grundlagen von SMT, die für diese Transformation benötigt werden,
werden im \chapRef{prelim} dargestellt. Wir werden dort auch eine
kurze Einführung in Modelchecking geben.

Abschließend wird ein Vergleich zwischen der bestehenden Software
und der in dieser Arbeit entwickelten durchgeführt.

\section{Ähnliche Arbeiten}
\label{sec:similar_work}

Die Arbeit von Hagen (\cite{Hagen08}) zusammen mit Tinelli
(\cite{HagenTinelli}) hat gezeigt, dass ein SMT-basierter
Ansatz für die Verifikation von synchronen Sprachen gute
Performanceeigenschaften hat. In der Arbeit ist das System
\emph{Kind} (später um nebenläufige Berechnung zu \emph{pKind}
in \cite{Kahsai2011}
bzw. mit automatischer Generierung von Invarianten zu \emph{Kind-Inv} 
erweitert) entwickelt worden. Dabei wurde die Verifikation auf
\lang{Lustre} betrieben. Der Fokus der Arbeit lag dabei auf
der Exploration von Optimierungs- und Abstraktionstechniken.

Eine Überlegung zu Beginn dieser Arbeit war, auf \emph{Kind}
aufzusetzen. Allerdings stellt \Scade einige Konstrukte bereit,
die dem Programmierer
Möglichkeiten der Abstraktion geben. Diese können ggf. genutzt werden,
um die Verifikation zu beschleunigen, z.B. indem für eine Eigenschaft
unnötige Programmteile ausgelassen werden. Durch die Einfachheit von
\lang{Lustre} geht diese Abstraktion allerdings verloren.
Daher haben wir uns für eine Neuentwicklung entschieden,
wobei hier darauf geachtet werden sollte, dass der entstehende
Code leicht wartbar ist.

Die Techniken zur Beschleunigung, die Hagen beschreibt,
können in einem weiteren Schritt auch in unser System
integriert werden.

Franzén hat in \cite{Franzen2006} mit dem Tool \emph{Rantanplan}
parallel dazu ebenfalls SMT-basierte Verifikation von \lang{Lustre}
ermöglicht. Hier wurden noch keine Optimierungen u.ä. umgesetzt.

Ein anderer Ansatz ist abstrakte Interpretation (\cite{Cousot1992}).
Dies wurde in \cite{Jeannet2003} in dem Tool \emph{NBAC}
für synchrone Sprachen umgesetzt.

In \cite{Schrammel11} wird „abstract Acceleration“ benutzt, um
erreichbare Zustände zu berechnen und somit die Verifikation darauf
zu beschränken. Da dies im Allgemeinen nicht entscheidbar ist,
wird abstrakte Interpretation verwendet, eine korrekte Approximation
zu bekommen.

In \cite{Roux} wird ebenfalls ein kombinierter Ansatz gewählt. Hier
wird die abstrakte Interpretation genutzt, um Invarianten für die
$k$-Induktion zu generieren.

\chapter{Grundlagen}
\label{chap:prelim}

\section{Scade und Datenfluss}
\label{sec:scade_dataflow}

\Scade ist eine synchrone Datenflusssprache mit einer Ergänzung um
Automaten und lokale Takte. \term{Synchron} bedeutet hier, dass ein
Programm in diskreten Zeitschritten ausgeführt wird.


Mit \term{Datenfluss} bezeichnen wir den Anteil der Sprache,
der aus Zuweisungen und Ausdrücken über einer Signatur $\Sigma_S$
besteht. Dabei entsteht $\Sigma_S$ aus einer Signatur $\Sigma$
durch Hinzufügen von \term[Stromoperator]{Stromoperatoren}.
Nur mit diesen kann
abhängig von der Zeit eine Berechnung beeinflusst werden.
$\Sigma$ ist in \Scade bzw. \lang{Lustre} die Menge
aller arithmetischen und logischen Operatoren usw.,
die unterstützt werden. $\Sigma_S$ erhält dann als zusätzliche
Operatoren z.B. \lstSc!pre! und \lstSc!->!. Dabei ist
\begin{align*}
  & (\pre M)_n = M_{n-1}, \quad n > 0 \text{ und} \\
  & (M \arr N)_n = \begin{cases}
    M_0, & n = 0 \\
    N_n, & n > 0
  \end{cases}.
\end{align*}

Dabei sind $M,N$ Ausdrücke. $M_n$ soll den Wert des Ausdrucks $M$
im Takt $n \in \N$ bezeichnen.\footnote{Aufgrund der Lesbarkeit, werden
wir häufig $\pre$/$\arr$ statt \lstSc!pre!/\lstSc!->! schreiben.}

Wie bereits angedeutet, baut \Scade auf \lang{Lustre} auf. Die Ideen
für Automaten entstammen z.B. \lang{Esterel} und \lang{StateChart}.
Mehr zur formalen Semantik kann z.B. in \cite{SynchFlowState}
oder in der \Scade-Language-Referenz \cite{ScadeRef} bzw.
informell in dem \Scade-Language-Primer \cite{ScadePrimer}
gefunden werden.

Wir werden uns hier einen kurzen Überblick über \Scade verschaffen.
In \chapRef{scade2lama} werden wir uns schwierige Sprachkonstrukte
bei der Übersetzung in eine Zwischensprache genauer ansehen.

\begin{figure}[h]
  \centering
  \includegraphics[width=0.75\textwidth]{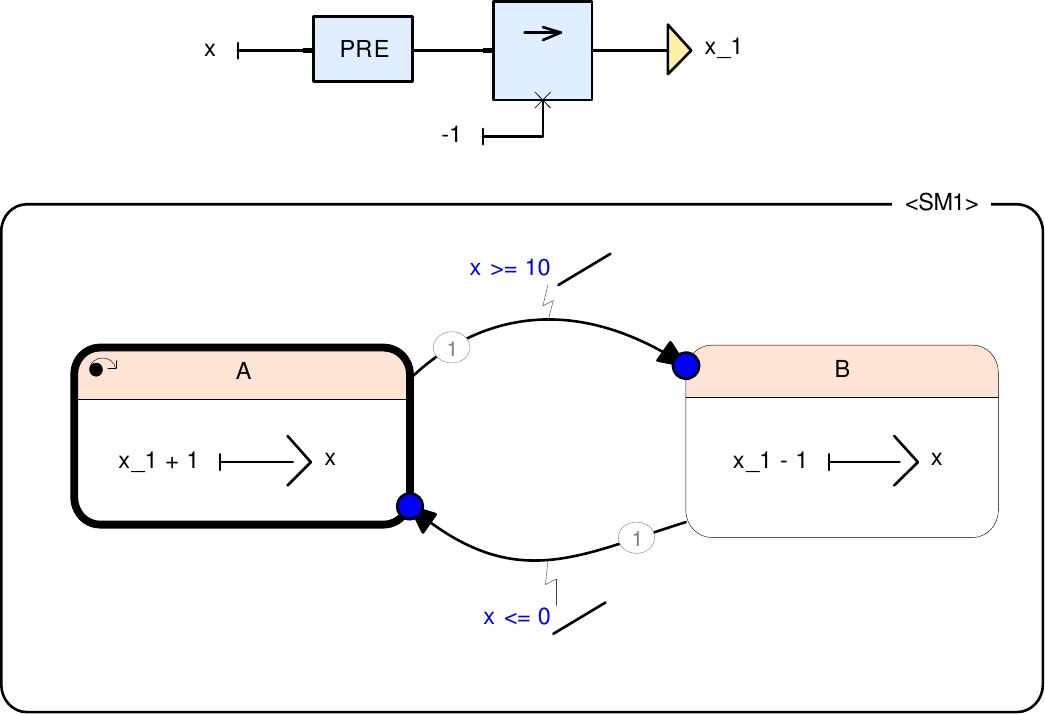}
  \caption{\Scade-Beispiel UpDownCounter}
  \label{fig:scade_updown_counter}
\end{figure}

Unser erstes wiederkehrendes Beispiel für diese Arbeit ist in
\figRef{scade_updown_counter} dargestellt. Dies ist eine graphische
Darstellung eines Modells, wie es in der \Scade-Suite normalerweise
erstellt wird. Alle Transformationen, die in dieser Arbeit
beschrieben werden, finden auf einer textuellen Darstellung statt.
Die graphischen Modelle können aber bei gleicher Semantik in eine
solche textuelle Darstellung überführt werden. Das Modell ist
in \lstRef{scade_updown_text} in textueller Form dargestellt.

\begin{listing}[h]
\begin{lstlisting}[language=scade]
node UpDownCounter() returns (x : int)
var
  x_1 : int;
let
  
  automaton SM1
    initial state A
      let
        x = x_1 + 1;
      tel
      until
        if x >= 10 do restart B;

    state B
      let
        x = x_1 - 1;
      tel
      until
        if x <= 0 do restart A;
  returns ..;
  
  x_1 = -1 -> pre x;
tel
\end{lstlisting}

  \caption{\Scade-Beispiel UpDownCounter in Textform}
  \label{lst:scade_updown_text}
\end{listing}

Das Beispiel umfasst einen \term{Knoten} \lstSc!UpDownCounter!,
der abwechselnd
von $0$ bis $10$ hoch und wieder herunter zählt. Ein solcher Knoten
kann Eingabe- und Ausgabevariablen, sowie lokale Variablen haben.
In der graphischen Darstellung sind \lstSc!x_1! durch den
gelb ausgefüllten Pfeil als lokale Variable und \lstSc!x!
durch den einfachen Pfeil als Ausgabevariable erkennbar. Das
Verhalten des Knotens wird durch Datenfluss und Automaten
beschrieben. In diesem Fall besitzt der Knoten einen Datenfluss
und einen Automaten mit zwei Zuständen $A$ und $B$,
sowie zwei Transitionen zwischen ihnen. Dabei ist $A$ als
Initialzustand (dicke Umrandung und kleiner Pfeil) deklariert.

Ein Zustand kann wiederum Datenfluss beinhalten. Hier ist das
\begin{itemize}
\item global: \lstSc!x_1 = -1 -> (pre x)!
\item in A: \lstSc!x = x_1 + 1!
\item in B: \lstSc!x = x_1 - 1!.
\end{itemize}
Eine Transition zwischen zwei Zuständen muss eine Bedingung und kann
eine Aktion haben, die bei Aktivierung ausgeführt wird. Hier gibt
es zwei Transitionen:
\begin{itemize}
\item $A \xrightarrow{\text{\lstSc!x>=10!}} B$ und
\item $B \xrightarrow{\text{\lstSc!x<=0!}} A$,
\end{itemize}
jeweils ohne Aktionen.

\begin{figure}[h]
  \centering
  \includegraphics[width=0.85\textwidth]{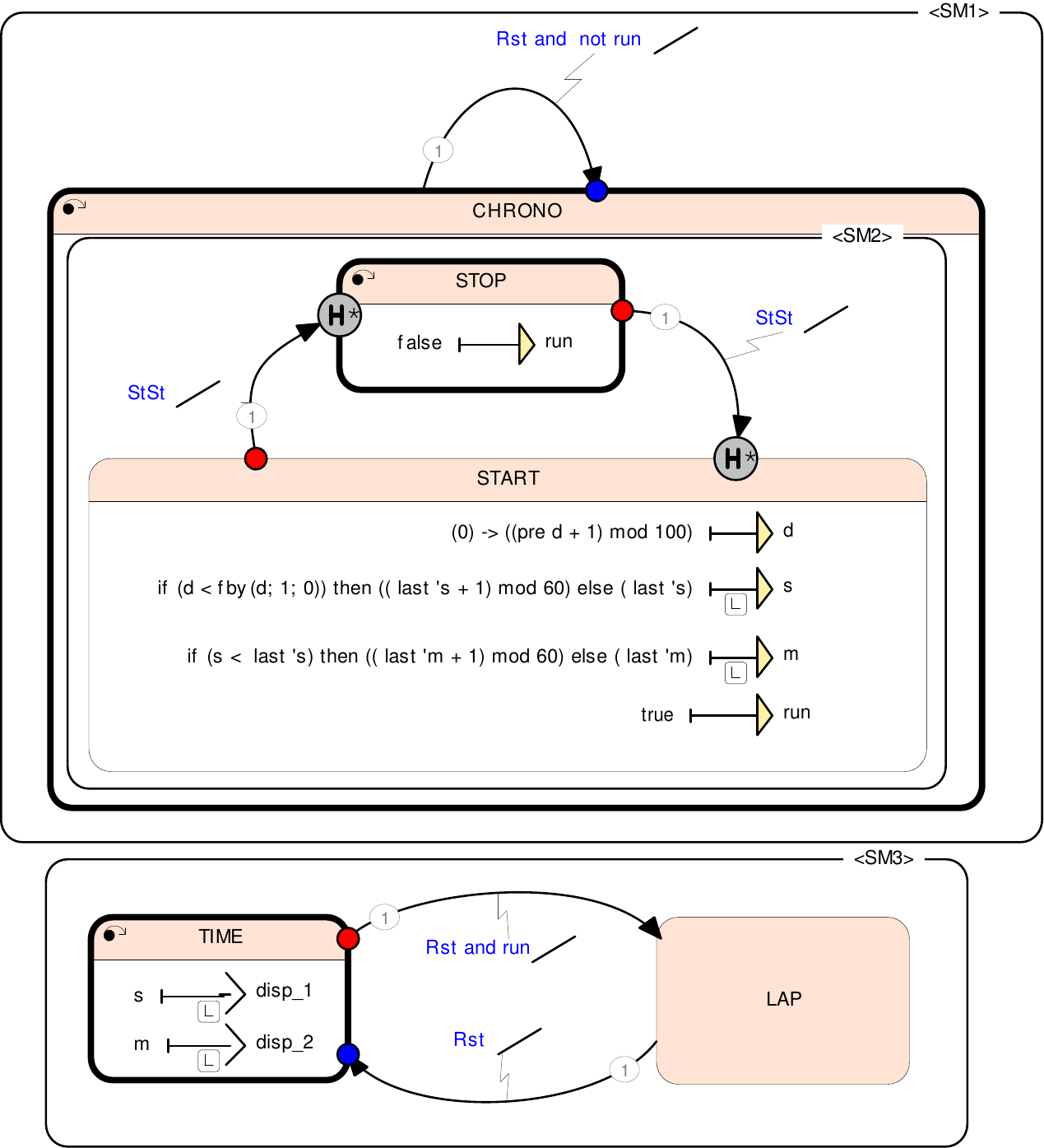}
  \caption{\Scade-Beispiel Chrono}
  \label{fig:scade_chrono}
\end{figure}

In \figRef{scade_chrono} ist ein weiteres \Scade-Beispiel
dargestellt. Dieses modelliert eine Stoppuhr mit einem
Start-/Stopp-Knopf und einem Reset-Knopf. Der Knoten hat zwei
ganzzahlige Ausgabevariablen für ein Display (\lstSc!disp_1, disp_2!).
In dem Beispiel werden die verschiedenen Arten von Transitionen
benutzt. Zum einen werden \term[Transition!Weak-]{Weak-Transitions}
(kleiner blauer Kreis
am Ende des Transitions-Pfeils) und
\term[Transition!Strong-]{Strong-Transitions}
(roter Kreis am Anfang des Transitions-Pfeils) unterschieden
und zum anderen \term[Transition!resume-]{resume}-
(H* am Ende des Transitions-Pfeils)
und \term[Transition!restart-]{restart-Transitions}
(keine Visualisierung).

Die erste Unterscheidung
betrifft den Zeitpunkt der Ausführung einer Transition.
\Scade unterscheidet zwischen
\term[Zustand!ausgewählter]{ausgewählten Zuständen}
und \term[Zustand!aktiver]{aktiven Zuständen}. Ein ausgewählter Zustand
kann erst im folgenden Takt aktiv werden. Der Datenfluss
eines Zustandes kann aber nur ausgeführt werden, wenn dieser aktiv ist.
Grob gesprochen legen Strong-Transitions den aktiven Zustand
und Weak-Transitions den ausgewählten Zustand eines Automaten
fest. Eine genauere Beschreibung wird in
\secRef{weak_strong_transitions} gegeben.

Die Unterscheidung von \lstSc!restart! und \lstSc!resume! betrifft
den Fluss und Automaten \emph{innerhalb} eines Zustandes,
sowie Bedingungen und
Aktionen an Transitionen. Im Falle einer restart-Transition
wird der initiale Datenfluss (durch \lstSc!->! spezifiziert)
ausgeführt, sowie Subautomaten in ihren Initialzustand versetzt.
Ein Beispiel für eine restart-Transition ist
$\text{\lstSc!CHRONO!} \longrightarrow \text{\lstSc!CHRONO!}$.
\lstSc!CHRONO! beinhaltet zwar keinen
Datenfluss, aber den Automaten \lstSc!SM2!. Dieser wird beim
Ausführen der Transition in den Initialzustand \lstSc!STOP! versetzt.
resume-Transitionen erhalten dagegen aktive Zustände von Subautomaten
und im betroffenen Datenfluss wird keine Initialisierung durchgeführt.

Eine Transition erhält außerdem eine Priorität (Nummerierung an
der Kante). Diese wird benutzt, um die Automaten deterministisch
zu machen. Wenn die Bedingung von zwei Transitionen mit
gleicher Quelle gilt, wird die mit der kleineren Priorität ausgeführt.

Eine weitere Besonderheit in dem Beispiel ist die Komposition von
Automaten. Es werden hierarchische Komposition
(\lstSc!SM2! in \lstSc!CHRONO!) und parallele Komposition
(\lstSc!SM3! neben \lstSc!SM1!) verwendet.

Zuletzt fällt auf, dass nicht alle Variablen in allen Zuständen
definiert sind. Die erste ist \lstSc!d!. Diese Variable
ist nur lokal in dem Zustand \lstSc!START!, braucht also
auch nur dort eine Definition. \lstSc!s, m, disp_1, disp_2!
dagegen sind nicht lokal. Sie haben aber ein kleines „L“ an
ihrem Namen. Dieses steht für \lstSc!last!. Wenn eine
Variable \lstSc!x! nicht in allen Zuständen definiert worden ist,
erhält sie eine Standard-Definition. Diese kann spezifiziert werden,
wenn das nicht geschieht, wird \lstSc!last 'x! verwendet.
Dieses Konstrukt gibt ähnlich wie \lstSc!pre!,
den Wert von \lstSc!x! aus dem letzten Taktzyklus zurück.\footnote{
\lstSc!last! kann nur auf Variablen
angewendet werden (syntaktisch durch das Häkchen vorm Namen
angedeutet). Semantisch ist der Unterschied, dass \lstSc!pre!
für jeden Zustand einen lokalen Speicher hat, während
\lstSc!last! geteilten Speicher erzeugt. Mehr dazu in
\chapRef{scade2lama}.}
Das erfordert eine Initialisierung im ersten Taktschritt. Diese
wird bei der Deklaration der Variablen angegeben und ist graphisch
durch besagtes „L“ angedeutet. \lstSc!last! wird auch explizit
bei der Definition von \lstSc!s! und \lstSc!m! benutzt.

\Scade hat noch eine Reihe weiterer Sprachkonstrukte.
Zum Beispiel gibt es die Möglichkeit,
Ausdrücke mit einem lokalen Takt, der langsamer als der
globale Takt ist, zu versehen. Um allerdings vollständig
definierte Ausgabevariablen zu haben,
dürfen diese auf keinem eigenen Takt liegen. Hier muss ein
getakteter Ausdruck mit einem komplementär getakteten zusammengeführt
werden. Dies werden wir aber in dieser Arbeit nicht näher behandeln.

\section{SMT}
\label{sec:smt}

Das \term{SAT}-Entscheidungsproblem (\term{Satisfiability}) fragt
nach der Erfüllbarkeit einer logischen Formel. Dies ist im
Allgemeinen ein schwieriges Problem ($\mathcal{NP}$-vollständig).
Allerdings gibt es bestimmte Unterklassen von SAT (z.B. 2-SAT oder die
Erfüllbarkeit von Hornklauseln) die in polynomieller Zeit gelöst
werden können. Außerdem existieren Algorithmen, die in der
Praxis bestimmte Probleme schnell lösen können. Siehe hierzu
z.B. \cite[Teil I]{HBSat}.

In dieser Arbeit soll der Geschwindigkeitsvorteil von
\term[SAT-Solver]{SAT-Solvern} gegenüber Systemen genutzt werden,
die eine vollständige Zustandssuche durchführen.
Allerdings werden wir eine Logik
benötigen, die nicht in Prädikatenlogik darstellbar ist. Wir
benötigen zusätzlich eine Peano-Arithmetik auf natürlichen Zahlen,
also $\N$ mit $0 \in \N$ und einer Funktion
$\func{\mathrm{succ}\,}{\N}{\N}$.
Weiterhin werden Algebren auf ganzen und auf reellen Zahlen
und \term{nicht interpretierte Funktionen} (s. unten) benötigt,
um Strom-Variablen $(x_n)_{n \in \N}$ darstellen zu können. Eine solche
Variable werden wir als Funktion über $\N$ interpretieren.

Auf dem Fortschritt in der Technologie von SAT-Solvern,
also Programmen, die SAT-Probleme lösen, baut die Idee von
\term{Satisfiability Modulo Theories} (\term{SMT}) auf. Dies ist eine
Erweiterung von SAT um Theorien für beispielsweise oben
genannte Arithmetik und nicht interpretierte Funktionen.
Letztere sind Prädikat- und Funktionssymbole, die nicht in der
Syntax der Logik vorhanden sind. Diese können z.B. genutzt werden,
um freie Variable zu modellieren. Für eine Einführung in
SMT siehe z.B. \cite{Biere2009}.

Es gibt eine Fülle von SMT-Solvern und entsprechend viele
Schnittstellen. Allerdings gibt es einen Ansatz, diese Schnittstellen
zu vereinheitlichen. Dazu wurde der \term{SMTLib}-Standard
entwickelt \cite{SMTLib}. Dieser liegt mittlerweile in der
Version 2 vor. Wir werden in der Implementierung diesen Standard
nutzen. Um in dieser Arbeit über Formeln und Terme zu sprechen,
wird allerdings die übliche Symbolik für Logik benutzt.
Diese wird in ein (sehr eingeschränktes) Lambda-Kalkül integriert.

Wir schreiben $\equiv$ für das Gleichheitsprädikat der SMT-Logik,
um von der Gleichheit in der Metasprache (d.h. wenn über die
Gleichheit von logischen Termen gesprochen wird) zu
unterscheiden. Ansonsten werden logische und arithmetische Symbole
mit ihrer üblichen Bedeutung verwendet. Weiterhin werden wir eine
Funktion $\ite$ (für „ite-then-else“) verwenden. Diese soll
abhängig vom ersten Parameter zum ersten bzw. zweiten Parameter
ausgewertet werden.\footnote{Die Logik würde auch ohne $\ite$
auskommen. Dazu wird ein Vorkommen von $\ite(c, a, b)$ durch
eine uninterpretierte Konstante $x$ ersetzt und die Aussage
$c \longrightarrow x \equiv a
\wedge \neg c \longrightarrow x \equiv b$ angenommen.}

\begin{figure}[h]
\centering
\begin{tabular}{l}
  
\begin{tabular}{lll}
{\nonterminal{Term}} & {\arrow}  &{\terminal{x}}
 {\delimit} {\terminal{c}} \\
 & {\delimit} &{\nonterminal{Form}} \\
 & {\delimit} &{\nonterminal{Term}} $\boxempty$ {\nonterminal{Term}} \\
 & {\delimit} &{$\ite$} \\
 & {\delimit} &{$\lambda$ \terminal{x} . \nonterminal{Term}} \\
 & {\delimit} &{\nonterminal{Term}}
   \terminal{(}} {\nonterminal{Term}} {\terminal{)} \\
 & {\delimit} &{\terminal{(}} {\nonterminal{Term}}
   {\terminal{,}} {$\dotsc$}
   {\terminal{,}} {\nonterminal{Term}} {\terminal{)}} \\
 & {\delimit} &{$p_i$} \\
\end{tabular}\\

$\boxempty \in \{+, -, *, / \}$ \\ \\

\begin{tabular}{lll}
{\nonterminal{Form}} & {\arrow}  &{\terminal{x}}
 {\delimit} {\terminal{c}} \\
 & {\delimit} &{$\neg$ \nonterminal{Form}} \\
 & {\delimit} &{\nonterminal{Form}}
   $\boxempty_1$ {\nonterminal{Form}} \\
 & {\delimit} &{\nonterminal{Term}}
   $\boxempty_2$ {\nonterminal{Term}} \\
\end{tabular}\\

$\boxempty_1 \in \{\wedge, \vee, \longrightarrow \}$,
$\boxempty_2 \in \{\equiv, <, >, \leq, \geq \}$.
\end{tabular}

\caption{Syntax der SMT-Logik}
\label{fig:syntax_smt_logic}
\end{figure}

Die Syntax der sich daraus ergebenden Logik ist in
\figRef{syntax_smt_logic} dargestellt. Dabei ist \terminal{x} ein
nicht interpretiertes Symbol und eine \terminal{c} Konstante.
$(M_1, \dotsc, M_k)$ beschreibt Tupel mit entsprechenden
Projektionen $p_i$.\footnote{Sollte der SMT-Solver keine Tupel
unterstützen, können diese durch uninterpretierte Konstanten
für jede Komponente dargestellt werden. Eine Projektion
$p_i(M)$ wird dann durch die entsprechende $i$-te Konstante
von $M$ ersetzt.}

In die Syntax ist auch ein einfacher Lambda-Kalkül integriert.
Dabei muss die Variable $x$ einen einfachen Typ (d.h. nur
Basistypen wie z.B. $\Z$, aber keine Funktionstypen) haben.
Die Applikation von Funktionen ist durch $f(\ldots)$ mit einer
Funktion $f$ gegeben. Wir werden ggf. Klammern weglassen oder
zusätzlich schreiben.

Jeder Term erhält, wie oben bereits angedeutet, einen Typ
(oder \term{Sorte} bzw. \term{Sort} im Englischen). Wir setzen voraus,
dass der Solver Theorien für $\Z$ und $\R$ und $\Z/n, n \in \N$ mit
entsprechender Arithmetik, sowie Peano-Axiome für $\N$ (statt 
$\mathrm{succ} \; n$ schreiben wir $n + 1$) besitzt.

$\mathrm{Term}_A$ bezeichnet alle Terme der Logik vom Typ
$A$. Formeln (Form) sind von Natur aus vom Typ $\B$
(also Formeln einer zweiwertigen Logik). Wenn wir
$f(x) = M$ schreiben, soll damit $f = \lambda x. \; M$ gemeint sein.

Diese Syntax gibt nun an, wie Terme der Logik erzeugt werden. Um
diese verwenden zu können, müssen nicht interpretierte Symbole der
Terme mit dem entsprechenden Typen beim Solver deklariert werden.
Danach werden die Terme als Aussagen hinzugefügt und eine
Anfrage auf Erfüllbarkeit gestellt. Wenn das System der Aussagen
erfüllbar ist, kann der Solver ein Modell mit Belegungen der
nicht interpretierten Symbole angeben.

\begin{example}
\label{ex:updown_smt}

  Wir wollen den UpDown-Knoten aus \figRef{scade_updown_counter}
  kodieren. Eine Variable $x$ muss einen Wert für jeden Zeitpunkt
  besitzen. Daher wird die Definition mit einem Parameter
  $n$ für die Zeit versehen. Weiterhin müssen
  die Zustände des Automaten kodiert werden. Wir werden dafür die
  ganzen Zahlen benutzen.
  \begin{align*}
    & \func{\mathrm{sm1}}{\N}{\Z} \\
    & \func{x}{\N}{\Z} \\
    & \func{x\_1}{\N}{\Z} \\
    \\
    & x\_1_{init} = x\_1(0) \equiv -1 \\
    & x\_1_{def} = \lambda n.\; x\_1(n+1) \equiv x(n) \\
    \\
    & x_{def} = \lambda n.\; x(n) \equiv
      \ite(\mathrm{sm1}(n) \equiv 1, x\_1(n) + 1, x\_1(n) - 1) \\
    \\
    & \mathrm{sm1}_{init} = \mathrm{sm1}(0) \equiv 1 \\
    & \mathrm{sm1}_{def} = \lambda n.\;
      \mathrm{sm1}(n+1) \equiv \ite(sm1(n) \equiv 1, \\
      & \quad ite(x(n) \geq 10, 2, 1), \\
      & \quad ite(x(n) \leq 0, 1, 2)) \\
    & \mathrm{sm1}_{range} = \lambda n.\;
      1 \leq \mathrm{sm1}(n) \leq 2
  \end{align*}
  
  Wenn nun ein Zeitschritt $k \in \N$ untersucht werden soll,
  muss dem Solver eine Aussage der Form
  \begin{equation*}
    T(k) = x_{def}(k) \wedge x\_1_{def}(k) \wedge \mathrm{sm1}_{def}(k)
      \wedge \mathrm{sm1}_{range}(k)
  \end{equation*}
  übergeben werden. Das Prädikat zur Initialisierung ist mit
  \begin{equation*}
    I = x\_1_{init} \wedge \mathrm{sm1}_{init}
  \end{equation*}
  gegeben.
\end{example}

Dies beschreibt den Takt $k$ allerdings unabhängig von den vorherigen.
Im nächsten Abschnitt werden wir dies ändern.

\section{SMT-basiertes Modelchecking}
\label{sec:model_checking}

\newcommand{\entailsL}{\vDash_{\mathcal{L}}}

Sei ein Modell $\mathcal{M}$ eines Systems gegeben.
Es wird von \term{Modelchecking} (oder Modellprüfung) gesprochen, wenn
es möglich ist, $\mathcal{M}$ automatisch gegen eine Eigenschaft
$P$ des Systems zu prüfen. Dies kann halbautomatisch z.B. durch
Theorembeweiser oder, wie in unserem Fall, vollautomatisch durch
z.B. SMT-Solver erfolgen.

Wir werden hier so genanntes
\term[Modelchecking!symbolisches]{symbolisches Modelchecking}
(s. z.B. \cite{SymbolicMC90}) betreiben. Statt den
Zustandsraum eines Systems vollständig zu kodieren (z.B. als
Transitionssystem), wird ein System mit Hilfe einer
entsprechenden Logik symbolisch beschrieben, d.h. in Formeln, die
das System repräsentieren. Da wir die Eigenschaft auch in dieser
Logik ableiten werden, spricht man auch von
\term[Modelchecking!Logik-basiertes]{Logik-basiertem Modelchecking}
(\cite{TinelliSMT2011}). Wenn $\mathcal{M}$ und $P$ in einer Logik
$\mathcal{L}$ kodiert werden können\footnote{Man beachte dass dieser
unscheinbare Punkt der Inhalt dieser Arbeit ist.}, soll
\begin{equation}
  \label{eq:model_prop}
  \mathcal{M} \entailsL P
\end{equation}
bedeuten, dass unter der Voraussetzung $\mathcal{M}$ die
Eigenschaft $P$ in der Logik gilt.

Entsprechend werden wir von
\term[Modelchecking!SMT-basiertes]{SMT-basiertem Modelchecking}
(ebenfalls \cite{TinelliSMT2011}) sprechen, wenn $\mathcal{L}$
durch einen SMT-Solver entschieden werden kann. Die Aussage
\eqRef{model_prop} kann geprüft werden, indem wir den Solver nach
der Erfüllbarkeit von
$\mathcal{M} \wedge \neg P = \neg (\mathcal{M} \longrightarrow P)$
fragen, sofern
$\mathcal{M} \entailsL P \text{ gilt, gdw. }
\entailsL \mathcal{M} \longrightarrow P \text{ gilt}$.
Denn wenn $\neg P$ nicht erfüllbar ist, dann ist $P$ gültig.

Da wir Systeme (Datenfluss) modellieren wollen, deren Semantik über
die Zeit festgelegt ist, muss das System entsprechend kodiert werden.
Nun sollen aber keine Formeln erzeugt werden, die Aussagen über
alle Zeitpunkte machen\footnote{Dies würde entweder zu undendlich
großen Formeln oder Quantoren führen. Ersteres ist offensichtlich
nicht möglich, wenn ein Programm diese Formeln analysieren soll.
Die zweite Möglichkeit führt dagegen zu Problemen beim
vollautomatischen Modelchecking. Es gibt hier auch Ansätze in
SMT-Solvern, aber das wollen wir nicht weiter betrachten. Siehe
z.B. \cite{MCMT10}.}, sondern nur Aussagen über einzelne Zeitpunkte
machen. Wir werden aber sehen, dass dies bereits für viele Probleme
ausreichend ist.

Dazu nehmen wir an, dass $\mathcal{M}$ in Prädikaten $I(s)$ und
$T(x,s,s',y)$ kodiert ist. Dabei sind $x,y$ Ein- bzw.
Ausgabe-Vektoren und $s,s'$ modellieren den Zustand des Systems.
$I$ beschreibt den Initialzustand und $T$ die Transitionen,
des Systems.
Für einen Zustands- bzw. Eingabe-/Ausgabevektor zu einem Zeitpunkt
$n$ werden wir $\deriv{s}{n}, \deriv{x}{n}, \deriv{y}{n}$ schreiben.
Damit ergeben sich verkürzte Schreibweisen für die Initialisierung
$I = I \left( \deriv{s}{0} \right)$, die Transitionen
$T^i = T \left( \deriv{x}{i}, \deriv{s}{i},
\deriv{s}{i+1}, \deriv{y}{i} \right)$ und die Invariante
$P^i = P \left( \deriv{x}{i}, \deriv{s}{i}, \deriv{y}{i} \right)$.

\begin{remark}
  Dies weicht etwas von der Notation in \exRef{updown_smt} ab,
  da hier die freien Variablen explizit angegeben werden sollen, statt
  implizit, wie es später beim SMT-Solver getan wird. Die verkürzten
  Notationen sollten aber den Zusammenhang deutlich machen.
\end{remark}

\subsection{Bounded Model Checking}
\label{sec:bmc}

\newcommand{\trans}[2]{T \left( {#1}, {#2} \right)}

Der erste Schritt während der Entwicklung von Programmen ist das
Finden von Fehlern. Hier werden wir
\term[Modelchecking!Bounded]{Bounded Modelchecking}
(BMC) verwenden. Für ein $k \geq 0$ wird dies durch
\begin{equation}
  \label{eq:bmc}
  I, T^0, \dotsc, T^{k} \entailsL P^0 \wedge \dotsb \wedge P^k
\end{equation}
realisiert. Das bedeutet, dass ausgehend von einer Initialkonfiguration
$k$ Transitionen durchgeführt werden. Dabei muss dann zu jedem
Zeitpunkt $P$ gelten.

\begin{example}[BMC, Fortsetzung von \ref{ex:updown_smt}]
  Wir wollen für den Knoten \lstSc!UpDown! zeigen, dass
  die Eigenschaft
  \begin{equation*}
    P = \lambda n. \; x(n) \geq 1
  \end{equation*}
  fehlerhaft ist. Dazu wird \eqRef{bmc} mit $I$ und $T$ aus
  \exRef{updown_smt} instantiiert:
  \begin{equation*}
    I, T(0) \entailsL P(0).
  \end{equation*}
  Nun ist aber
  \begin{equation*}
    I = x\_1(0) \equiv -1 \wedge \mathrm{sm1}(0) \equiv 1
  \end{equation*}
  und
  \begin{align*}
    T(0) & = x_{def}(0) \wedge x\_1_{def}(0) \wedge \mathrm{sm1}_{def}(0)
      \wedge \mathrm{sm1}_{range}(0) \\
    & = x(0) \equiv \ite(\mathrm{sm1}(0) \equiv 1,
        x\_1(0)+1, x\_1(1)-1) \\
      & \quad \wedge  x\_1_{def}(0) \wedge \ldots \\
    & = x(0) \equiv x\_1(0)+1 \wedge \ldots \\
    & = x(0) \equiv 0 \wedge \ldots
  \end{align*}
  Also $I \wedge T(0) \not\entailsL P(0)$, wie erwartet.
\end{example}

\subsection{(k-)Induktion}
\label{sec:induction}

Wie bereits erwähnt, eignet sich BMC zwar zum Finden von Fehlern,
allerdings ist $k$ in \eqRef{bmc} fest. Wenn ein Fehler also erst
bei $k+1$ auftritt, kann das so nicht festgestellt werden. Wir müssten
die Aussage für alle $k$ prüfen. BMC ist als keine
Entscheidungsprozedur.

Für eine gewisse Klasse von Problemen lässt sich das Verfahren
verallgemeinern. Dazu werden wir \eqRef{bmc} um einen
Induktionsschritt erweitern.

\begin{align}
  I^0, T^0 & \entailsL P^0
    \label{eq:induction_start} \\
  P^n, T^n, T^{n+1} & \entailsL P^{n+1}
    \label{eq:induction_step}
\end{align}

Dies entspricht normaler Induktion. Gilt nun \eqRef{induction_start},
aber nicht \eqRef{induction_step}, kann keine Aussage über
$\mathcal{M} \entailsL P$ gemacht werden, da die
Induktionsvoraussetzung evtl. nicht stark genug gewesen ist.
Wir sagen dann, dass $P$ nicht induktiv ist.

Ein Verfahren, um die Induktionsvoraussetzung automatisiert
stärker zu machen, ist \term{k-Induktion}. Dies ist eine
Erweiterung der Voraussetzung auf $k$ Schritte.

\begin{align}
  I^0, T^0, \dotsc, T^{k}
    & \entailsL P^0 \wedge \dotsb \wedge P^k
    \label{eq:k_induction_start} \\
  P^n, \dotsc, P^{n+k}, T^n, \dotsc, T^{n+k+1}
    & \entailsL P^{n+k+1}
    \label{eq:k_induction_step}
\end{align}

Für $k=0$ entspricht dies der normalen Induktion von oben.

Auch hier kann es passieren, dass der Induktionsschritt
\eqRef{k_induction_step} nicht gilt, während aber der
Induktionsanfang \eqRef{k_induction_start} gilt. Dann kann $k$
erhöht werden, bis auch der Induktionschritt erfüllt ist.
Wir sprechen davon, dass $P$ \term{k-induktiv} ist. Es kann aber auch
sein, dass dies nie passiert. Daher ist auch dies keine
Entscheidungsprozedur. Viele Eigenschaften lassen sich aber
mit k-Induktion nachweisen.

Für bestimmte Systeme ist es möglich, die Voraussetzungen so zu
Ergänzen, dass die k-Induktion vollständig wird. Für einfache
Induktion wurde dies in \cite{Sheeran2000,Een2003} untersucht. Die Idee ist,
dass man ein Prädikat als Voraussetzung ergänzt, dass minimale
Pfade in dem Transitionssystem beschreibt. Damit kann beim
Erreichen einer Schleife in einem endlichen Transitionssystem
mit der Verlängerung des Pfades aufgehört werden. Die Schwierigkeit
dabei ist,
ein Prädikat ohne Quantoren zu Finden. Für k-Induktion und SMT ist
dies ähnlich als \term{Pfadkompression} in \cite{MouraInduct03}
und \cite{Hagen08,HagenTinelli} beschrieben.

k-Induktion und auch BMC haben die Eigenschaft, dass sie sehr
gut mit inkrementellen SMT-Solvern (oder auch SAT-Solvern)
funktioniert. Beim erhöhen von $k$ bleiben alle Voraussetzungen
erhalten, lediglich die neue muss ergänzt werden. Siehe hierzu
\cite[S.468]{HBSat} oder \cite{Een2003}.

\begin{example}[(0)-Induktion, Fortsetzung von \ref{ex:updown_smt}]
  Wir wollen hier beweisen, dass das System aus \exRef{updown_smt}
  die Eigenschaft
  \begin{equation*}
    P = \lambda n. \; 0 \leq x(n) \leq 10
  \end{equation*}
  hat. Der Basisfall ist analog zu BMC:
    \begin{equation*}
    I = x\_1(0) \equiv -1 \wedge \mathrm{sm1}(0) \equiv 1
  \end{equation*}
  und
  \begin{equation*}
    T(0) = \ldots = x(0) \equiv 0
  \end{equation*}
  und damit $I, T(0) \entailsL P(0)$.

  Die Induktionsannahme ist
  \begin{align*}
    T(n) & = x\_1(n+1) \equiv x(n) \\
      & \wedge \mathrm{sm1}(n+1) \equiv \ite(\mathrm{sm1}(n) \equiv 1,
        \ite(x(n) \geq 10, 2, 1), \ite(x(n) \leq 0, 1, 2)) \\
      & \wedge 1 \leq \mathrm{sm1}(n) \leq 2 \\
    T(n+1) & = x(n+1) \equiv \ite(\mathrm{sm1}(n+1) \equiv 1,
      x\_1(n+1)+1, x\_1(n+1)-1) \\
    P(n) & = 0 \leq x(n) \leq 10
  \end{align*}
  Wir können nun die drei möglichen Fälle für $x(n)$ untersuchen:
  \begin{itemize}
  \item $x(n) = 10$. Dann ist
    $\mathrm{sm1}(n+1) \overset{T(n)}{=} 2$ und damit
    \[x(n+1) \overset{T(n+1)}{=} x\_(n+1)-1
      \overset{T(n)}{=} x(n)-1 = 9\]
  \item $x(n) = 0$. Dann ist
    $\mathrm{sm1}(n+1) \overset{T(n)}{=} 1$ und damit
    \[x(n+1) \overset{T(n+1)}{=} x\_(n+1)+1
      \overset{T(n)}{=} x(n)+1 = 9\]
  \item $0 < x(n) < 10$. Dann ist auch $0 < x\_1(n+1) < 10$. Nun
    müssen wir den aktiven Zustand unterscheiden:
    \begin{itemize}
    \item $\mathrm{sm1}(n+1) = 1$. Daraus folgt
      \[0 < x(n+1) = x\_1(n+1)+1 = x(n)+1 \leq 10.\]
    \item $\mathrm{sm1}(n+1) = 2$. Hier folgt
      \[0 \leq x(n+1) = x\_1(n+1)-1 = x(n)-1 < 10.\]
    \end{itemize}
  \end{itemize}
  Insgesamt folgt also $0 \leq x(n+1) \leq 10$ und somit
  $T(n),T(n+1),P(n) \entailsL P(n+1)$.
\end{example}


\chapter{Die Sprache \Lama}
\label{chap:lama}

\lstset{language=lama,mathescape=true}

Um eine strikte Trennung zwischen der Komplexität von \Scade und der
Verifikation auf Basis von SMT zu erreichen, wurde eine Zwischensprache
entwickelt. Diese sollte einerseits einen kleinen Sprachumfang
(im Gegensatz zu \Scade) haben und strukturell nah an den zu
erzeugenden Systemen logischer Formeln sein.
Andererseits sollte sie auch einige der Abstraktionsmechanismen
erhalten, mit denen ein Programmierer die Struktur eines Programms
vorgeben kann. Diese Strukturen können bei der Erzeugung von
SMT-Formeln genutzt werden, um Optimierungen durchzuführen. Hierbei ist
insbesondere die Abstraktion von Programmteilen gemeint, die nicht zur
Verifikation einer Eigenschaft notwendig sind.

Bei der Abwägung dieser gegenläufigen Ziele wurden folgende
Eigenschaften ausgewählt, die die Sprache haben sollte:

\begin{itemize}
\item Datenflusssprache
\item Beschreibung von vorgegebenen (Vorbedingungen) und zu
  verifizierenden Eigenschaften
\item Erhaltung von Abstraktionsmöglichkeiten
  \begin{itemize}
  \item Datentypen: Zweiwertige Logik, ganze und reelle Zahlen,
    Arrays/Produkttypen/Structs, Enums
  \item Automaten mit Datenfluss die hierarchisch und parallel
    komponiert werden können
  \end{itemize}
\item Einfach zu Parsen und Analysieren
  \begin{itemize}
  \item Deklaration vor Benutzung, möglichst tabellenartig
  \item keine Module/Imports u.ä. syntaktischer Zucker
  \end{itemize}
\item einfache Semantik
  \begin{itemize}
  \item Vermeidung von Initialisierungsproblemen
  \item keine Komposition von Automaten durch Interleaving
    mit geteilten Variablen o.ä.
  \item möglichst nah an \Scade/Lustre
  \item deterministisch
  \end{itemize}
\end{itemize}

Diese Eigenschaften mündeten in eine Sprache, die auf Sprache
\lang{NBAC} \cite{NBAC} aufbaut und diese um Knoten
und Produkttypen erweitert. \lang{NBAC} ist
eine Sprache die ebenfalls zur Verifikation entwickelt worden ist.
Sie ist ein
Low-Level-Format zur Spezifikation von Datenflüssen und optionalem
kontinuierlichem Verhalten (hybride Automaten). Daher kommt das
LA („Low Abstraction“) in \Lama.

Die Art der Automaten wurde allerdings geändert. Statt lediglich einen
Ausdruck zuzulassen, der durch eine \term{Location} (so werden Zustände
eines Automaten in \lang{NBAC} genannt) erzwungen wird, kann in \Lama
eine
solche Location wieder einen Datenfluss enthalten. Auch wenn dies
äquivalent ist, macht es die Sprache einheitlicher. Diese Art
von Automaten lehnt sich an \term[Mode-Automat]{Mode-Automaten}
(\cite{ModeAutomata}) an.
Daher kommt das MA in LAMA. Wir werden eine Location im Folgenden
als \term{Modus} bezeichnen. 

Eine weitere Unterscheidung zu \lang{NBAC} ist die Syntax von
Ausdrücken. Diese werden in \Lama in Form von S-Expressions geschrieben
(\lang{NBAC} benutzt eine Infix-Notation). Dadurch wird das Parsen
stark vereinfacht.

Entfernt worden sind Sprachkonstrukte für hybride Systeme.

\section{Syntax}
\label{sec:lama_syntax}

Hier soll lediglich ein Überblick über die Syntax von \Lama gegeben
werden. Die Grammatik findet sich in Backus-Naur-Form im Anhang
(\secRef{lama_grammar}). Dazu betrachten wir einige Beispiele
und erläutern an diesen die Syntax.

Das erste Beispiel in \lstRef{lama_updown} stammt aus
\cite{ModeAutomata} bzw. \cite{ModeAutomataVerimag}. Wobei in
letzterem Artikel und hier das Beispiel etwas gegenüber dem
Original aus \cite{ModeAutomata} geändert worden ist
(s. Kommentar im Quelltext). Dies ist eine mögliche Umsetzung des
UpDown-Knotens aus \figRef{scade_updown_counter} in \Lama. Zusätzlich
wurde eine Sicherheitseigenschaft ergänzt.

Zunächst besteht ein \Lama-Programm aus einer Reihe von
Deklarationen (Knoten und Variablen), einem Datenfluss,
Initialisierungen, Vorbedingungen und einer Invariante.
Knoten sind sehr ähnlich aufgebaut,
können aber zusätzlich noch Automaten enthalten, dafür aber
keine Invariante.

Deklarationen bestehen aus einer Menge von Knoten (\lstLm!nodes!),
lokalen Variablen (\lstLm!local!) und Zustandsvariablen
(\lstLm!state!).

Der Datenfluss wiederum besteht aus Definitionen lokaler Variablen
(\lstLm!definition!) und Zustandsübergängen
(\lstLm!transition!). Die Initialisierung von Zustandsvariablen
wird durch \lstLm!initial! eingeleitet. Weiterhin können
Vorbedingungen festgelegt werden (\lstLm!assertion!).
Letztere kommen in dem obigen Beispiel nicht vor.

Syntaktisch unterscheiden sich Zuweisungen an lokale und
Zustandsvariablen durch ein „'“ hinter der Variable. Dies soll
andeuten, dass der nächste Schritt eines \term[Strom]{Ströme}
definiert wird.

Ein Knoten besitzt einen Namen, Parameter (hier leer) und
Ausgaben. Benutzt werden kann ein Knoten mit der Syntax
\lstLm!(use N $e_1$ $\ldots$ $e_k$)!.
Wobei $e_k$ Ausdrücke sind, die für die jeweiligen Parameter eingesetzt
werden.

\begin{listing}[H]
\begin{lstlisting}[linerange={7-43}]
-- Notice that x is initialised with -1 instead of 0 here.
-- With that xo is 0 at the beginning and not 1.
nodes
  node UpDown () returns (xo : int) let
    local x_ : int;
    state x : int;
    definition xo = x_;
    transition x' = x_;

    automaton let
      location A let
        definition x_ = (+ x 1);
      tel
      location B let
        definition x_ = (- x 1);
      tel
      initial A;
      edge (A, B) : (= x 10);
      edge (B, A) : (= x 0);
      edge (A, A) : (not (= x 10));
      edge (B, B) : (not (= x 0));
    tel

    initial x = (- 1);
  tel
local x : int;
state x_1 : int;
definition x = (use UpDown);
transition x_1' = x;
initial x_1 = (- 1);

invariant (and
  -- ranges only from 0 to 10
  (and (>= x 0) (<= x 10))
  -- always moves just one step
  (or (= (- x_1 x) 1) (= (- x_1 x) (- 1)))
);
\end{lstlisting}
  \caption{UpDown-Knoten in \Lama}
  \label{lst:lama_updown}
\end{listing}

Automaten bestehen aus einer Menge von Modi
(\lstLm!location!)\footnote{Wir werden von Modi sprechen,
auch wenn in \Lama \lstLm!location! benutzt wird.
Dies stammt aus \lang{Nbac}, hat aber keine sinnvolle Übersetzung.
Außerdem ist bei Modus klarer, was gemeint ist.}
mit zugeordnetem Datenfluss, einem Initialzustand (\lstLm!initial!)
und Kanten zwischen je zwei Modi (\lstLm!edge!) mit einer
zugeordneten Bedingung. Außerdem kann in einem \lstLm!default!-Block
ein Standardverhalten für Variablen festgelegt werden (nicht in
diesem Beispiel). Dies greift dann, wenn eine Variable nicht in allen
Modi definiert worden ist.

Wir werden von \term[Fluss!globaler]{globalem Fluss} sprechen,
wenn es sich um
einen Fluss außerhalb von Automaten handelt, von
\term[Fluss!lokaler]{lokalem Fluss}
dagegen wenn dieser innerhalb eines Modus definiert wird.

Ein Programm hat auf der obersten Ebene noch einen globalen Block
\lstLm!constants!, in dem Bezeichner für Konstanten eingeführt
werden können.
Nach \lstLm!constants! können in dem Block \lstLm!input!
Parameter für das Programm deklariert werden.

Die Benutzung des Typsystems von \Lama ist in \lstRef{lama_enum_prod}
dargestellt. In der globalen Sektion \lstLm!typedef! können
Enumerations definiert werden. Jede Enumeration ist ein neuer Typ
mit benannten Konstanten dieses Typs. Um einen Enum-Typen zu
eliminieren, gibt es das \lstLm!match!-Konstrukt (Zeile 23-30).
\lstLm!match! benötigt als Argument einen Ausdruck und eine
Menge von Patterns. Ein Pattern hat die Form \lstLm!p.N!,
wobei \lstLm!p! eine Enum-Konstante oder \lstLm!_! und \lstLm!N!
ein Ausdruck ist. Wenn also ein Ausdruck
\lstLm!match M {$p_1.N_1, \dotsc, p_k.N_k$}! gegeben ist, dann
wird dieser zu einem der $N_i$ ausgewertet. Wobei dann $M = p_i$
oder $p_i = \_$ und $i$ minimal ist.

In Zeile 18 wird eine Variable $s\_$ mit einem Produkttypen deklariert.
Dabei wird der oben deklarierte Enum-Typ verwendet. Außerdem handelt
es sich um ein geschachteltes Produkt. Der Typ \lstLm!T^n! ist
eine Kurzschreibweise:
\[
\text{\lstLm!T^n!} = 
\text{\lstLm!(# $\, \underbrace{\text{T T} \ldots \text{T}}_{n-mal}$)!.}
\]
Um ein Produkt zu konstruieren wird die gleiche Syntax
verwendet (z.B. Zeile 8-9). Um ein Produkt zu zerlegen, gibt es eine
indizierte Projektion. Diese erwartet einen Bezeichner und eine
natürliche Zahl als Index.

Auf das Typsystem wird in \secRef{lama_types} näher eingegangen.

\begin{listing}[H]
  \lstset{numbers=left, firstnumber=last,xleftmargin=25pt}

\begin{lstlisting}[linerange={2-3}]
typedef
  enum E1 = {Fill, Sum};
\end{lstlisting}
  \hspace{25pt} $\vdots$
\begin{lstlisting}[linerange={7-15}]
nodes
  node Write1 (a : int^5, i : int, x : int)
  returns (b : int^5) let
    definition b =
      (ite (= i 0)
        (# x (project a 1) (project a 2)
          (project a 3) (project a 4)
        )
      (ite (= i 1)
\end{lstlisting}
  \hspace{25pt} $\vdots$
\begin{lstlisting}[linerange={27-32}]
      (ite (= i 4)
        (# (project a 0) (project a 1)
          (project a 2) (project a 3) x
        )
        a )))));
  tel
\end{lstlisting}
  \hspace{25pt} $\vdots$
\begin{lstlisting}[linerange={67-67,68-70}]
  s_ : (# E1 int int^5);
state
  s : (# E1 int int^5);
definition
\end{lstlisting}
  \hspace{25pt} $\vdots$
\begin{lstlisting}[linerange={76-76,78-85}]
  a1 = (use Write1 a i (+ 1 ai));
  s_ = (match (project s 0) {
          Fill. (#
                  (ite (= i l) Sum Fill)
                  i1 a1),
          Sum. (#
                (ite (= i l) Fill Sum)
                i1 a2)
        });
\end{lstlisting}

  \lstset{numbers=none, firstnumber=auto,xleftmargin=0pt}

  \caption{Enums und Produkte in \Lama}
  \label{lst:lama_enum_prod}
\end{listing}

\section{Statische Semantik}
\label{sec:lama_static_semantics}

In diesem Abschnitt werden wir die statische Semantik von \Lama
festlegen. Diese umfasst die Gültigkeitsbereiche von Variablen
und das Typsystem. Außerdem wird eine Prüfung der Abhängigkeiten von
Variablen eingeführt, da es eine strikte Ordnung der
Auswertungsreihenfolge von Zuweisungen geben muss.

\subsection{Gültigkeitsbereiche}
\label{sec:lama_scope}
Die Gültigkeitsbereiche sind sehr eng gefasst. Variablen sind
immer nur im umschließenden Block gültig (nicht in untergeordneten),
dasselbe gilt für Knoten. Die einzige Ausnahme bilden hier
Enums und Konstanten. Diese werden für das gesamte Programm deklariert.

\subsection{Typsystem}
\label{sec:lama_types}

\begin{figure}[h]
  \ExecuteMetaData[content/Grammar.tex]{types}
  \caption{Syntax des \Lama-Typsystems}
  \label{fig:syntax_lama_types}
\end{figure}

In \figRef{syntax_lama_types} ist die Syntax der \Lama-Typen
dargestellt. Es gibt eine Reihe von Basistypen und den Typkonstruktor
für Produkte. Außerdem können Namen von Enums als Typen verwendet
werden. Wir geben nun die Semantik der Typen an und danach die
wird die Prüfung der Typen beschrieben.

\begin{definition}[Typsemantik]
Im Folgenden bezeichnet $\llbracket T \rrbracket \Sigma$ die Semantik
des Typen $T$ in der Umgebung $\Sigma$. Diese enthält die deklarierten
Enums. 

\begin{align*}
  \llbracket \ty{bool} \rrbracket \Sigma & = \B \\
  \llbracket \ty{int} \rrbracket \Sigma & = \Z \\
  \llbracket \ty{real} \rrbracket \Sigma & = \R \\
  \llbracket \ty{sint}[n] \rrbracket \Sigma
  & = \{-2^{n-1}, \ldots, 2^{n-1}-1\} \\
  \llbracket \ty{uint}[n] \rrbracket \Sigma
  & = \{0, \ldots, 2^n-1\} \\
  \llbracket \ty{x} \rrbracket \Sigma & = \Sigma(x) \\
  \llbracket T^\wedge n \rrbracket \Sigma
  & = \llbracket (\# \underbrace{T \ldots T}_n) \rrbracket \Sigma \\
  \llbracket \ty{(\#} T_0 \ldots T_n) \rrbracket \Sigma
  & = \prod_{i=0}^n \left( \llbracket T_i \rrbracket \Sigma \right)
\end{align*}

Den Produkten werden jeweils Projektion
$\func{p_i}{\prod_{j=0}^n A_j}{A_i}$ zugeordnet.
\end{definition}

\begin{remark}
  Zwar wird $\ty{real}$ mit $\R$ interpretiert, aber ein
  \Lama-Programm kann zu jedem Zeitpunkt trotzdem nur rationale
  Zahlen berechnen.
\end{remark}

\subsubsection*{Typberechnung und -prüfung}
Für die Berechnung der Typen ergänzen wir die Typsyntax um $\ty{ok}$.
Dies ist ein Meta-Typ, der angibt, dass eine Deklaration
korrekt getypt worden ist. Außerdem werden ein Funktionstyp
und polymorphe Typen hinzugefügt. Dies vereinfacht die Berechnung
der Typen von Funktionen, die auf unterschiedliche Typen angewendet
werden können (z.B. Addition). Die neue Typsyntax ist in
\figRef{lama_intermediate_types} dargestellt.

\begin{figure}[h]
\begin{tabular}{lll}
  {\nonterminal{IntermediateT$_0$}} & {\arrow}
  & {\nonterminal{Type}} \\
  & {\delimit} & {\terminal{ok}}  \\
  & {\delimit} & {\terminal{x}} \\
  & {\delimit} & {\nonterminal{IntermediateT$_0$}}
  {\terminal{$\Rightarrow$}} {\nonterminal{IntermediateT$_0$}}
\end{tabular}\\

\begin{tabular}{lll}
  {\nonterminal{IntermediateT}} & {\arrow}
  & {\nonterminal{IntermediateT$_0$}} \\
  & {\delimit} & {\terminal{$\forall$}}
  {\terminal{x}} {\terminal{:}} {\nonterminal{Universe}} {\terminal{.}}
  {\nonterminal{IntermediateT}}
\end{tabular}\\

\begin{tabular}{lll}
  {\nonterminal{Universe}} & {\arrow}
  & {\terminal{Type}} \\
  & {\delimit} & {\terminal{Num}}
\end{tabular}\\
  
\caption{Zwischentypen der Typberechnung}
\label{fig:lama_intermediate_types}
\end{figure}

Dabei stellt \nonterminal{IntermediateT$_0$} die quantorenfreie
Ebene dar. Nur \nonterminal{IntermediateT} darf Quantoren einführen.
Das verhindert, dass man Typen erzeugen kann, die polymorphe
Argumente haben. Jeder Typvariable $x$ wird ein Universum zugeordnet,
aus der eine konkrete Belegung stammen darf. Es gibt zwei Universen:
$Num \subset Type$. Dabei enthält $Num$ alle numerischen Typen.

Wir wollen uns hier kurz einige Ableitungsregeln der Typberechnung
ansehen. Daran sollen die verwendeten Bezeichnungen erläutert werden.
Das vollständige Typsystem findet sich im Anhang
(\ref{sec:lama_type_rules}).

Bei der Typberechnung werden drei Umgebungen benötigt:
\begin{itemize}
\item $\Sigma$ ordnet einem Bezeichner einen Typen zu
  (Enum oder Basistyp)
\item $\Delta$ ordnet einem Bezeichner eine Konstante zu
\item $\Gamma$ ordnet einem Bezeichner eine deklarierte
  Variable oder einen Knoten zu
\end{itemize}

Der Aufbau der Typregeln ist ähnlich zu
\cite[S. 24]{CardelliTypeSystems}. Wir benutzen dabei die folgende
Notation. $\Sigma \vdash D \therefore \Delta$ bedeutet, dass in der
Umgebung $\Sigma$ die Deklaration $D$ wohlgeformt ist und die
Signatur $\Delta$ ergibt. Wir werden
$\Sigma,\Delta,\Gamma \vdash M : A$ schreiben, wenn $M$ in der
Umgebung $\Sigma \cup \Delta \cup \Gamma$ den Typ $A$ hat. Wenn
alle diese Umgebungen leer sind, schreiben wir auch $\vdash M : A$
bzw. $\vdash D \therefore \Delta$.

Zunächst zur Typisierung von \Lama-Programmen. Diese bestehen aus:
Typdeklarationen ($T$), Konstantendeklarationen ($C$),
Variablendeklarationen ($D$), einem Datenfluss ($F$),
einem Anfangszustand ($I$), einer Assertion ($A$)
und Invarianten ($P$). Diese finden sich alle in der folgenden
Regel wieder.

\ExecuteMetaData[content/Types.tex]{progrule}

Für Enumerations, Konstanten und Deklarationen wird der
Operator $\therefore$ verwendet. Daraus ergeben sich jeweils die
drei oben genannten Umgebungen. Diese werden dann verwendet,
um die Typen für die anderen Teile eines Programms abzuleiten.

Beispielhaft wollen wir uns die Deklaration von Typen (d.h. Enums)
ansehen.

\ExecuteMetaData[content/Types.tex]{typedefrules}

Jeder Enum-Typ definiert einen Typen und zugehörige Konstruktoren,
diese werden durch die $enum$-Regel erzeugt. Die Regel
$\tyRule{typedef}$ sammelt diese in $\Sigma$. Die Typisierung der
Konstanten und Deklarationen erfolgt analog.

Bei Flüssen, Initialisierungen und Vorbedingungen/Invarianten werden
jeweils die Typen der definierenden Ausdrücke berechnet und mit
dem erwarteten Typen zusammengeführt (Typ der definierten Variable
oder $\ty{bool}$). Dazu beispielhaft die Regel für Invarianten:

\ExecuteMetaData[content/Types.tex]{invariantrule}

Das polymorphe Typsystem wird bei der Berechnung der Typen von
Ausdrücken genutzt.
Zunächst aber zu einfachen Funktionen. \lstLm!div! und
\lstLm!mod! sind Funktionen, die nur auf ganze Zahlen
angewendet werden können und jeweils zwei Parameter erwarten.
Das sagt die folgende Regel $\tyRule{int-arith}$ aus.

\ExecuteMetaData[content/Types.tex]{intarith}

Um eine solche Funktion nun verwenden zu können, gibt es eine
Regel zur Funktionsapplikation. Man beachte dabei, dass die
Funktionsapplikation links-assoziativ ist:
$(M N_1 ... N_n) = (((M N_1) N_2) ... N_n)$.

\ExecuteMetaData[content/Types.tex]{funapp}

Interessanter sind die Regeln zur Behandlung polymorpher Funktionen,
z.B. für Arithmetik:

\ExecuteMetaData[content/Types.tex]{arith}

Diese Regel sagt aus, dass $+, -, *$ auf jeden Typen angewendet
werden kann, der im Universum $Num$ liegt und jeweils zwei Argumente
vom diesem Typen erwarten. Offensichtlich kann hier die Regel
$\tyRule{app}$ nicht direkt angewendet werden, da kein Funktionstyp,
sondern ein quantifizierter Typ vorliegt. Um dies zu ermöglichen
existiert eine Regel zur Eliminierung von $\forall$:

\ExecuteMetaData[content/Types.tex]{forallelim}

Diese sagt aus, dass immer, wenn wir einen quantifizierten Typen
haben, dessen Typvariable $t$ in dem Universum $U$ liegt und
wir weiterhin einen Typen $T \in U$ haben, kann $t$ durch
$T$ ersetzt werden.
Diese Regel muss nun implizit überall da angewendet werden, wo ein
konkreter Typ erwartet wird, aber ein quantifizierter
vorhanden ist.

Wird nun ein Typ aus dem Universum $Type$ erwartet, ist es nicht
möglich, direkt einen Typen aus dem Universum $Num$ einzusetzen
(z.B. \lstLm!ite! mit einem \lstLm!int!-Argument).
Um eine Einsetzung aber doch zu ermöglichen, führen wir eine
Regel zur Generalisierung von Universen ein. Da wir nur zwei
Universen mit einer einzigen Inklusion $Num \hookrightarrow Type$
haben, ist die einzige Regel hierfür:

\ExecuteMetaData[content/Types.tex]{univgen}

\pagebreak
Der Typ den ein Knoten
\begin{lstlisting}
node N ($x_1 : T_1, \dotsc, x_n : T_n$)
  returns ($y_1 : R_1, \dotsc, y_m : R_m$)
  let $\ldots$ tel
\end{lstlisting}
bei der Typberechnung zugewiesen bekommt, ist
$(\# \; T_1 \dotsc T_n) \Rightarrow (\# \; R_1 \dotsc R_m)$.
Dass ein Aufruf \lstLm!(use N M)! nicht frei von Seiteneffekten ist,
ist somit aber auch nicht in dem Typ enthalten. Daher
wird diese Benutzung bereits syntaktisch unterschieden. Dies
vereinfacht aber das Typsystem und die Semantik.

\begin{example}
Wir wollen uns die Ableitung des Typs für den Ausdruck
\begin{lstlisting}
  (ite a (+ x 1) (- x 1))
\end{lstlisting}
ansehen. Dabei ist $\Gamma = \{a : \ty{bool}, x : \ty{int} \}$,
$\Sigma$ und $\Delta$ sind leer.

Der Beweis erfolgt in mehreren Schritten. Zunächst zeigen wir, dass
$\vdash \text{\lstLm!-!} : \ty{int} \Rightarrow \ty{int}
  \Rightarrow \ty{int}$ gilt:

\begin{equation}
  \AxiomC{}
  \LeftLabel{(arith)}
  \UnaryInfC{$\vdash \text{\lstLm!-!}
    : \forall t : Num. \; t  \Rightarrow t \Rightarrow t$}
  \AxiomC{}
  \RightLabel{(num-univ)}
  \UnaryInfC{$\vdash \ty{int} : Num$}
\LeftLabel{($\forall$-E)}
\BinaryInfC{$\vdash \text{\lstLm!-!}
  : \ty{int} \Rightarrow \ty{int} \Rightarrow \ty{int}$}
\DisplayProof
\label{proof:minus_int}
\end{equation}

Dies wird genutzt, um 
$\Gamma \vdash \text{\lstLm!(- x 1)!} : \ty{int}$ zu zeigen.

\begin{equation}
    \AxiomC{}
    \LeftLabel{(\ref{proof:minus_int})}
    \UnaryInfC{$\vdash \text{\lstLm!-!}
      : \ty{int} \Rightarrow \ty{int} \Rightarrow \ty{int}$}
      \AxiomC{$\Gamma(x) = \ty{int}$}
    \UnaryInfC{$\Gamma \vdash
      \text{\lstLm!x!} : \ty{int}$}
  \LeftLabel{(app)}
  \BinaryInfC{$\Gamma \vdash
    \text{\lstLm!(- x)!} : \ty{int} \Rightarrow \ty{int}$}
  \AxiomC{$\!\!\!\! \Gamma \vdash
    \text{\lstLm!1!} : \ty{int}$}
\LeftLabel{(app)}
\BinaryInfC{$\Gamma \vdash
  \text{\lstLm!(- x 1)!} : \ty{int}$}
\DisplayProof
\label{proof:x_minus_1}
\end{equation}

Um \lstLm!ite! mit $\ty{int}$ instantiieren zu können, muss die
Regel \textit{univ-gen} angewendet werden:

\begin{equation*}
  \AxiomC{$\vdash \text{\lstLm!ite!} : \forall t:Type. \;
    \ty{bool} \Rightarrow t \Rightarrow t \Rightarrow t$}
  \AxiomC{$\vdash \ty{int} : Num$}
  \RightLabel{(univ-gen)}
  \UnaryInfC{$\vdash \ty{int} : Type$}
\RightLabel{($\forall$-E)}
\BinaryInfC{$\vdash \text{\lstLm!ite!}
  : \ty{bool} \Rightarrow \ty{int}
  \Rightarrow \ty{int} \Rightarrow \ty{int}$}
\DisplayProof
\end{equation*}

Das benutzen wir, um zu zeigen, dass $\Gamma \vdash
\text{\lstLm!(ite a (+ x 1))!} : \ty{int} \Rightarrow \ty{int}$
gilt. Wobei $\Gamma \vdash \text{\lstLm!(+ x 1)!} : \ty{int}$
ganz analog wie oben abgeleitet wird.

\begin{equation}
  \AxiomC{$\vdash \text{\lstLm!ite!}
    : \ty{bool} \Rightarrow \ty{int}
    \Rightarrow \ty{int} \Rightarrow \ty{int}$}
  \AxiomC{$\Gamma(a) = \ty{bool}$}
  \UnaryInfC{$\Gamma \vdash a : \ty{bool}$}
\BinaryInfC{$\Gamma \vdash \text{\lstLm!(ite a)!}
  : \ty{int} \Rightarrow \ty{int} \Rightarrow \ty{int}$}
\AxiomC{$\Gamma \vdash
  \text{\lstLm!(+ x 1)!} : \ty{int}$}
\BinaryInfC{$\Gamma \vdash
  \text{\lstLm!(ite a (+ x 1))!} : \ty{int} \Rightarrow \ty{int}$}
\DisplayProof
\label{proof:long_ite}
\end{equation}

Und zuletzt setzen wir die beiden Beweise zusammen:

\begin{equation*}
\AxiomC{}
\LeftLabel{(\ref{proof:long_ite})}
\UnaryInfC{$\Gamma \vdash
  \text{\lstLm!(ite a (+ x 1))!} : \ty{int} \Rightarrow \ty{int}$}
\AxiomC{}
\RightLabel{(\ref{proof:x_minus_1})}
\UnaryInfC{$\Gamma \vdash
  \text{\lstLm!(- x 1)!} : \ty{int}$}
\BinaryInfC{$\Gamma \vdash
  \text{\lstLm!(ite a (+ x 1) (- x 1))!} : \ty{int}$}
\DisplayProof
\end{equation*}  
\end{example}

\begin{example}
  Dem folgenden Programmausschnitt soll ein Typ gegeben werden.
  \begin{lstlisting}
typedef
  enum E = {E1, E2};
constants
  n = 5;
$\vdots$
  x = (match y {E1.n, E2.4});
$\vdots$
  \end{lstlisting}

  Zunächst beginnen wir mit den Definitionen:
  \begin{prooftree}
    \AxiomC{}
    \UnaryInfC{$\vdash \syn{enum } E \syn{ = \{} E1, E2 \syn{\}}
      \therefore \Sigma = \{(E1, E) (E2, E)\}$}
    \UnaryInfC{$\vdash \syn{typedef enum E = \{E1, E2\}; }
      \therefore \Sigma$}
  \end{prooftree}

  \begin{prooftree}
    \AxiomC{$\Sigma \vdash 5 : \ty{int}$}
    \UnaryInfC{$\Sigma \vdash \syn{constants n  = 5; }
      \therefore \Delta = \{(n, \ty{int})\}$}
  \end{prooftree}

  Um der Zuweisung einen Typ zu geben, nehmen wir
  $\Gamma = \{x : \ty{int}, y : E\}$ an.
  Für den Ausdruck ergibt sich dann (die Regel für $\syn{match}$
  findet sich in \secRef{lama_type_rules}):
  \begin{equation}
      \AxiomC{$\Gamma(y) = E$}
    \UnaryInfC{$\Sigma,\Delta,\Gamma \vdash y : E$}
      \AxiomC{$\Sigma(E1) = E, \Sigma(E1) = E$}
    \UnaryInfC{$\Sigma \vdash E1 : E, E1 : E$}
      \AxiomC{$\Delta(n) = \ty{int}, 4 : \ty{int}$}
    \UnaryInfC{$\Delta \vdash n, 4 : \ty{int}$}
    \TrinaryInfC{$\Sigma,\Delta,\Gamma \vdash
      \syn{(match y \{E1.n, E2.4\})} : \ty{int}$}
    \DisplayProof
  \label{eq:proof_match_y}
  \end{equation}

  Und damit die Zuweisung:
  \begin{prooftree}
      \AxiomC{$\Gamma(x) = \ty{int}$}
    \UnaryInfC{$\Sigma,\Delta,\Gamma \vdash x : \ty{int}$}
      \AxiomC{}
    \RightLabel{\eqRef{proof_match_y}}
    \UnaryInfC{$\Sigma,\Delta,\Gamma \vdash
      \syn{(match y \{E1.n, E2.4\})} : \ty{int}$}
    \BinaryInfC{$\Sigma,\Delta,\Gamma \vdash
      \syn{x = (match y \{E1.n, E2.4\})} : \ty{ok}$}
  \end{prooftree}
\end{example}

\subsection{Abhängigkeiten}
\label{sec:lama_dependencies}
Die zweite statische Prüfung, der ein \Lama-Programm unterzogen
wird, ist eine Analyse der Abhängigkeiten der Variablen. Diese
definiert eine Ordnung der Variablen, in der deren Definitionen
ausgewertet werden müssen. Diese Ordnung kann in der Implementierung
eines Simulators zur Auswertung genutzt werden.

Die Berechnung der Abhängigkeiten erfolgt für jeden Knoten einzeln
und abschließend für den äußersten Fluss des Programms.
Dabei wird ein Graph aufgebaut, der für jedes Setzen und
Lesen einer Variable einen Knoten enthält. Einem solchen Knoten
wird ein Kontext zugeordnet, in dem eine Variable gesetzt wurde
(globaler bzw. lokaler Fluss) und um was für eine Art von Variable
es sich handelt (Knotenparameter, lokale Variable, etc.).

Bei der Analyse einer Definition der Form $x = M$ werden Kanten
von $x$ (in dem aktuellen Kontext) zu allen Variablen eingefügt,
die in $M$ verwendet werden. Wenn sich diese Definition außerdem
im Kontext eines Modus befindet, werden Kanten von $x$ zu allen
Variablen eingefügt, die in Bedingungen an Modus-Kanten
benutzt werden. Der entstehende Graph muss ein einfacher zyklenfreier
gerichteter Graph sein. Zyklische Definitionen sind unzulässig.
Dieser induziert eine strikte Ordnung der Definitionen (transitiv
und asymmetrisch).


Im Folgenden sei $P$ ein \Lama-Programm und
\begin{itemize}
\item $X$ -- deklarierte Variablen im aktuellen Block
\item $\Usage = \{ I, O, L, SIn, SOut \}$
  (Input, Output, lokal, Zustand dieser/nächster Takt)
\item $\Mode = \{ Global \} \cup
  (\bigcup_{A \text{ Automat in P}}
    \setDef{(A, m)}{m \in A \text{ Modus}})$.
\end{itemize}

\algRef{lama_dep_analysis} wird für den globalen Programmfluss
und alle Knoten von $P$ ausgeführt. Der entstehende Graph besitzt
als Knoten die jeweiligen Variablen mit ihrem Kontext, in dem
sie benutzt werden. Dieser wird benötigt, um die Definition einer
Variablen in unterschiedlichen Modi und als Abhängigkeit die
Herkunft unterscheiden zu können.

$\Usage$ gibt dabei an, ob eine Variable schreibbar/lesbar ist
(\mathlist{O, L, SOut \text{ bzw. } I, L, SIn}). Außerdem muss
unterschieden werden, ob eine Zustandsvariable gelesen oder
geschrieben wird. Sonst würde ein Zyklus entstehen,
wenn beispielsweise die Zuweisung $x' = x$ gegeben wäre.

$\Mode$ gibt an, ob eine Variable in einem Modus eines Automaten
definiert wird und wenn ja, in welchem. Dadurch können die
einzelnen Definitionen in verschiedenen Modi unterschieden werden.

\begin{algorithm}[H]
\caption{Berechnung der \Lama-Abhängigkeiten}
\label{alg:lama_dep_analysis}
\begin{algorithmic}[1]
  \REQUIRE $F$ Fluss, $\mathcal{A}$ Automaten
  \ENSURE $G = (V, E)$ gerichteter Graph wobei
    $V \subset X \times \Mode \times \Usage$
  \STATE $G \colonequals (\emptyset, \emptyset)$
  \STATE $G \colonequals flowDep(G, Global, F)$
  \FOR {$A \in \mathcal{A}$}
    \FOR {$(m, F') \in A$}
      \STATE $G \colonequals flowDep(G, (A, m), F')$
    \ENDFOR
  \ENDFOR
\end{algorithmic}
\end{algorithm}

In \algRef{lama_flow_dep} benutzen wir die folgenden Graphoperationen:
\begin{itemize}
\item $G + v$ -- Hinzufügen eines Knotens, wenn noch nicht in
  $G$ vorhanden
\item $G + e$ -- Hinzufügen einer Kante und der inzidenten Knoten,
  sofern noch nicht in $G$ vorhanden
\end{itemize}

\begin{algorithm}[H]
\caption{$flowDep$ -- Berechnung der Abhängigkeiten eines Flusses}
\label{alg:lama_flow_dep}
\begin{algorithmic}[1]
  \REQUIRE $G$ gerichteter Graph, $m \in \Mode$, $F$ Fluss
  \ENSURE $G'$ gerichteter Graph
  \STATE $G' \colonequals G$
  \FOR[lokale Definitionen] {$x = M \in F$}
    \STATE $u \colonequals
      \begin{cases}
        L & \text{, x lokale Variable} \\
        O & \text{, x Rückgabevariable}
      \end{cases}$
    \STATE $v \colonequals (x, m, u)$
    \STATE $G' \colonequals G' + v$
    \IF {$m \not= Global$}
      \STATE $G' \colonequals G' + ((x, Global, u), v)$
      \COMMENT{Kante zu globalem  $x$ einfügen}
      \label{flow_dep_non_global1}
    \ENDIF
    \STATE $G'
      \colonequals G' + (\{v\} \times (deps(M) \times \{Global\}))$
      \COMMENT{Kanten zu Abhängigkeiten}
  \ENDFOR
  \FOR[Zustandsübergänge] {$x' = M \in F$}
    \STATE $v \colonequals (x, m, SOut)$
    \STATE $G' \colonequals G' + v$
    \IF {$m \not= Global$}
      \STATE $G' \colonequals G' + ((x, Global, SOut), v)$
      \COMMENT{Kante zu globalem $x$ einfügen}
      \label{flow_dep_non_global2}
    \ENDIF
    \STATE $G'
      \colonequals G' + (\{v\} \times (deps(M) \times \{Global\}))$
      \COMMENT{Kanten zu Abhängigkeiten}
  \ENDFOR
\end{algorithmic}
\end{algorithm}

Dabei bestimmt
$\func{deps}{Expr}{X \times (\Usage \setminus \{O, SOut\}) }$
alle benutzten Variablen in einem Ausdruck.
$flowDep$ fügt die Abhängigkeiten eines Flusses ein
(d.h. für lokale Definitionen und Zustandsübergänge). Dabei kann
ein Modus vorgegeben werden. Ist dieser nicht global (d.h. ein
Modus eines Automaten), wird zusätzlich ein Referenzknoten eingefügt,
der die Abhängigkeiten der Modi zusammenfasst
(Zeile \ref{flow_dep_non_global1} und \ref{flow_dep_non_global2}).
Über diesen kann eine Variable $x$ als Abhängigkeit angegeben werden.

Danach werden Kanten zu allen Abhängigkeiten des definierenden Terms
hinzugefügt. Hier kommt der obige Referenzknoten zum Tragen: wir
müssen lediglich eine Abhängigkeit zum globalen Modus einfügen.

Nach Berechnung des Abhängigkeitsgraphen, muss noch geprüft werden, ob
\begin{itemize}
\item alle Rückgabevariablen eine Definition besitzen.
\item in jedem Modus eine Definition vorliegt, sofern die
  Variable nicht global definiert wurde.
\item eine Variable in höchstens einem Automaten definiert wurde.
\end{itemize}

Ein solches Programm bezeichnen wir (im Sinne der Systemtheorie)
als \term{kausal}.

\begin{example}
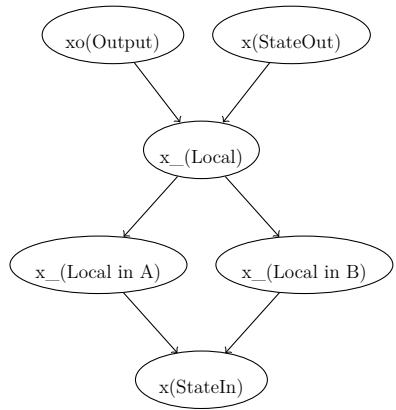
\begin{figure}[h]
  \centering
  \begin{tikzpicture}[scale=0.6, every node/.style={scale=0.6},baseline=(current bounding box.center)]
    \pgfsetcolor{black}
  \draw [->] (70.493bp,72.571bp) .. controls (78.549bp,63.507bp) and (88.528bp,52.281bp)  .. (104.09bp,34.773bp);
  \draw [->] (167.51bp,72.571bp) .. controls (159.45bp,63.507bp) and (149.47bp,52.281bp)  .. (133.91bp,34.773bp);
  \draw [->] (161.44bp,216.57bp) .. controls (154.63bp,207.82bp) and (146.25bp,197.04bp)  .. (132.43bp,179.27bp);
  \draw [->] (133.84bp,145.3bp) .. controls (141.77bp,136.38bp) and (151.69bp,125.23bp)  .. (167.34bp,107.61bp);
  \draw [->] (77.314bp,216.57bp) .. controls (84.002bp,207.82bp) and (92.232bp,197.04bp)  .. (105.81bp,179.27bp);
  \draw [->] (104.16bp,145.3bp) .. controls (96.231bp,136.38bp) and (86.312bp,125.23bp)  .. (70.656bp,107.61bp);
\begin{scope}
  \definecolor{strokecol}{rgb}{0.0,0.0,0.0};
  \pgfsetstrokecolor{strokecol}
  \draw (64bp,234bp) ellipse (44bp and 18bp);
  \draw (64bp,229bp) node {xo(Output)};
\end{scope}
\begin{scope}
  \definecolor{strokecol}{rgb}{0.0,0.0,0.0};
  \pgfsetstrokecolor{strokecol}
  \draw (175bp,234bp) ellipse (49bp and 18bp);
  \draw (175bp,229bp) node {x(StateOut)};
\end{scope}
\begin{scope}
  \definecolor{strokecol}{rgb}{0.0,0.0,0.0};
  \pgfsetstrokecolor{strokecol}
  \draw (119bp,162bp) ellipse (36bp and 18bp);
  \draw (119bp,157bp) node {x\_(Local)};
\end{scope}
\begin{scope}
  \definecolor{strokecol}{rgb}{0.0,0.0,0.0};
  \pgfsetstrokecolor{strokecol}
  \draw (119bp,18bp) ellipse (41bp and 18bp);
  \draw (119bp,13bp) node {x(StateIn)};
\end{scope}
\begin{scope}
  \definecolor{strokecol}{rgb}{0.0,0.0,0.0};
  \pgfsetstrokecolor{strokecol}
  \draw (55bp,90bp) ellipse (55bp and 18bp);
  \draw (55bp,85bp) node {x\_(Local in A)};
\end{scope}
\begin{scope}
  \definecolor{strokecol}{rgb}{0.0,0.0,0.0};
  \pgfsetstrokecolor{strokecol}
  \draw (183bp,90bp) ellipse (55bp and 18bp);
  \draw (183bp,85bp) node {x\_(Local in B)};
\end{scope}
  \end{tikzpicture}
  \caption{Abhängigkeitsgraph für UpDownCounter-Knoten}
  \label{fig:dep_graph_updown}
\end{figure}
In \figRef{dep_graph_updown} ist der Abhängigkeitsgraph
für den Knoten \lstLm!UpDown! aus \lstRef{lama_updown} dargestellt.
Hier sind allerdings die Elemente von $\Usage$ ausgeschrieben.

\lstLm!x_! liegt wie beschrieben in drei Varianten vor: global
als Referenz bei der Benutzung (keine weitere Kennzeichnung)
und für die beiden Zustände \lstLm!A! und \lstLm!B!.

Es liegt ein DAG vor und \lstLm!xo! und \lstLm!x'! haben eine
Definition. Also ist das Programm kausal.
\end{example}

\section{Dynamische Semantik}
\label{sec:lama_dynamic_semantics}

Wir gehen im Folgenden davon aus, dass ein \Lama-Programm lediglich
aus Knoten besteht, die nur einen einzigen Automaten und keinen
weiteren Fluss enthalten. Es ist offensichtlich möglich, jedes
Programm in so eine Form zu bringen, indem der globale Fluss
in einem einzelnen Modus definiert wird und dann das
Produkt aller Automaten in dem Knoten gebildet wird. Zusätzlich
müssen lokale Variablen, die in Bedingungen an Kanten
verwendet werden, durch ihre Definition ersetzt werden.
Ansonsten würde die Bedingung einer Kante von dem Fluss innerhalb
von Zuständen abhängen und das wäre ein Abhängigkeitsfehler.

\begin{example}
  In \figRef{lama_trans_deps_error} ist ein Beispiel für
  eine solche fehlerhafte Transformation dargestellt.
  \begin{figure}[H]
    \centering
    \begin{tikzpicture}[anchor=base,scale=0.7, every node/.style={scale=0.7},baseline=(current bounding box.center)]
      \pgfsetcolor{black}
  \draw [->] (54.038bp,72bp) .. controls (64.342bp,72bp) and (76.28bp,72bp)  .. node[auto] {$c$} (97.71bp,72bp);
\begin{scope}
  \definecolor{strokecol}{rgb}{0.0,0.0,0.0};
  \pgfsetstrokecolor{strokecol}
  \draw (125bp,72bp) ellipse (27bp and 18bp);
  \draw (125bp,67bp) node {$s_1$};
\end{scope}
\begin{scope}
  \definecolor{strokecol}{rgb}{0.0,0.0,0.0};
  \pgfsetstrokecolor{strokecol}
  \draw (27bp,72bp) ellipse (27bp and 18bp);
  \draw (27bp,67bp) node {$s_1$};
\end{scope}
\begin{scope}
  \definecolor{strokecol}{rgb}{0.0,0.0,0.0};
  \pgfsetstrokecolor{strokecol}
  \draw (27bp,13bp) node {$c = M$};
\end{scope}
    \end{tikzpicture}
    $\leadsto$
    \begin{tikzpicture}[anchor=base,scale=0.7, every node/.style={scale=0.7},baseline=(current bounding box.center)]
      \pgfsetcolor{black}
  \draw [->] (54.038bp,74bp) .. controls (64.342bp,74bp) and (76.28bp,74bp)  .. node[auto] {$c$} (97.71bp,74bp);
\begin{scope}
  \definecolor{strokecol}{rgb}{0.0,0.0,0.0};
  \pgfsetstrokecolor{strokecol}
  \draw (125bp,74bp) ellipse (27bp and 18bp);
  \draw (125bp,69bp) node {$s_1$};
\end{scope}
\begin{scope}
  \definecolor{strokecol}{rgb}{0.0,0.0,0.0};
  \pgfsetstrokecolor{strokecol}
  \draw (27bp,74bp) ellipse (27bp and 18bp);
  \draw (27bp,69bp) node {$s_1$};
\end{scope}
\begin{scope}
  \definecolor{strokecol}{rgb}{0.0,0.0,0.0};
  \pgfsetstrokecolor{strokecol}
  \draw (12bp,0bp) -- (42bp,0bp);
  \draw (42bp,0bp) .. controls (48bp,0bp) and (54bp,6bp)  .. (54bp,12bp);
  \draw (54bp,12bp) -- (54bp,26bp);
  \draw (54bp,26bp) .. controls (54bp,32bp) and (48bp,38bp)  .. (42bp,38bp);
  \draw (42bp,38bp) -- (12bp,38bp);
  \draw (12bp,38bp) .. controls (6bp,38bp) and (0bp,32bp)  .. (0bp,26bp);
  \draw (0bp,26bp) -- (0bp,12bp);
  \draw (0bp,12bp) .. controls (0bp,6bp) and (6bp,0bp)  .. (12bp,0bp);
  \draw (0bp,19bp) -- (54bp,19bp);
  \draw (27bp,23bp) node {$g$};
  \draw (27bp,4bp) node {$c = M$};
\end{scope}
    \end{tikzpicture}
    $\leadsto$
    \begin{tikzpicture}[anchor=base,scale=0.7, every node/.style={scale=0.7},baseline=(current bounding box.center)]
      \pgfsetcolor{black}
  \draw [->] (83.558bp,19bp) .. controls (94.512bp,19bp) and (106.47bp,19bp)  .. node[auto] {$c$} (128.21bp,19bp);
\begin{scope}
  \definecolor{strokecol}{rgb}{0.0,0.0,0.0};
  \pgfsetstrokecolor{strokecol}
  \draw (13bp,0bp) -- (72bp,0bp);
  \draw (72bp,0bp) .. controls (78bp,0bp) and (84bp,6bp)  .. (84bp,12bp);
  \draw (84bp,12bp) -- (84bp,26bp);
  \draw (84bp,26bp) .. controls (84bp,32bp) and (78bp,38bp)  .. (72bp,38bp);
  \draw (72bp,38bp) -- (13bp,38bp);
  \draw (13bp,38bp) .. controls (7bp,38bp) and (1bp,32bp)  .. (1bp,26bp);
  \draw (1bp,26bp) -- (1bp,12bp);
  \draw (1bp,12bp) .. controls (1bp,6bp) and (7bp,0bp)  .. (13bp,0bp);
  \draw (1bp,19bp) -- (84bp,19bp);
  \draw (42bp,23bp) node {$g \times s_1$};
  \draw (42bp,4bp) node {$c = M$};
\end{scope}
\begin{scope}
  \definecolor{strokecol}{rgb}{0.0,0.0,0.0};
  \pgfsetstrokecolor{strokecol}
  \draw (141bp,0bp) -- (200bp,0bp);
  \draw (200bp,0bp) .. controls (206bp,0bp) and (212bp,6bp)  .. (212bp,12bp);
  \draw (212bp,12bp) -- (212bp,26bp);
  \draw (212bp,26bp) .. controls (212bp,32bp) and (206bp,38bp)  .. (200bp,38bp);
  \draw (200bp,38bp) -- (141bp,38bp);
  \draw (141bp,38bp) .. controls (135bp,38bp) and (129bp,32bp)  .. (129bp,26bp);
  \draw (129bp,26bp) -- (129bp,12bp);
  \draw (129bp,12bp) .. controls (129bp,6bp) and (135bp,0bp)  .. (141bp,0bp);
  \draw (129bp,19bp) -- (212bp,19bp);
  \draw (170bp,23bp) node {$g \times s_2$};
  \draw (170bp,4bp) node {$c = M$};
\end{scope}
    \end{tikzpicture}
    \caption{Entstehung eines Abhängigkeitsfehler bei der
      Transformation eines \Lama-Programms in einen Automaten}
    \label{fig:lama_trans_deps_error}
  \end{figure}
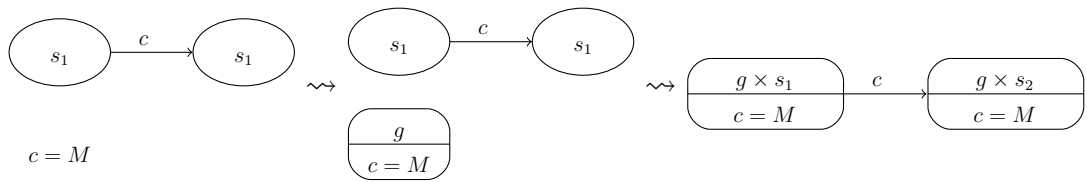

  Korrekt wäre der Automat in \figRef{lama_trans_correct}.
  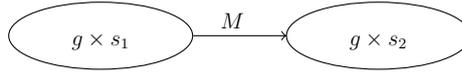
\begin{figure}[h]
    \centering
    \begin{tikzpicture}[anchor=base,scale=0.7, every node/.style={scale=0.7},baseline=(current bounding box.center)]
      \pgfsetcolor{black}
  \draw [->] (98.282bp,18bp) .. controls (110.89bp,18bp) and (124.58bp,18bp)  .. node[auto] {$M$} (147.93bp,18bp);
\begin{scope}
  \definecolor{strokecol}{rgb}{0.0,0.0,0.0};
  \pgfsetstrokecolor{strokecol}
  \draw (49bp,18bp) ellipse (49bp and 18bp);
  \draw (49bp,13bp) node {$g \times s_1$};
\end{scope}
\begin{scope}
  \definecolor{strokecol}{rgb}{0.0,0.0,0.0};
  \pgfsetstrokecolor{strokecol}
  \draw (197bp,18bp) ellipse (49bp and 18bp);
  \draw (197bp,13bp) node {$g \times s_2$};
\end{scope}
    \end{tikzpicture}
    \caption{Korrekt transformierter Automat}
    \label{fig:lama_trans_correct}
  \end{figure}
\end{example}

Weiterhin wird erwartet, dass die Definitionen so sortiert sind,
dass syntaktisch eine Definition erst nach der Definition all
ihrer Abhängigkeiten auftritt. Dies ist für ein gültiges
\Lama-Programm möglich (\ref{sec:lama_dependencies}).

Wir werden hier die Semantik von Flüssen, Knoten und Automaten
informell beschreiben.

Wir werden die Semantik von \Lama auf zwei Arten kennenlernen.
In diesem Abschnitt werden wir ein System beschreiben, dass die Form
\begin{equation}
  \label{eq:symbolic_trans_system}
  \begin{cases}
    I(\vect{s},q), & (\vect{s},q) \in S \times Q \\
    A(\vect{s},q, \vect{x}) \longrightarrow
      (\vect{s'},q',\vect{y}) = f(\vect{s},q,\vect{x}),
    & (\vect{s},q), (\vect{s'},q') \in S \times Q,
      \vect{x} \in X, \vect{y} \in Y
  \end{cases}  
\end{equation}
hat. Dabei ist $I$ ein Prädikat, dass den Anfangszustand festlegt,
$A$ ist eine Vorbedingung („assertion“) und $f$ ist ein
Zustandsübergang mit Ausgabe. Die genaue Bedeutung werden wir
im Folgenden festlegen.

\begin{definition}
  $S$ sei der \term{Zustandsraum} mit
  \[ S = \prod_{i=1, \dotsc, n_S} A_i \]
  wobei $A_i$ ein \Lama-Typ und $n_s$ die Zahl der Zustandsvariablen
  ist. Analog sind $X$ (\term{Eingaberaum}) und $Y$
  (\term{Ausgaberaum}) definiert (mit $n_X$ bzw. $n_Y$ der Zahl
  der Eingaben/Ausgaben.

  $Q$ ist das Produkt über die Modi der einzelnen Automaten
  \[
    Q = \prod_{A \text{ Automat in P}}
      \setDef{m_A}{m_A \text{ Modus in } A}.
  \]
\end{definition}

\begin{definition}
  In der Situation von \eqRef{symbolic_trans_system} definieren
  wir einen \term{Lauf} eines \Lama-Programms als ein
  (potentiell unendlich langes) Wort über
  $S \times Q \times X \times Y$:
  \begin{equation}
    t = (\vect{s_0},q_0,\vect{x_0},\vect{y_0})
      (\vect{s_1},q_1,\vect{x_1},\vect{y_1})
      \ldots \in (S \times Q \times X \times Y)^\infty.
  \end{equation}
  
  Dabei muss $I(\vect{s_0},q_0)$ und für alle
  $i \geq 0$ muss $A(\vect{s_i},q_i,\vect{x_i})$ und
  $(\vect{s_{i+1}},q_{i+1},\vect{y_{i}})
    = f(\vect{s_i},q_i,\vect{x_i})$ gelten.

  Wenn $t$ endliche Länge $k > 0$ hat, dann ist $k-1$ der
  minimale Index, so dass $\neg A(s_{k-1},x_{k-1})$.
\end{definition}

Die zweite, aber sehr eng verwandte, Möglichkeit ist, ein Programm
durch eine Stromtransformation zu beschreiben, d.h. durch eine
Funktion $\func{F}{X^\N}{Y^\N}$. Diese lässt sich aus den oben
beschriebenen Läufen erzeugen (wir gehen hier nicht auf Details ein).
Diese Art der Semantik werden wir in der Transformation in ein
SMT-Modell benutzen.

\subsubsection{Fluss}
\label{sec:lama_flow_semantics}

Ein \term{Fluss} unterteilt sich in zwei Teile: lokale Definitionen
und Zustandsübergänge.

Bei lokalen Definitionen werden wiederum drei Arten unterschieden:
\begin{enumerate}
\item Definition von nicht-Ausgabevariablen durch Ausdrücke
\item Definition von Ausgabevariablen
\item Definition durch die Benutzung eines Knotens.
\end{enumerate}

Ausdrücke sind frei von Seiteneffekten und können für
die jeweiligen Variablen substituiert werden. Das heißt, die erste
Art ist nicht außerhalb des Programms sichtbar.

Handelt es sich im zweiten Fall um Ausgabevariablen eines Knotens,
ist dies ebenfalls nicht außerhalb des Programms sichtbar
(s. Semantik von Knoten). Im Falle von Ausgabevariablen des Programms,
sind diese Variablen sehr wohl nach außen sichtbar und liegen
damit in $Y$.

Die dritte Art allerdings bewirkt eine Änderung
des Zustandes eines Knotens. Hier kann also keine Substitution
vorgenommen werden (siehe bei der Semantik von Knoten).

Zustandsübergänge sind nach außen hin sichtbar ($S$).
Ein solcher definiert (was die Syntax andeuten soll) $s' \in S$.
Es wird also die rechte Seite ausgewertet und abschließend an
$s'$ zugewiesen. Die Definition von $I(s)$ ist nicht notwendig,
sofern $s$ von $f$ im ersten Schritt nicht benutzt wird.

$f$ setzt sich also zum einen aus Ausgaben des Programms und
zum anderen aus Zustandsübergängen zusammen.

\subsubsection{Automaten}
\label{sec:lama_automaton_semantics}

Die Automaten in \Lama sind von der Idee her Mode-Automaten
\cite{ModeAutomata}.
Das bedeutet, dass sich ein Automat immer in genau einem Modus
befindet. Dieser bestimmt den ausgeführten Datenfluss. Die
Ausdrücke in allen anderen Modi werden \emph{nicht} ausgewertet.

Damit unterscheiden sich Modi signifikant von
\lstLm!ite! und \lstLm!match!. Letztere werten
immer alle beteiligten Ausdrücke aus. Das bedeutet, dass Automaten
durch diese Konstrukte nicht vollständig simuliert werden können.

\begin{example}
  Die Zuweisung
  \begin{lstlisting}
  x = (ite (= y 0) 1 (/ z y));
  \end{lstlisting}
  verursacht einen Laufzeitfehler, wenn $y = 0$. Dagegen
  kann das richtige Verhalten mit dem Automaten in
  \lstRef{lama_correct_division} erreicht werden.

  \begin{listing}
  \begin{lstlisting}
automaton let
  location A let x = (/ z y); tel
  location B let x = 1; tel
  initial A;
  edge (A, B) : (= y 0);
  edge (B, A) : (not (= y 0));
tel
  \end{lstlisting}
  \caption{Korrektes Abfangen einer Division durch 0 in \Lama}
  \label{lst:lama_correct_division}
  \end{listing}
\end{example}

Die Übergänge zwischen zwei Modi werden durch Kanten bestimmt. Die
Semantik von diesen unterscheidet sich von den original Mode-Automaten
von Maraninchi und Rémond (\cite{ModeAutomata}). Die Semantik dort
entspricht den \term[Transition!Weak-]{Weak-Transitions} in \Scade,
wir setzen allerdings die \term[Transition!Strong-]{Strong-Transitions}
um. Dies ist nötig, da sich, aus dem
gleichen Grund wie bei \lstLm!ite!, Strong- nicht durch
Weak-Transitions simulieren lassen. Umgekehrt ist das aber sehr wohl
möglich. Ein Nebeneffekt ist, dass sich damit auch lokale \Scade-Takte
etc. in \Lama simulieren lassen.

Wir sehen uns nun diese Unterschiede in der Semantik kurz an. Für
die Transformation von Weak- in Strong-Transitions siehe
\secRef{weak_strong_transitions}. Unter einer Weak-Transition
wollen wir einen Übergang zwischen zwei Modi verstehen, der erst
im nächsten Taktschritt wirksam wird. Das bedeutet, wenn
zum Zeitpunkt $n$ ein Modus $m$ aktiv ist und eine Kante
$m \xrightarrow{c} m'$ mit geltender Bedingung $c$ existiert,
dann wird zum Zeitpunkt $n$ der Fluss aus $m$ ausgewertet
und bei $n+1$ der Fluss von $m'$.

Dagegen wird im Falle von Strong-Transitions zum Zeitpunkt $n$ bereits
der Fluss von $m'$ ausgewertet. In \Lama wird diese zweite Semantik
verwendet.

Um die durch die Kanten definierte Transitionsrelation als Funktion
interpretieren zu können, muss sie linkstotal und -eindeutig sein.
Dies erreichen wir, indem wir
\begin{enumerate}
\item Kanten abhängig von ihrer Position eine Priorität zuordnen und
\item eine Kante $A \xrightarrow{\syn{true}} A$ mit größter
  Priorität für jeden Modus $A$ hinzufügen.
\end{enumerate}
Dabei wird die Kante mit der niedrigsten Priorität benutzt, deren
Bedingung gilt und Quellmodus aktiv ist.

\subsubsection{Knoten}
\label{sec:lama_node_semantics}

Ein Knoten dient der hierarchischen Abstraktion eines Programms.
Knoten sind aber keine echte Erweiterung, denn ein Knoten $N$ darf
in seinem Gültigkeitsbereich höchstens einmal benutzt werden.
Das bedeutet,
dass die Variablen (mit entsprechender Umbenennung) in den
umgebenden Gültigkeitsbereich übernommen werden dürfen und der
Fluss in die Stelle des Aufrufes substituiert werden darf.
Dabei muss ein Modus $A$, in dem der Knoten benutzt wird,
durch ein Produkt mit dem Automaten des Knotens ersetzt
werden. Außerdem muss in allen anderen Modi der aktive Modus von
$N$ kodiert werden. Dabei entstehen zwei parallele Kopien des
Automaten, zwischen denen nur bei $A$ gewechselt werden kann.
Siehe \figRef{lama_flatten_subautom}.

\begin{figure}[h]
  \centering
  \begin{minipage}{0.95\textwidth}
    \centering
    \begin{tikzpicture}[anchor=base,scale=0.7, every node/.style={scale=0.7},baseline=(current bounding box.center)]
      \begin{scope}
  \pgfsetstrokecolor{black}
  \definecolor{strokecol}{rgb}{0.0,0.0,0.0};
  \pgfsetstrokecolor{strokecol}
  \draw (37bp,8bp) -- (37bp,124bp) -- (270bp,124bp) -- (270bp,8bp) -- cycle;
  \draw (154bp,106bp) node {$N$};
\end{scope}
  \pgfsetcolor{black}
  \draw [->] (204.47bp,53bp) .. controls (178.01bp,53bp) and (138.47bp,53bp)  .. node[auto] {$c_2$} (99.311bp,53bp);
  \draw [->] (212.7bp,167.56bp) .. controls (204.69bp,163.33bp) and (195.2bp,159.13bp)  .. (186bp,157bp) .. controls (175.85bp,154.65bp) and (165.09bp,153.8bp)  .. node[auto] {$c_3$} (144.18bp,154.36bp);
  \draw [->] (144.18bp,176.8bp) .. controls (150.21bp,177.25bp) and (156.21bp,177.66bp)  .. (162bp,178bp) .. controls (172.91bp,178.64bp) and (184.79bp,179.07bp)  .. node[auto] {$c_3$} (205.67bp,179.62bp);
  \draw [->] (99.248bp,43.486bp) .. controls (122.39bp,36.653bp) and (156.45bp,29.534bp)  .. (186bp,35bp) .. controls (188.86bp,35.529bp) and (191.77bp,36.224bp)  .. node[auto] {$c_1$} (204.4bp,40.153bp);
\begin{scope}
  \definecolor{strokecol}{rgb}{0.0,0.0,0.0};
  \pgfsetstrokecolor{strokecol}
  \draw (12bp,151bp) -- (132bp,151bp);
  \draw (132bp,151bp) .. controls (138bp,151bp) and (144bp,157bp)  .. (144bp,163bp);
  \draw (144bp,163bp) -- (144bp,177bp);
  \draw (144bp,177bp) .. controls (144bp,183bp) and (138bp,189bp)  .. (132bp,189bp);
  \draw (132bp,189bp) -- (12bp,189bp);
  \draw (12bp,189bp) .. controls (6bp,189bp) and (0bp,183bp)  .. (0bp,177bp);
  \draw (0bp,177bp) -- (0bp,163bp);
  \draw (0bp,163bp) .. controls (0bp,157bp) and (6bp,151bp)  .. (12bp,151bp);
  \draw (0bp,170bp) -- (144bp,170bp);
  \draw (72bp,174bp) node {$A$};
  \draw (72bp,155bp) node {$x = (use \; N \dotsm)$};
\end{scope}
\begin{scope}
  \definecolor{strokecol}{rgb}{0.0,0.0,0.0};
  \pgfsetstrokecolor{strokecol}
  \draw (217bp,34bp) -- (250bp,34bp);
  \draw (250bp,34bp) .. controls (256bp,34bp) and (262bp,40bp)  .. (262bp,46bp);
  \draw (262bp,46bp) -- (262bp,60bp);
  \draw (262bp,60bp) .. controls (262bp,66bp) and (256bp,72bp)  .. (250bp,72bp);
  \draw (250bp,72bp) -- (217bp,72bp);
  \draw (217bp,72bp) .. controls (211bp,72bp) and (205bp,66bp)  .. (205bp,60bp);
  \draw (205bp,60bp) -- (205bp,46bp);
  \draw (205bp,46bp) .. controls (205bp,40bp) and (211bp,34bp)  .. (217bp,34bp);
  \draw (205bp,53bp) -- (262bp,53bp);
  \draw (233bp,57bp) node {$s_2$};
  \draw (233bp,38bp) node {$x = g(b)$};
\end{scope}
\begin{scope}
  \definecolor{strokecol}{rgb}{0.0,0.0,0.0};
  \pgfsetstrokecolor{strokecol}
  \draw (57bp,34bp) -- (87bp,34bp);
  \draw (87bp,34bp) .. controls (93bp,34bp) and (99bp,40bp)  .. (99bp,46bp);
  \draw (99bp,46bp) -- (99bp,60bp);
  \draw (99bp,60bp) .. controls (99bp,66bp) and (93bp,72bp)  .. (87bp,72bp);
  \draw (87bp,72bp) -- (57bp,72bp);
  \draw (57bp,72bp) .. controls (51bp,72bp) and (45bp,66bp)  .. (45bp,60bp);
  \draw (45bp,60bp) -- (45bp,46bp);
  \draw (45bp,46bp) .. controls (45bp,40bp) and (51bp,34bp)  .. (57bp,34bp);
  \draw (45bp,53bp) -- (99bp,53bp);
  \draw (72bp,57bp) node {$s_1$};
  \draw (72bp,38bp) node {$x = f(a)$};
\end{scope}
\begin{scope}
  \definecolor{strokecol}{rgb}{0.0,0.0,0.0};
  \pgfsetstrokecolor{strokecol}
  \draw (233bp,180bp) ellipse (27bp and 18bp);
  \draw (233bp,175bp) node {$B$};
\end{scope}
    \end{tikzpicture}
    $\leadsto$
    \begin{tikzpicture}[anchor=base,scale=0.7, every node/.style={scale=0.7},baseline=(current bounding box.center)]
      \pgfsetcolor{black}
  \draw [->] (106.16bp,91.801bp) .. controls (95.565bp,79.021bp) and (93.613bp,66.241bp)  .. node[auto] {$$} (105.95bp,45.062bp);
  \definecolor{strokecol}{rgb}{0.0,0.0,0.0};
  \pgfsetstrokecolor{strokecol}
  \draw (71bp,53bp) node[auto] {$c_2 \wedge \neg c_3$};
  \draw [->] (246.8bp,6.7562bp) .. controls (229.54bp,1.8314bp) and (208.32bp,-1.8043bp)  .. (189.3bp,1.8043bp) .. controls (186.47bp,2.3411bp) and (183.6bp,3.0063bp)  .. node[auto] {$c_4$} (171.02bp,6.6962bp);
  \draw [->] (170.94bp,113.84bp) .. controls (177.1bp,114.21bp) and (183.36bp,114.55bp)  .. (189.3bp,114.8bp) .. controls (199.58bp,115.25bp) and (210.51bp,115.6bp)  .. node[auto] {$c_3$} (231.15bp,116.11bp);
  \draw [->] (248.3bp,103.24bp) .. controls (237.35bp,99.389bp) and (224.98bp,95.727bp)  .. (213.3bp,93.804bp) .. controls (202.83bp,92.082bp) and (191.68bp,92.597bp)  .. node[auto] {$c_4$} (171bp,96.141bp);
  \draw [->] (150.64bp,45.062bp) .. controls (161.12bp,57.842bp) and (162.96bp,70.622bp)  .. node[auto] {$$} (150.43bp,91.801bp);
  \draw (186bp,73bp) node[auto] {$c_1 \wedge \neg c_3$};
  \draw [->] (170.91bp,24.133bp) .. controls (186.44bp,23.524bp) and (204.29bp,22.824bp)  .. node[auto] {$c_3$} (231.32bp,21.764bp);
\begin{scope}
  \definecolor{strokecol}{rgb}{0.0,0.0,0.0};
  \pgfsetstrokecolor{strokecol}
  \draw (98bp,92bp) -- (159bp,92bp);
  \draw (159bp,92bp) .. controls (165bp,92bp) and (171bp,98bp)  .. (171bp,104bp);
  \draw (171bp,104bp) -- (171bp,118bp);
  \draw (171bp,118bp) .. controls (171bp,124bp) and (165bp,130bp)  .. (159bp,130bp);
  \draw (159bp,130bp) -- (98bp,130bp);
  \draw (98bp,130bp) .. controls (92bp,130bp) and (86bp,124bp)  .. (86bp,118bp);
  \draw (86bp,118bp) -- (86bp,104bp);
  \draw (86bp,104bp) .. controls (86bp,98bp) and (92bp,92bp)  .. (98bp,92bp);
  \draw (86bp,111bp) -- (171bp,111bp);
  \draw (128bp,115bp) node {$A \times s_2$};
  \draw (128bp,96bp) node {$x = g(b)$};
\end{scope}
\begin{scope}
  \definecolor{strokecol}{rgb}{0.0,0.0,0.0};
  \pgfsetstrokecolor{strokecol}
  \draw (98bp,7bp) -- (159bp,7bp);
  \draw (159bp,7bp) .. controls (165bp,7bp) and (171bp,13bp)  .. (171bp,19bp);
  \draw (171bp,19bp) -- (171bp,33bp);
  \draw (171bp,33bp) .. controls (171bp,39bp) and (165bp,45bp)  .. (159bp,45bp);
  \draw (159bp,45bp) -- (98bp,45bp);
  \draw (98bp,45bp) .. controls (92bp,45bp) and (86bp,39bp)  .. (86bp,33bp);
  \draw (86bp,33bp) -- (86bp,19bp);
  \draw (86bp,19bp) .. controls (86bp,13bp) and (92bp,7bp)  .. (98bp,7bp);
  \draw (86bp,26bp) -- (171bp,26bp);
  \draw (128bp,30bp) node {$A \times s_1$};
  \draw (128bp,11bp) node {$x = f(a)$};
\end{scope}
\begin{scope}
  \definecolor{strokecol}{rgb}{0.0,0.0,0.0};
  \pgfsetstrokecolor{strokecol}
  \draw (281bp,20bp) ellipse (50bp and 18bp);
  \draw (281bp,14bp) node {$B \times s_1$};
\end{scope}
\begin{scope}
  \definecolor{strokecol}{rgb}{0.0,0.0,0.0};
  \pgfsetstrokecolor{strokecol}
  \draw (281bp,117bp) ellipse (50bp and 18bp);
  \draw (281bp,111bp) node {$B \times s_2$};
\end{scope}
    \end{tikzpicture}
  \end{minipage}
  \caption{Abgeflachter Subautomat in \Lama}
  \label{fig:lama_flatten_subautom}
\end{figure}

Um die Benutzung eines Knotens direkt auszuwerten, werden zunächst
alle Parameter ausgewertet (strikt). Danach wird der Automat
für alle im Knoten deklarierten Variablen
ausgewertet (unter Beachtung der Parameter). Abschließend werden die
Rückgabevariablen gesetzt und beim Punkt des Aufrufs an die linke
Seite der Zuweisung übernommen.

\chapter{Transformation \Lama nach SMT}
\label{chap:lama_smt}

\lstset{language=lama,mathescape=true}

In diesem Kapitel wird die Transformation eines \Lama-Programms
in ein System von SMT-Formeln beschrieben. Dies soll ausführlich
geschehen, da dies ein wichtiger Transformationsschritt ist.
Wir beginnen zunächst mit der Definition der Begriffe, die
verwendet werden und beschreiben Besonderheiten in der
Transformation.

\section{Begriffe und Notationen}
\label{sec:lama_smt_prelim}

\begin{definition}
  Unter einem \term{Strom} (oder auch Folge oder Sequenz) vom Typ $A$
  verstehen wir eine Funktion $\func{x}{\N}{A}$. $\stream{A}$
  bezeichnet die Menge aller Ströme vom Typ $A$. Eine Abbildung
  $\func{F}{\stream{A}}{\stream{B}}$ ist eine
  \term{Stromtransformation}.

  Zur Schreibweise: zu jedem Strom $x$ gibt es einen Strom $x'$ mit
  $x'(n+1) = x(n), n \geq 0$. Dabei ist $x'(0)$ zunächst undefiniert.
\end{definition}

Wie in \secRef{lama_dynamic_semantics} angedeutet,
interpretieren wir ein \Lama-Programm als eine solche
Stromtransformation. Dabei wird das Programm
als ein System von Strömen und Formeln über diesen übersetzt.
Das bedeutet, dass beispielsweise aus einer Deklaration
\lstLm!x : int! ein Strom $\func{x}{\N}{\Z}$ wird. Aus einer
entsprechenden Definition \lstLm!x = M! wird ein Prädikat
$D_x(n) = (x(n) \equiv \overline{M}(n)), n \geq 0$
(dabei ist $\overline{M}$ die Fortsetzung von $M$ zu einem
Strom, s. u.). Ein solches Prädikat kann dann als Transitionsrelation
genutzt werden (s. \secRef{model_checking}).

Die Übersetzung wird in Form einer Funktion $Tr$ beschrieben.
$Tr$ erzeugt Deklaration von Strömen ($\Delta$) und ein Prädikat,
dass das Programm beschreibt. Dabei ist $\Delta$ eine Funktion
$\Ident \rightarrow \seqTerm + (\seqTerm^* \times \seqTerm^*)$,
die für einen Bezeichner entweder den deklarierten Strom
(im Falle einer Variablen) oder Eingabe- und Ausgabeströme
eines Knotens zurückgibt.

Der Funktion $Tr$ soll kein Typ zugewiesen werden. Diese ist als
Meta-Funktion zu verstehen. Tatsächlich müsste man für die einzelnen
Programmteile Funktionen
$Tr_{\mathrm{Node}}$, $Tr_{\mathrm{Flow}}$, $Tr_{\mathrm{Transition}}$ usw.
definieren. Wir beschränken uns darauf, nur $Tr$ zu schreiben und
den Typ implizit durch Parameter und Resultat anzugeben.

Die entstehenden Prädikate haben, bis auf die Zeitvariable $n$,
ausschließlich freie Variable, die durch die Deklarationen in
$\Delta$ vom SMT-Solver gebunden werden.

\begin{definition}
  Mit $Var$ wird die Menge der Variablen eines
  \Lama-Programms und mit $\Ident$ die Menge der Bezeichner notiert.
  Weiterhin bezeichnen wir mit $\seqTerm$ die Menge der Terme
  der SMT-Logik, die eine natürliche Zahl als Parameter haben
  und etwas vom Typ $A$ zurückgeben (dies sind Ströme in
  der Logik ausgedrückt).
  Wir brauchen weiterhin eine Funktion
  \begin{align*}
    \func{\delta}{Var}{(\Ident \rightarrow \seqTerm)}
    & \text{ und eine Fortsetzung zu} \\
    \func{\delta}{\powSet{Var}}{(\Ident \rightarrow \seqTerm)}
  \end{align*}
  die für die Variablen jeweils Ströme deklarieren. Deren
  Namen müssen überall eindeutig sein, da die Gültigkeitsbereiche für
  das entstehende Prädikat keine Bedeutung mehr haben.

  Der Typ $A$ ergibt sich dabei aus der Semantik aus
  \secRef{lama_types}.\footnote{Formal kann die Semantik von
    $\ty{uint[n]}$ auch durch $\Z/2^n$ und die von $\ty{sint[n]}$ durch
    $\Z/2^{n-1} \coprod \Z/2^{n-1}$ dargestellt werden. Daher kann
    beides in der SMT-Logik repräsentiert werden. Praktisch wird man
    aber bspw. Bit-Vektoren benutzen.}
\end{definition}

Wenn wir eine Menge von Deklarationen
$\Delta' \subset \Ident \rightarrow \seqTerm$,
die mit $\delta$ erzeugt worden sind, mit einem
$\Delta \subset \Ident
\rightarrow \seqTerm + (\seqTerm^* \times \seqTerm^*)$ vereinigen,
soll implizit mit der Inklusion
\begin{equation*}
\func{inl}{\Ident}{\seqTerm + (\seqTerm^* \times \seqTerm^*)}
\end{equation*}
komponiert werden. Für die rechte Seite und $\delta_{\mathrm{node}}$
soll analog mit $inr$ komponiert werden. Dies sollte aus dem Kontext
immer klar sein.

Bei der Definition von $Tr$ werden einige Notationen überladen
oder besonders geschrieben, um die Definition kürzer bzw. lesbarer
zu machen:
\begin{itemize}
\item Funktionen mit endlicher Definitionsmenge
  (meist Variablen\-umgebungen)
  werden ggf. als Relationen interpretiert, so dass
  man sie vereinigen und Paare $(x, y) \in f$ mit $f(x) = y$
  auswählen kann.
\item Wir benutzen $\lambda x.M$, um Funktionen und auch Prädikate
  zu erzeugen, d.h. Prädikate werden als Funktionen
  $\N \rightarrow \B$ interpretiert.
\item Wenn $D_i$ Prädikate über Streams sind, soll
  $\bigwedge_i D_i$ für $\lambda n. \; \bigwedge_i D_i(n)$ stehen.
\item Geschwungene Buchstaben sollen Mengen von Objekten darstellen,
  z.B. ist $\mathcal{V}$ eine Menge von Variablen.
\item Die Notation $\overline{M}$ soll andeuten, dass der Term $M$
  zu einem Stream hochgehoben worden ist. Dies ergibt sich aber
  auch aus ihren Definitionen.
\end{itemize}

\section{Aktivierung von Knoten}
\label{sec:node_activation}

Wir sehen uns nun die Erzeugung dieses Systems von Formeln aus
einem \Lama-Programm an. Der schwierige Teil dabei ist die Übersetzung
der Automaten und Knoten, da die Struktur der Flüsse bereits sehr
nahe an einem solchen System von Formeln ist.

In einem Modus gibt es zwei Arten von Definitionen, die wir
unterscheiden müssen: Definitionen, die Knoten benutzen und
die die keine benutzen. Die Ersteren sind dabei problematisch.
Wir sehen uns dazu \lstRef{lama_node_in_mode} an.

\begin{listing}[h]
\begin{lstlisting}
  $\ldots$
  node N ($\ldots$) returns ($\ldots$) let
    $\ldots$
  tel
  $\ldots$
  automaton let
    location m let
      x = (use N M);
    tel
    $\ldots$
  tel
  $\ldots$
\end{lstlisting}
\caption{Benutzung eines Knotens innerhalb eines Modus}
\label{lst:lama_node_in_mode}
\end{listing}

Bei der Übersetzung des Knotens $N$ entsteht ein Prädikat $D_N$,
zusammen mit Prädikaten $D_M$ und $D_x$. $D_M$ stellt die Belegung
der Eingabvariablen mit $M$ und $D_x$ die von $x$ mit den Ausgaben dar.
Sie sind Teil einer Konjunktion des Prädikates $D$.
Dadurch werden sie also zu jedem Zeitpunkt als gültig
vorausgesetzt, unabhängig davon, in welchem Modus sich der Automat
befindet. Als Lauf  dieses Systems interpretiert, würde das bedeuten,
dass der Knoten unabhängig, ob sich der Automat in $m$ befindet,
seinen nächsten Zustand berechnet. Die Parameter sind dabei
undefiniert. Dies ist offensichtlich ein falsches Verhalten.
Bei der Übersetzung wird dies so gelöst, dass ein Prädikat $E_m$
(enable) eingeführt wird. Dieses gilt nur dann,
wenn der Modus $m$ aktiv ist.

Sei $D_L$ ein Prädikat, das die lokalen Variablen von $N$ festlegt
und $states(N)$ die Menge der Zustandsvariablen von $N$. $D_N$ hat
dann die Form:
\[
D_N(n) = D_L(n) \wedge
\left( \bigwedge_{y \, \in \, \states(N)} D_y(n) \right).
\]
Wobei $D_y$ die Variable $y$ definiert. Wir ändern dies zu
\begin{align*}
  D_N'(n) & = \left( E_m(n) \longrightarrow D_N(n) \right) \\
  & \wedge \left( \neg E_m(n) \longrightarrow
    \bigwedge_{y \, \in \, \states(N)}(y(n+1) \equiv y(n)) \right).
\end{align*}
Wenn also $m$ aktiv ist, wird die Definition von $N$ ausgeführt.
Wenn nicht, bleiben alle Zustände erhalten und die anderen
Variablen sind undefiniert. Die im Folgenden definierte
Funktion $Tr$ konstruiert die Prädikate direkt auf diese Art.

\section{Transformation}
\label{sec:smt_transformation}

\begin{remark}
  Um den Unterschieden in der Semantik von \lstLm!ite! und
  Automaten gerecht zu werden (s. \secRef{lama_dynamic_semantics}),
  müssen mögliche Fehler, die in Ausdrücken vorkommen können,
  behandelt werden. Hierbei handelt
  es sich um Fehler durch Überläufe und Divisionen durch 0.
  Derartige Fehler werden in dieser Arbeit nicht betrachtet.
  Um aber anzudeuten, an welchen Stellen diese relevant sind,
  verwenden wir Funktionen $\iteL$ und $\iteS$ statt
  einem einfachen $\ite$ (s. \secRef{smt}).

  Um ein Programm auf derartige Fehler zu untersuchen,
  müssen entsprechende zusätzliche Invarianten generiert werden.
  $\iteL$ aktiviert nur die Invarianten für die Bedingung
  und, abhängig von der Bedingung, für eines der beiden Argumente.
  $\iteS$ aktiviert immer alle Invarianten.
\end{remark}

\begin{figure}[H]
\begin{align}
  & Tr(\text{\lstLm!node N x returns y let
      $\; \mathcal{N} \; \mathcal{V} \;$ F I
      $\; \mathcal{A} \; A \; $ tel!}, E)
    & & = \; (\Delta, D) \\
  & \quad \begin{tabular}{l}
    \text{wobei} \\
    $(\Delta_{\mathcal{N}}, D_{\mathcal{N}})
      = \delta_{\mathrm{node}}(\mathcal{N}, \mathcal{A}, E)$ \\
    $\Delta_x = \delta(x)$ \\
    $\Delta_y = \delta(y)$ \\
    $\Delta_N = \Delta_{\mathcal{N}} \cup \Delta_x \cup \Delta_y
       \cup \delta(\mathcal{V})$ \\
    $D_F = Tr(F, \Delta_N, E)$ \\
    $(\Delta_{\mathcal{A}}, D_{\mathcal{A}})
      = Tr(\mathcal{A}, \Delta_N, E)$ \\
    $\Delta = \Delta_N \cup \Delta_{\mathcal{A}}
      \cup (N \mapsto (\Delta_x(x), \Delta_y(y)))$ \\
    $D = D_F \wedge D_{\mathcal{A}} \wedge Tr(I, \Delta_N)
      \wedge Tr(A, \Delta_N)$
  \end{tabular} \notag
\end{align}
\caption{Transformation von \Lama-Knoten in Prädikate}
\label{fig:lama_node_predicate}
\end{figure}

Zunächst beginnen wir mit der Erzeugung von Prädikaten für
Knoten. Das Prädikat für ein Programm wird analog erzeugt.
Zusätzlich müssen dort noch (sofern vom Solver unterstützt), die
Enum-Typen deklariert werden. Außerdem muss das Prädikat $P$ für die
Invariante erzeugt werden.

Die Transformation eines Knotens ist in \figRef{lama_node_predicate}
dargestellt. Zunächst werden die Subknoten deklariert. Dabei erzeugt
$\delta_{\mathrm{node}}$ (\figRef{lama_subnodes_predicates})
die oben genannten Aktivierungsbedingungen und generiert damit die
Prädikate für die Subknoten. Die Aktivierungsbedingung wird in den
Knoten bei der Erzeugung des Flusses benutzt, um diesen bedingt
zu aktivieren (s. Gleichung
\eqRef{pred_flow_def}--\eqRef{pred_flow_trans}).
Danach wird die Schnittstelle ($x$ und $y$) und die Variablen
deklariert. Diese Deklarationen werden in den Prädikaten für den
Fluss des Knotens und der Automaten verwendet.

Zusätzlich werden Deklarationen für die Zustände der Automaten (s.u.)
und das Interface des Knotens zurückgegeben.

\begin{figure}[H]
\begin{align}
  & Tr(\text{\lstLm!definition D transition T!}, \Delta, E)
    \notag \\
    & \qquad = \; Tr(D, \Delta, E) \wedge Tr(T, \Delta, E) \\
  & Tr(\text{\lstLm!x = M!}, \Delta, E) & & = \; D
    \label{eq:pred_flow_def} \\
  & \quad \begin{tabular}{l}
    \text{wobei} \\
    $D(n) = E(n) \longrightarrow \overline{x}(n) \equiv D_M(n)$ \\
    $\overline{x} = \Delta(x)$ \\
    $D_M = Tr(M, \Delta)$
  \end{tabular} \notag \\
  & Tr(\text{\lstLm!x = (use N M)!}, \Delta, E)
    & & = \; D_I \wedge D_O \\
  & \quad \begin{tabular}{l}
    \text{wobei} \\
    $(I, O) = \Delta(N)$ \\
    $D_I(n) =
      E(n) \longrightarrow I(n) \equiv D_M(n)$ \\
    $D_O(n) =
      E(n) \longrightarrow \overline{x}(n) \equiv O(n)$ \\
    $\overline{x} = \Delta(x)$ \\
    $D_M = Tr(M, \Delta)$
  \end{tabular} \notag \\
  & Tr(\text{\lstLm!x' = M!}, \Delta, E) & & = \; D
    \label{eq:pred_flow_trans} \\
  & \quad \begin{tabular}{l}
    \text{wobei} \\
    $D(n) =
      (E(n) \longrightarrow
        \overline{x}(n+1) \equiv \overline{M}(n))$ \\
      $\qquad \!\!\; \wedge \: (\neg E(n) \longrightarrow
        \overline{x}(n+1) \equiv \overline{x}(n))$ \\
    $\overline{x} = \Delta(x)$ \\
    $\overline{M} = Tr(M, \Delta)$
  \end{tabular} \notag \\
  & Tr(\text{\lstLm!initial! } x_1 = c_1, \dotsc, x_k = c_k, \Delta)
    & & = \; \bigwedge_{i=1}^k I_i \\
  & \quad I_i = \overline{x}_i(0) \equiv c_i,
    \quad \overline{x}_i = \Delta(x_i) \notag
\end{align}
\caption{Transformation von \Lama-Fluss in Prädikate}
\label{fig:lama_flow_predicate}
\end{figure}

\begin{figure}[H]
\begin{align*}
  & \delta_{\mathrm{node}}(\mathcal{N}, \mathcal{A}, E)
    = \left( \bigcup_N \Delta_N, \bigwedge_N D_N \right),
  \quad (\Delta_N, D_N)
    = \delta_{\mathrm{node}}(N, \mathcal{A}), \, N \in \mathcal{N} \\
  & \delta_{\mathrm{node}}(N, \mathcal{A}, E) = Tr(N, E'),\\
  & \quad \begin{tabular}{l}
    \text{wobei} \\
    $ E'(n) =
    \begin{cases}
      E(n), & N \text{ in keinem $A \in \mathcal{A}$ benutzt} \\
      E(n) \wedge s_A(n) \equiv m,
        & N \text{ in Modus $m \in A \in \mathcal{A}$ benutzt}
    \end{cases}$ \\
  \end{tabular}
\end{align*}
\caption{Deklaration von Sub-Knoten}
\label{fig:lama_subnodes_predicates}
\end{figure}

Dabei ist $\mathcal{A}$ in \figRef{lama_subnodes_predicates}
eine Menge von Automaten und $s_A$ die Variable,
die den aktuellen Modus von $A \in \mathcal{A}$ repräsentiert.
Diese wird bei der folgenden Übersetzung von Automaten erzeugt.

\begin{figure}[H]
\begin{align*}
  & Tr(\text{\lstLm!automaton let
      $\; \mathcal{M} \;$ IM $\; \mathcal{E} \;$ O!}, \Delta, E)
    = (\{s,s_1\},
      D_{Inp} \wedge D_L \wedge D_S \wedge D_{\mathcal{E}}) \\
  & \quad \begin{tabular}{l}
    \text{wobei} \\
    $s, s_1 : \mathrm{enum}(\mathcal{M}), \; s, s_1 \not\in \Delta$ \\
    $(\mathcal{D}_L \cup \mathcal{D}_S, D_{Inp})
      = \mathrm{gather}(\mathcal{M}, \Delta, E)$ \\
    $D_L(n) = \bigwedge_{(x, \mathcal{D}) \in \mathcal{D}_L}$
      $E(n) \longrightarrow (\overline{x}(n)
          \equiv \mathrm{match}(s, \mathcal{D}, O)(n)$ \\
    $D_S(n) = \bigwedge_{(x, \mathcal{D}) \in \mathcal{D}_S}$
      $(E(n) \longrightarrow (\overline{x}(n)
          \equiv \mathrm{match}(s, \mathcal{D}, O)(n))$ \\
        \hspace{95pt}
        $\wedge \; \neg E(n) \longrightarrow
          (\overline{x}(n) \equiv \overline{x}(n)))$ \\
   $\overline{x} = \Delta(x)$ \\
   $D_{\mathcal{E}} = D_s \wedge D_{s_1}$ \\
   $D_s = \mathrm{next}(s_1, \mathcal{E}) $ \\
   $I_{s_1} = s_1(0) \equiv \mathrm{IM}$ \\
   $D_{s_1} = I_{s_1} \wedge (\lambda n. \; s_1(n+1) \equiv s(n)$)
  \end{tabular}
\end{align*}
\caption{Transformation von \Lama-Automaten in Prädikate}
\label{fig:lama_automaton_predicates}
\end{figure}

Für einen Automaten werden zunächst Variablen $s, s_1$ angelegt,
die den aktiven Modus im aktuellen bzw. letzten Zyklus repräsentieren.
$\mathrm{enum}(\mathcal{M})$ generiert dabei einen Enum-Typen,
der als Konstruktoren die Modi $\mathcal{M}$ hat. Der Datenfluss
des Automaten wird nach lokalen Definitionen und Zustandsübergängen
aufgeteilt und den jeweils definierten Variablen zugeordnet.
$\mathrm{gather}$ generiert dazu eine Funktion
$\Ident \rightarrow (\mathcal{M} \rightarrow \Term{\stream{\B}}$
(s. \figRef{lama_automaton_helper}). $\mathrm{match}$ erzeugt daraus
eine Kette von $ite$, die abhängig vom aktiven Modus den jeweiligen
Ausdruck einer Variable zurückgibt. Die Bedingung $E$ wird genutzt,
um die gesamte Kette ggf. zu deaktivieren. Zuletzt generiert
aus den Kanten $\mathrm{next}$ ein Prädikat,
das den nächsten Modus festlegt.
$s_1$ speichert lediglich den Modus für den nächsten Zyklus.

\begin{figure}[H]
\begin{align*}
& \mathrm{gather}(\mathcal{M}, \Delta, E, s)
  = \left( \bigcup_{M \in \mathcal{M}} \mathcal{D}_M,
    \bigwedge_{M \in \mathcal{M}} D_{M, Inp} \right), \\
  & \quad (\mathcal{D}_M, D_{M, Inp}) = \mathrm{gather}(M, \Delta, E, s)
    \notag \\ \notag \\
& \mathrm{gather}(\text{\lstLm!location M let F tel!},
  \Delta, E, s)
  = \left( \bigcup_{(x, D) \in \mathcal{D}} x \mapsto M \mapsto D,
    \bigwedge D_{Inp} \right), \label{eq:gather_location_match} \\
  & \quad E'(n) = E(n) \wedge s(n) \equiv M \notag \\
  & \quad (\mathcal{D}, D_{Inp})
    = \mathrm{unzip} (\mathrm{gather}(F, \Delta, E'))
    \notag \\ \notag \\
& \mathrm{gather}(\text{\lstLm!x = M!}, \Delta, E')
  = x \mapsto Tr(M, \Delta) \\
& \mathrm{gather}(\text{\lstLm!x = (use N M)!}, \Delta, E')
  = (x \mapsto O, D_I) \\
  & \quad \text{wobei} \notag \\
  & \qquad \begin{tabular}{l}
    $(I, O) = \Delta(N)$ \\
    $D_I(n) =
      E'(n) \longrightarrow I(n) \equiv \overline{M}(n)$ \\
    $\overline{M} = Tr(M, \Delta)$
  \end{tabular} \notag \\
& \mathrm{gather}(\text{\lstLm!x' = M!}, \Delta, E')
  = x' \mapsto Tr(M, \Delta) \\
\notag \\
  & \mathrm{match}(s, \{(M, \overline{D})\} \cup \mathcal{D}, Default)
    \notag \\
    & \quad = \lambda n. \; \iteL(s(n) \equiv M,
      \overline{D}(n), \mathrm{match}(s, \mathcal{D}, Default)) \\
  & \mathrm{match}(s, \{(M, \overline{D})\}, Default) \notag \\
    & \quad = \lambda n. \; \iteL(s(n) \equiv M,
      \overline{D}(n), Default(M)(n))) \\
\notag \\
  & \mathrm{next}(s_1, \mathcal{E})
    = \mathrm{next}(s_1, \mathrm{byHead}(\mathcal{E})) \\
  & \mathrm{next}(s_1, \{\mathcal{E}_m\} \cup \widetilde{\mathcal{E}})
    = \lambda n. \; \iteL(s_1(n) \equiv m,
      \mathrm{next}(\mathcal{E}_m)(n),
      \mathrm{next}(s_1, \mathcal{E})(n)) \\
  & \mathrm{next}(\{m \xrightarrow{C} m'\} \cup \mathcal{E}_m)
    = \lambda n. \;
      \iteL(Tr(C)(n), m', \mathrm{next}(\mathcal{E}_m)(n))
\end{align*}
\caption{Hilfsfunktionen für die Transformation von \Lama-Automaten}
\label{fig:lama_automaton_helper}
\end{figure}

Dabei ist in \figRef{lama_automaton_helper}
$\func{\mathrm{unzip}}{\powSet{A \times B}}
{\powSet{A} \times \powSet{B}}$ die kanonische Abbildung mit
$x \mapsto (\powSet{\pi_0}(x), \powSet{\pi_1}(x))$
(allgemeiner: $\func{\mathrm{unzip}}{F(A \times B)}{F(A) \times F(B)}$,
für einen Funktor $F$).

Die Funktion $\mathrm{byHead}(\mathcal{E})$ partitioniert die
Kantenmenge in die Familie $\widetilde{\mathcal{E}}$ der Mengen
\[
\mathcal{E}_m = \{m \xrightarrow{C} m' \in \mathcal{E}\}
  \subseteq \mathcal{E}
\]
die durch Modi $m$ indiziert sind. Außerdem nehmen wir an, dass in der
letzten Definition von $\mathrm{next}$ die Kanten in der Ordnung ihrer
Priorität entnommen werden.

Die Transformation von \Lama-Ausdrücken in
\figRef{lama_expr_predicate} erfordert lediglich die Transformation
der Operatoren ($\boxempty \leadsto \boxempty'$)
und die Auswahl der richtigen
$ite$-Funktion. $\mathrm{match}$ erzeugt auch hier wieder eine Ketten
von $ite$.

\begin{figure}[H]
\begin{align*}
  & Tr(\text{\lstLm!x!}, \Delta) & & = \Delta(x) \\
  & Tr(\text{\lstLm!c!}, \Delta) & & = \lambda n. \; c \\
  & Tr(\text{\lstLm!(not M)!}, \Delta)
    & & = \lambda n. \; \neg Tr(M, \Delta)(n)) \\
  & Tr(\text{\lstLm!($\boxempty \;$ M N)!}, \Delta)
    & & = \lambda n.
      Tr(M, \Delta)(n) \; \boxempty' \; Tr(N, \Delta)(n),\\
    & & & \quad \boxempty \in
      \{\text{\lstLm!or, and, xor, =>, <, >,!} \dotsc \} \\
  & Tr(\text{\lstLm!(ite P M N)!}, \Delta)
    & & = \lambda n. \; \iteS \left( \overline{P}(n), \,
      \overline{M}(n), \, \overline{N}(n) \right), \\
    & & & \quad \overline{P} = Tr(P, \Delta), \,
      \overline{M} = Tr(M, \Delta), \,
      \overline{N} = Tr(N, \Delta)\\
  & Tr(\text{\lstLm!(\# $\, M_1 \ldots M_k$)!}, \Delta)
    & & = \lambda n. \;
      (Tr(M_1, \Delta)(n), \dotsc, Tr(M_k, \Delta)(n)) \\
  & Tr(\text{\lstLm!(project x i!}, \Delta)
    & & = \lambda n. \; p_i (\Delta(x)(n)) \\
  & Tr(\text{\lstLm!(match M \{$P_1, \dotsc, P_k$\})!}, \Delta)
    & & = \mathrm{match}(Tr(M,\Delta), P_1, \dotsc, P_k)
\end{align*}
\begin{align*}
  & \mathrm{match}(M, \text{\lstLm!P.N!})
    = Tr(N)n \\
  & \mathrm{match}(M, (\text{\lstLm!P.N!}, \mathbf{R}))
    = \lambda n. \; \iteS((M(n) \equiv P)
      , \, Tr(N)(n), \, \mathrm{match}(M, \mathbf{R})(n)) \\
  & \mathrm{match}(M, (\text{\lstLm!\_.N!}, \mathbf{R}))
    = Tr(N)
\end{align*}
\caption{Transformation von \Lama-Ausdrücken in Prädikate}
\label{fig:lama_expr_predicate}
\end{figure}


\chapter{Transformation \Scade nach \Lama}
\label{chap:scade2lama}

In diesem Kapitel wird die Transformation von \Scade nach
\Lama beschrieben. Wir werden dabei den Datenfluss
außerhalb (\secRef{nonstate_dataflow}) und innerhalb
(\secRef{state_dataflow}) von \Scade-Zuständen getrennt
betrachten. Hier ergeben sich jeweils eigene Schwierigkeiten.

Weiterhin werden Termersetzungen beschrieben, mit denen einige
Sprachkonstrukte auf andere zurückgeführt werden können
(\secRef{derived}). Zuletzt werden wir beispielhafte Optimierungen
angeben, die auf den \Scade-Termen durchgeführt werden können.

Die hier angegebene Übersetzung stellt nicht den kompletten
Sprachumfang von \Scade dar. Für eine Übersicht der fehlenden
Elemente ist in \secRef{missing_scade} angegeben.

\section{Zustandsunabhängiger Datenfluss}
\label{sec:nonstate_dataflow}

Jeder Knoten kann einen Datenfluss außerhalb von Zuständen
haben. Wir betrachten dazu den Ausschnitt in
\lstRef{scade_updown_flow} den Beispielknotens aus der
Einführung.

\begin{listing}[h]
\begin{lstlisting}[language=scade,linerange={1-6,20-23},
    numbers=left,xleftmargin=25pt]
node UpDownCounter() returns (x : int)
var
  x_1 : int;
let
  
  automaton SM1
  returns ..;
  
  x_1 = -1 -> pre x;
tel
\end{lstlisting}

  \caption{Datenfluss des Knotens UpDownCounter}
  \label{lst:scade_updown_flow}
\end{listing}

Der Datenfluss ist hier in Zeile 9. Dieser kann unabhängig von
dem Automaten \lstSc!SM1! übersetzt werden. Dabei
wird (wie erwartet) aus einem solchem Datenfluss ein
globaler \Lama-Datenfluss.

In diesem Abschnitt werden wir das Beispiel aus \lstRef{scade_counter}
verwenden.

\begin{listing}[h]
\begin{lstlisting}[language=scade, linerange={7-14},
    numbers=left,xleftmargin=25pt]
node Count(x : bool; reset : bool) returns (c : int)
let
  c = if not reset
      then 0 ->
        (pre c + if x then 1 else 0)
      else 0;
tel
\end{lstlisting}

  \caption{\Scade-Beispiel Count}
  \label{lst:scade_counter}
\end{listing}

Der Knoten zählt, wie viele Takte die Variable $x$ gesetzt war.
Mit der Eingabe $reset$ ist dieser Zähler auf 0 zurücksetzbar.

Das Ergebnis der Transformation ist in \lstRef{lama_counter}
dargestellt.

\begin{listing}[h]
\begin{lstlisting}[language=lama,linerange={1-29,31-31},
    numbers=left,xleftmargin=25pt]
input
  x : bool;
  reset : bool;
nodes
  node Count (x : bool, reset : bool)
  returns (c_out : int) let
    local
      c_33 : int;
      c : int;
    state
      c_32 : int;
    definition
      c = (ite (not reset) c_33 0);
      c_out = c;
    transition
      c_32' = c ;
    automaton let
      location dummy_34 let tel
      location init_35 let tel
      location running_36 let
        definition c_33 = (+ c_32 (ite x 1 0));
      tel
      initial dummy_34 ;
      edge (dummy_34, init_35) : true ;
      edge (init_35, running_36) : true ;
      default c_33 = 0;
    tel
  tel
local c : int;
definition c = (use Count x reset);
\end{lstlisting}

  \caption{Transformation von Count in \Lama}
  \label{lst:lama_counter}
\end{listing}

Wie wir zu dem Datenfluss kommen, wird in den folgenden Abschnitten
beschrieben. Klar ist allerdings, wie die Struktur entsteht:
zunächst wird der Knoten \lstLm!Count! in einen \Lama-Knoten
übersetzt (Schnittstelle in Zeile 5-6). Bei den Knoten muss
noch eine technische Besonderheit beachtet werden: in \Lama
sind Ausgabevariablen nicht lesbar. Daher müssen diese einen neuen
Namen(\lstLm!c_out!) und eine extra Zuweisung (Zeile 14) erhalten.

\lstLm!Count! ist auch der oberste Knoten des Programms und
bestimmt deshalb auch die Eingabevariablen des \Lama-Programms
(Zeile 1-3). Die Ausgaben des Knotens werden an lokale Variablen
gebunden (Zeile 29-30).
Der oberste Knoten wird dabei von außen (durch einen
Programm-Parameter) festgelegt.

\subsection{Basisoperationen}
\label{sec:basic_functions}

Dies ist der einfachste Teil der Übersetzung. Als Basisfunktionen
betrachten wir alle logischen und arithmetischen Operatoren, sowie
Variablen, Konstanten, Array-Funktionen und
\lstSc!if-then-else!. Diese haben eine
direkte Entsprechung in \Lama.

Eine Kleinigkeit ist noch bei der Verwendung von Operatoren
zu beachten. In \Scade erzeugt jede Verwendung eines Operators
$N$ eine Instanz von diesem. Diese Instanz erhält einen Speicher,
der unabhängig von allen anderen Instanzen ist.
In \Lama ist es nun so, dass ein Knoten einen Zustand hat, der
bei jeder Benutzung verwendet wird. Daher muss für jede
Verwendung von $N$ eine Kopie mit einem neuen Namen
des erzeugten \Lama-Knotens angelegt werden.

\subsection{Streamoperationen}
\label{sec:stream_functions}

Hier soll es um die Übersetzung von \lstSc!->!
und \lstSc!pre! gehen.

Dieser Teil der Übersetzung birgt ein paar Fallstricke. Wir
werden mehrere Arten von Ausdrücken mit $\arr$ und $\pre$
betrachten.

\begin{enumerate}
\item $x = c \arr \pre M \text{ mit }
  \arr, \pre \not\in \mathrm{Sub}(M)$
\item $x = \pre M \text{ mit } \arr, \pre \not\in \mathrm{Sub}(M)$
\item $x = M \arr N \text{ mit }
  \arr, \pre \not\in \mathrm{Sub}(M) \cup \mathrm{Sub}(N)$
\item $x = M$, wobei $M$ keine der obigen Formen hat.
\end{enumerate}

Dabei sind $M,N$ jeweils \Scade-Ausdrücke und $c$ eine Konstante
und $\mathrm{Sub}(M)$ soll die Menge der Teilausdrücke vom $M$
darstellen.

\begin{remark}
  Der erste Fall
  wird dabei bereits von den anderen abgedeckt, ist aber
  eine signifikante Optimierung und auch am leichtesten zu
  übersetzen. Beispielsweise \lstSc!fby!
  liefert, wie wir in \secRef{fby} sehen werden, Ketten
  von solchen Ausdrücken. Daher lohnt sich diese Optimierung.  
\end{remark}

\begin{remark}
  Wir behandeln hier $\arr$ und $\pre$ getrennt,
  da es Fälle gibt, in denen sich der Datenfluss nicht ohne
  Weiteres mit \lstSc!fby! darstellen lässt.
  Dies tritt z.B. in der Definition von \lstSc!d!
  in \figRef{scade_chrono} auf. Hier sind $\arr$ und $\pre$ durch
  nicht-Stream Operatoren getrennt.
  Eine mögliche Ersetzung wäre: $(-1 \arr \pre d) + 1 \mod 100$.
  Dies ist aber im Allgemeinen schwierig zu lösen.
\end{remark}

\begin{remark}
  Der letzte Fall deckt im Wesentlichen zwei Dinge ab: zum einen
  $x = M \arr \pre N$ wobei $M$ nicht konstant ist und zum anderen
  geschachtelte temporale Ausdrücke. Beides wird durch Abrollen
  behandelt.  
\end{remark}

\begin{translation}[Übersetzung $c \arr \pre N$]
Gegeben sei die \Scade-Zuweisung $x = c \arr \pre N$.
Zum Zeitpunkt $0$ hat $x$ dabei den Wert $c$. Zu einem
Zeitpunkt $n > 0$, hat $x$ den Wert von $N$ zum Zeitpunkt $n-1$.
Wir werden hierfür $x_0 = c, x_n = N_{n-1}$ schreiben.

Eine Verschiebung der Indizes ergibt: $x_0 = c$ und $x_{n+1} = N_n$
für alle $n \geq 0$. Das heißt wir
können dies direkt in einen \Lama-Fluss umwandeln:
$x' = N, x(0) = c$. Dies soll für ein Teilprogramm der Form
\begin{lstlisting}[language=lama]
  transition x' = N;
  initial x = c;
\end{lstlisting}
stehen. Dabei wird $N$ implizit übersetzt.

\end{translation}

\begin{translation}[Übersetzung $\pre M$]
Um eine Zuweisung der Form $x = \pre M$ zu übersetzen,
machen wir uns zunutze, dass es in \Lama erlaubt ist,
Zustandsvariablen nicht zu initialisieren, sofern sie
niemals im ersten Takt ausgewertet werden. Bei einem
korrekten \Scade-Programm muss dies gegeben sein.
Wie die Auswertung verhindert wird, wird im nächsten
Schritt, bei der Übersetzung von $\arr$ beschrieben.

Damit erfolgt die Übersetzung analog zum ersten Fall. Wir erzeugen
also die Definitionen $x' = N, x(0) = \bot$. Dabei soll $\bot$
andeuten, dass keine Initialisierung angegeben wird. In \Lama
entfällt damit der Initialwert für $x$.
\end{translation}

\begin{translation}[Übersetzung $M \arr N$]
\label{trans:lone_arrow}
Um die Initialisierung mit $\arr$ umzusetzen, kann
nicht das \lstLm!initial!-Konstrukt verwendet werden.
Dies hat zwei Gründe:
\begin{enumerate}
\item Das Konstrukt kann nur für konstante Ausdrücke verwendet
  werden. Wenn wir also einen Ausdruck $M \arr \pre N$ mit
  nicht-konstantem $M$ übersetzen wollen, wäre 
  \lstLm!initial x = M! nicht zulässig.
\item Wenn $\arr$ kein $\pre$ zugeordnet ist, wird keine
  Zustandsvariable generiert. Dies wäre aber nötig, um
  \lstLm!initial! verwenden zu können.
\end{enumerate}

Die Transformation macht sich zunutze, dass der Fluss in einem
Modus nur ausgewertet wird, wenn dieser aktiv ist. Wir generieren
dafür einen Automaten, der einen Modus zur Initialisierung (init)
und einen für den eigentlichen Fluss (run) hat. Dabei wird
eine $\varepsilon$-Kante
$\mathrm{init} \xrightarrow{true} \mathrm{run}$ erzeugt.
Zusätzlich wird noch ein Initialmodus (dummy) mit einer Kante
$\mathrm{dummy} \xrightarrow{true} \mathrm{init}$ benötigt.
Dieser ist aber niemals aktiv, da durch die Semantik der Kanten,
sofort der Folgemodus betreten wird.
Die Initialisierung soll aber einmal ausgeführt werden. Hätten
wir also diesen Modus nicht, würde das gleiche mit dem
Modus init passieren. Man kann dies auch anders lösen,
beispielsweise durch eine Bedingung, die erst im zweiten Takt
wahr ist. Der Automat mit Fluss ist in \figRef{lama_init_autom}
dargestellt.

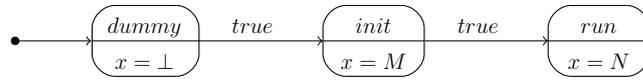
\begin{figure}[h]
  \centering
  \begin{tikzpicture}[anchor=base,scale=0.7, every node/.style={scale=0.7},baseline=(current bounding box.center)]
    \pgfsetcolor{black}
  \draw [->] (99.604bp,19bp) .. controls (116.32bp,19bp) and (137.62bp,19bp)  .. node[auto] {$true$} (165.83bp,19bp);
  \draw [->] (3.9906bp,19bp) .. controls (8.4789bp,19bp) and (19.953bp,19bp)  .. node[auto] {$$} (42.456bp,19bp);
  \draw [->] (220.23bp,19bp) .. controls (236.65bp,19bp) and (257.78bp,19bp)  .. node[auto] {$true$} (285.88bp,19bp);
\begin{scope}
  \definecolor{strokecol}{rgb}{0.0,0.0,0.0};
  \pgfsetstrokecolor{strokecol}
  \draw (55bp,0bp) -- (88bp,0bp);
  \draw (88bp,0bp) .. controls (94bp,0bp) and (100bp,6bp)  .. (100bp,12bp);
  \draw (100bp,12bp) -- (100bp,26bp);
  \draw (100bp,26bp) .. controls (100bp,32bp) and (94bp,38bp)  .. (88bp,38bp);
  \draw (88bp,38bp) -- (55bp,38bp);
  \draw (55bp,38bp) .. controls (49bp,38bp) and (43bp,32bp)  .. (43bp,26bp);
  \draw (43bp,26bp) -- (43bp,12bp);
  \draw (43bp,12bp) .. controls (43bp,6bp) and (49bp,0bp)  .. (55bp,0bp);
  \draw (43bp,19bp) -- (100bp,19bp);
  \draw (71bp,23bp) node {$dummy$};
  \draw (71bp,4bp) node {$x = \bot$};
\end{scope}
\begin{scope}
  \definecolor{strokecol}{rgb}{0.0,0.0,0.0};
  \pgfsetstrokecolor{strokecol}
  \draw (178bp,0bp) -- (208bp,0bp);
  \draw (208bp,0bp) .. controls (214bp,0bp) and (220bp,6bp)  .. (220bp,12bp);
  \draw (220bp,12bp) -- (220bp,26bp);
  \draw (220bp,26bp) .. controls (220bp,32bp) and (214bp,38bp)  .. (208bp,38bp);
  \draw (208bp,38bp) -- (178bp,38bp);
  \draw (178bp,38bp) .. controls (172bp,38bp) and (166bp,32bp)  .. (166bp,26bp);
  \draw (166bp,26bp) -- (166bp,12bp);
  \draw (166bp,12bp) .. controls (166bp,6bp) and (172bp,0bp)  .. (178bp,0bp);
  \draw (166bp,19bp) -- (220bp,19bp);
  \draw (193bp,23bp) node {$init$};
  \draw (193bp,4bp) node {$x = M$};
\end{scope}
\begin{scope}
  \definecolor{strokecol}{rgb}{0.0,0.0,0.0};
  \pgfsetstrokecolor{strokecol}
  \draw (298bp,0bp) -- (328bp,0bp);
  \draw (328bp,0bp) .. controls (334bp,0bp) and (340bp,6bp)  .. (340bp,12bp);
  \draw (340bp,12bp) -- (340bp,26bp);
  \draw (340bp,26bp) .. controls (340bp,32bp) and (334bp,38bp)  .. (328bp,38bp);
  \draw (328bp,38bp) -- (298bp,38bp);
  \draw (298bp,38bp) .. controls (292bp,38bp) and (286bp,32bp)  .. (286bp,26bp);
  \draw (286bp,26bp) -- (286bp,12bp);
  \draw (286bp,12bp) .. controls (286bp,6bp) and (292bp,0bp)  .. (298bp,0bp);
  \draw (286bp,19bp) -- (340bp,19bp);
  \draw (313bp,23bp) node {$run$};
  \draw (313bp,4bp) node {$x = N$};
\end{scope}
\begin{scope}
  \definecolor{strokecol}{rgb}{0.0,0.0,0.0};
  \pgfsetstrokecolor{strokecol}
  \definecolor{fillcol}{rgb}{0.0,0.0,0.0};
  \pgfsetfillcolor{fillcol}
  \filldraw [opacity=1.0] (2bp,19bp) ellipse (2bp and 2bp);
\end{scope}
  \end{tikzpicture}
  \caption{\Lama-Automat zur Initialisierung einer Variable}
  \label{fig:lama_init_autom}
\end{figure}

\end{translation}

\subsubsection{Abrollen nicht konstanter Initialisierungen und
geschachtelter Streamoperatoren}
Um diese Art von Ausdrücken übersetzen zu können, wird in
$x = M$ die rechte Seite $M$ aufgespalten und damit auf einen
der oberen Fälle zurückgeführt. Dabei sollen Ausdrücke der ersten
Form erhalten bleiben. Diese Aufspaltung ist natürlich nur notwendig,
wenn $M$ mindestens einen der Operatoren $\arr, \pre$ enthält und
dieser nicht als Wurzel eines entsprechenden Syntaxbaums von
$M$ vorkommt.

Sei $N \in \mathrm{Sub}(M)$, so dass $N$ eine der Formen aus 1-3 hat,
$M \not= N$ (echter Teilausdruck) und
$c \arr N \not\in \mathrm{Sub}(M)$ (maximaler Teilausdruck im
Sinne von 1). Diesen werden wir an eine neue Variable $y$ zuweisen
und $N$ in $M$ durch $y$ ersetzen.

Wir definieren also das folgende Ersetzungssystem. Ein \Scade-Programm
wird dabei als ein Term angesehen.
Zunächst die Relation, um Ausdrücke an der Wurzel zu schützen:
\begin{align*}
  & x = M; \; \leadsto \; y:A; \; x = M'; \; y = N; \\
  & \quad \text{ wobei } M \not= c \arr N, \pre N, N \arr P \\
  & \quad \text{ und } M \Rightarrow (M', y:A, N)
\end{align*}

Die Relation $\Rightarrow$ ist dabei folgendermaßen definiert:
\begin{align*}
c \arr M & \Rightarrow (y, y:A, c \arr M), & & c:A, M:A \\
\pre M & \Rightarrow (y, y:A, \pre M), & & M:A \\
N \arr M & \Rightarrow (y, y:A, N \arr M), & & N:A, M:A
\end{align*}
Dabei ist $y$ ein Name, der nicht im aktuellen Gültigkeitsbereich
vorkommt und $A$ jeweils ein \Scade-Typ.

Wir bilden also $S \leadsto^* S'$ mit einem \Scade-Programm $S$, um
ein Programm $S'$ zu erhalten, dass mit den einfachen
Regeln transformiert werden kann.

\begin{example}[Count]
Wir wollen uns nun die obigen Übersetzungschritte an dem
Beispiel aus \lstRef{scade_counter} ansehen. Zunächst wird
die Termersetzung durchgeführt. Das Ergebnis ist
in \lstRef{scade_counter_rewritten} zu sehen.

\begin{listing}[h]
\begin{lstlisting}[language=scade,
    numbers=left,xleftmargin=25pt]
node Count (x : bool; reset : bool) returns (c : int)
var
  c_33 : int;
  c_32 : int;
let
  c = if not reset then c_33 else 0;
  c_33 = 0 -> c_32 + (if x then 1 else 0);
  c_32 = pre c;
tel
\end{lstlisting}

  \caption{\Scade-Beispiel Count nach Termersetzung}
  \label{lst:scade_counter_rewritten}
\end{listing}

Hier ist die Relation $\leadsto$ zweimal angewendet worden.
Zunächst wird der Ausdruck \lstSc!pre c!
aus der Zuweisung an $c$ herausgezogen. Danach
wird aus dem verbleibenden
\begin{lstlisting}[language=scade]
  if not $\ldots$ 0 -> c_32 + (if x then 1 else 0) $\ldots$
\end{lstlisting}
der Ausdruck
\lstSc!0 -> c_32 + (if x then 1 else 0)!
herausgezogen.\footnote{\lstSc!->! besitzt eine
niedrige Bindungstärke und daher ist der Ausdruck
\lstSc!0 -> c_32 + (if x then 1 else 0)!
als \lstSc!0 -> (c_32 + (if x then 1 else 0))! zu lesen.}

Der nächste Schritt ist dann die Transformation in das
\Lama-Programm aus \lstRef{lama_counter}. Der erste Fall
der Streamoperatoren tritt hier nicht auf. Die anderen beiden
aber sehr wohl.

Die Definition von $c\_32$ hat ein $\pre$ auf der rechten Seite,
erzeugt also einen Speicher. Daher wird $c\_32$ zu einer
Zustandsvariable (Zeile 11 und 16 in \lstRef{lama_counter}).

Die Variable $c\_33$ benötigt dagegen eine Initialisierung
und muss daher oben beschriebenen Automaten aufbauen (Zeile 17-27).
Eine Besonderheit der Implementierung ist hier, dass durch
die Standardzuweisung (Zeile 26) $c\_33$ auch im Dummy-Modus
definiert ist. Das ist lediglich eine kleine Optimierung, dadurch
muss bei der SMT-Übersetzung nicht zwischen dummy und init
unterschieden werden.
\end{example}

\section{Zustandsabhängiger Datenfluss}
\label{sec:state_dataflow}

Dieser Abschnitt behandelt die Transformation von Automaten
und Flüssen in diesen. Im Wesentlichen wird aus einem
\Scade-Automaten ein \Lama-Automat erzeugt. Hierarchische
Automaten werden dabei in Knoten übersetzt
(\seeR{sec:subautomata}). Allerdings birgt das Zusammenspiel
von Zuständen und den Streamoperatoren $\arr$ und $\pre$
einiges an Schwierigkeiten
(\seeR{sec:pre_last_in_states} und \ref{sec:restart_resume}).
Außerdem müssen die Weak-Transitions korrekt transformiert
werden (\ref{sec:weak_strong_transitions}).

\subsection{Subautomaten}
\label{sec:subautomata}

\Scade unterstützt die hierarchische Komposition von Automaten,
das bedeutet hier, dass jeder Zustand eines Automaten wieder
eine Menge von Automaten (als Subautomaten bezeichnet)
enthalten kann. Ist der Zustand nicht
aktiv, sind auch die untergeordneten Automaten nicht aktiv.

In \Lama wird dieses Verhalten mit Hilfe von Knoten modelliert.

\begin{translation}
  Sei $A$ ein \term{Subautomat} eines Zustandes $s$. Wir
  deklarieren einen Knoten $NA$, der als Eingaben alle verwendeten
  und als Ausgaben alle geschriebenen Variablen hat. Siehe
  \figRef{scade_subautomaton}.
  Wenn \lstSc!restart!-Transitionen in
  $s$ führen, erhält $NA$ einen zusätzliche Eingabevariable,
  zum Zurücksetzen des Automaten (s. \ref{sec:restart_resume}).
  $NA$ wird in $s$ statt $A$ verwendet.

\begin{figure}[h]
\centering


\begin{minipage}{0.95\textwidth}
 \centering
  $\vcenter{\hbox{\includegraphics[width=0.8\textwidth]
      {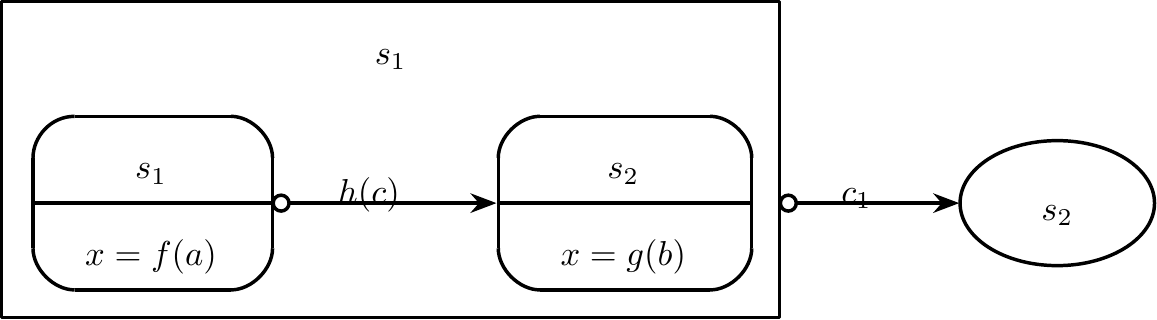}}}$

  \vspace{15pt}
  $\downarrow$
  \vspace{15pt}

  $\vcenter{\hbox{\includegraphics[width=0.8\textwidth]
      {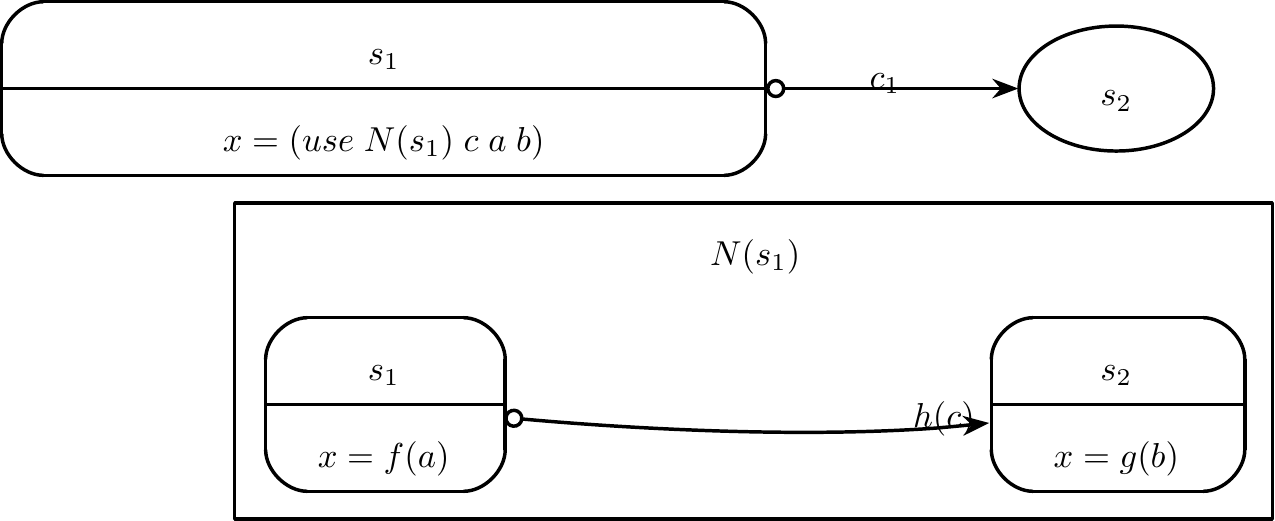}}}$
\end{minipage}
\caption{Übersetzung von Subautomaten}
\label{fig:scade_subautomaton}
\end{figure}

\end{translation}

Die Semantik von \Lama-Knoten sorgt nun dafür, dass $A$ nur aktiv
ist, wenn auch $s$ es ist.

\subsection{Weak- und Strong-Transitions}
\label{sec:weak_strong_transitions}

Strong-Transitions können durch die Semantik von \Lama direkt
übersetzt werden. Weak-Transitions benötigen dagegen eine
besondere Behandlung. Allerdings ergibt die Kombination der
beiden Transitionsarten einige Schwierigkeiten, auf die
wir hier näher eingehen werden.

\begin{figure}[h]
\centering
\begin{minipage}{0.95\textwidth}
  \centering
  $\vcenter{\hbox{\includegraphics[width=0.4\textwidth]
      {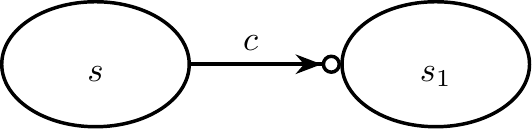}}}$
  $\leadsto$
  $\vcenter{\hbox{\includegraphics[width=0.4\textwidth]
      {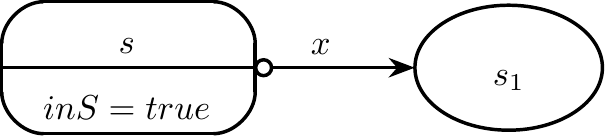}}}$
\end{minipage}
\caption{Übersetzung einer einfachen Weak-Transition}
\label{fig:scade_simple_weak}
\end{figure}

Der einfachste Fall ist natürlich eine einzelne Weak-Transition.
Die Transformation ist in \figRef{scade_simple_weak} dargestellt.

Dabei ist $x$ folgendermaßen definiert:
\begin{definition}
  $x$ ist ein Strom mit
  \begin{equation*}
    x(0) = false, \; x' = c \wedge inS.
  \end{equation*}
  In allen Zuständen außer $s$ ist $inS = false$ (default).
  Weiterhin definieren wir einen Strom
  \begin{equation*}
    wasS(0) = \mathrm{isInitial}(s), \; wasS' = inS.
  \end{equation*}
\end{definition}

Die Bedingung wird also
um einen Takt verzögert und wir müssen prüfen, ob die Bedingung
galt, während $s$ aktiv war. In der Abbildung wird allerdings nicht
sichtbar, dass die entstehende Strong-Transition eine
höhere Priorität als alle anderen Strong-Transitions von $s$
haben muss (s. 1.).

\begin{remark}
  Dies ist eine lokale Übersetzung,
  es wird also angenommen, dass nur eine Weak-Transition
  ersetzt wird. Nun ist es aber so, dass das Ziel die
  Eliminierung aller Weak-Transitions ist. Die
  Ergänzung von der Bedingung $x$ um $inS$ wird dann
  aber überflüssig, da resultierende Automat nur noch
  Strong-Transitions hat. Es gibt dann zwei Möglichkeiten,
  wenn zum Zeitpunkt $n$ die Bedingung $c$ wahr ist und
  der Automat sich in einem Zustand $s_2 \not= s$ befindet.
  Es gibt eine Transition $s_2 \xrightarrow{c_2} s$ mit
  \begin{enumerate}
  \item $c_2 = T$ oder
  \item $c_2 = F$.
  \end{enumerate}

  Im ersten Fall wird die Transition ausgeführt, der Automat
  befindet sich in Zustand $s$ und $inS(n) = T$. Hier
  ist also $c = c \wedge inS$.

  Im zweiten Fall muss $c_2(n+1) = T$. Dann wird die
  Transition bei $n+1$ ausgeführt. Dann ist
  $x(n+1) = F$, da $inS(n) = F$. Nun wurde aber bereits
  eine bei $n+1$ Transition ausgeführt, und obwohl $c(n) = T$,
  bleibt der Automat in $s$.

  Wenn also während der Übersetzung
  ungültige Zwischenschritte entstehen dürfen,
  kann diese Bedingung weggelassen werden.
\end{remark}

Kritisch sind die Kombinationen von Weak- und Strong-Transitions.
Dazu sehen wir uns die Semantik von \Scade-Automaten an, wie sie
in \cite[S.74-76]{ScadeRef} beschrieben ist.
Ein Automat hat in einem Takt zwei Arten von Zuständen:
\term[Zustand!ausgewählter]{ausgewählter Zustand} und
\term[Zustand!aktiver]{aktiver Zustand}. Dabei
werden vom ausgewählten Zustand die Strong- und vom aktiven
die Weak-Transitions ausgewertet. Strong-Transitions wiederum
legen dann den nächsten \emph{aktiven} Zustand fest, während
Weak-Transitions den nächsten \emph{ausgewählten} festlegen.
Der ausgeführte Datenfluss ist immer der des aktiven Zustandes.

Ein Automat wird nun in jedem Takt folgendermaßen ausgewertet:
\begin{enumerate}
\item Wenn der ausgewählten Zustand Strong-Transitions mit
  geltender Bedingung hat, wird die mit der höchsten Priorität
  ausgeführt. Der Zielzustand wird zum aktiven Zustand. Wenn
  keine solche Transition existiert, wird der ausgewählte auch
  zum aktiven Zustand.
\item Der Datenfluss des aktiven Zustands wird ausgeführt.
\item Wenn in diesem Takt keine Strong-Transition ausgeführt
  worden ist und der aktive Zustand Weak-Transitions mit
  geltender Bedingung hat, wird die mit der höchsten Priorität
  ausgeführt. Der Zielzustand wird zum ausgewählten Zustand. Wenn
  keine Transition ausgeführt worden ist, wird der aktive
  Zustand zum ausgewählten Zustand.
\end{enumerate}

Daraus ergeben sich nun drei Kombinationen an einem Zustand $s$:

\subsubsection{1. Weak- und Strong-Transition verlassen $s$ parallel}
\label{sec:weak_strong_par}

\begin{figure}[h]
\centering
\begin{minipage}{0.95\textwidth}
  \centering
  $\vcenter{\hbox{\includegraphics[width=0.45\textwidth]
      {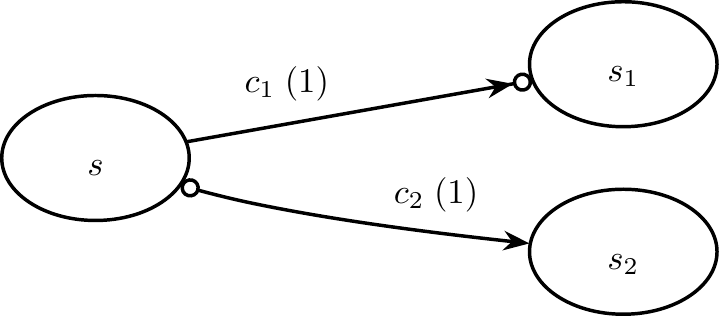}}}$
  $\leadsto$
  $\vcenter{\hbox{\includegraphics[width=0.45\textwidth]
      {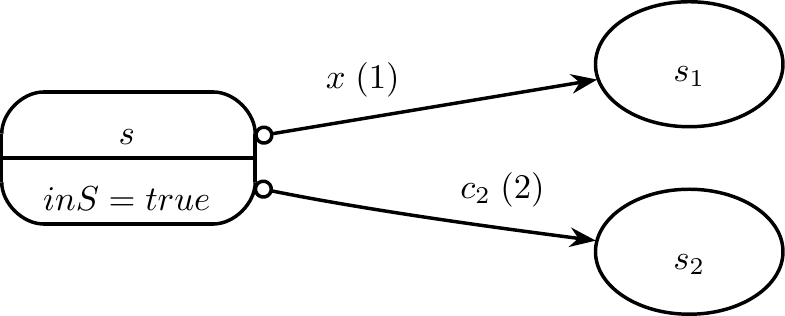}}}$
\end{minipage}
\caption{Parallele Weak- und Strong-Transition}
\label{fig:scade_weak_strong_par}
\end{figure}

Der erste Fall ist dabei bereits durch die obige Übersetzung
abgedeckt. Davon wollen wir uns kurz überzeugen. Dabei sollen
die Flüsse $x$ und $inS$ für beide Automaten existieren.
$sel$ bezeichne den ausgewählten und $act$ den aktiven
Zustand. $T$ und $F$ seien die Wahrheitswerte Wahr und Falsch. In
\figRef{scade_weak_strong_par} ist in Klammern die Priorität
einer Transition gekennzeichnet.

\begin{remark}
  Weak- und Strong-Transitions können jeweils
  getrennt geordnet werden, da diese nach der obigen
  Semantik nicht im Konflikt stehen. Daher erhalten
  die beiden Transitionen in \figRef{scade_weak_strong_par}
  syntaktisch die gleiche Priorität.
\end{remark}

Wir betrachten einen Takt $n$ in einem beliebigen Lauf dieses
Automaten. Dabei können weitere Weak-Transitions in $s$
auftreten. Es soll aber der Zustand $s$ ausgewählt sein,
d.h. $sel(n) = s$. Zunächst unterscheiden wir zwei Fälle. Die
Bezeichnungen beziehen sich auf \figRef{scade_weak_strong_par}.
\begin{enumerate}
\item Im Takt $n-1$ sind $c_1$ wahr und $s$ aktiv
  gewesen. Dann ist im ursprünglichen Automaten $act(n) = s_1$.
  Da dann ist aber auch $inS(n-1) = T$ und $x(n) = T$. Damit gilt
  auch im transformierten Automaten $act(n) = s_1$.  
\item Ist bei $n-1$ der Zustand $s$ nicht aktiv gewesen,
  hängen $act(n)$ und $sel(n+1)$ nur noch von $c_1,c_2$ ab.
  Im transformierten Automaten ist $x(n) = F$, hier
  hängt $act(n)$ also auch nur noch von $c_1,c_2$ ab.
\end{enumerate}

Für den zweiten Fall betrachten wir die beiden folgenden Tabellen.
Beide stellen die möglichen Kombinationen von $c_1,c_2$ zum
Zeitpunkt $n$ dar. Dabei ist links der Lauf in dem Ausgangsautomaten
und rechts im erzeugten Automaten dargestellt.
Wenn wir „$-$“ schreiben, ist der Wert der Variablen an der Stelle
nicht von belang.
\begin{align*}
\begin{array}[h]{c|c|c|c|c}
  t   & sel & c_1 & c_2 & act \\
  \hline
  n   & s   & T   & F   & s \\
  n+1 & s_1 & -   & -   & s_1 \\
  \hline \hline
  n   & s   & F   & T   & s_2 \\
  \hline \hline
  n   & s   & T   & T   & s_2 \\
  \hline \hline
  n   & s   & F   & F   & s \\
\end{array}
& \quad
\begin{array}[h]{c|c|c|c|c|c|c|c}
  t   & sel & c_1 & c_2 & \pre c_1 & wasS & x & act \\
  \hline
  n   & s   & T   & F   & - & - & F & s \\
  n+1 & s   & -   & -   & T & T & T & s_1 \\
  \hline \hline
  n   & s   & F   & T   & - & - & F & s_2 \\
  \hline \hline
  n   & s   & T   & T   & - & - & F & s_2 \\
  \hline \hline
  n   & s   & F   & F   & - & - & F & s \\
\end{array}
\end{align*}

Die Einträge der Tabelle sind leicht zu prüfen. Zu beachten ist,
dass in der ersten Zeile der Wert von $c_2$ irrelevant ist, weil
im ursprünglichen Automaten $s_1$ bereits bei $n+1$ ausgewählt
worden ist. Im anderen Automaten hat die übersetzte Transition
eine höhere Priorität, und wird daher vorgezogen. Darum ist
der aktive Zustand dann $s_1$.

Dies ist offensichtlich ein induktives Argument. Im Basisfall
muss unterschieden werden, ob $s$ ein Initialzustand ist oder
nicht. Dies übernimmt die Initialisierung von
$wasS(0) = \mathrm{isInitial}(s)$ (s. oben).

\subsubsection{2. Eine Weak- führt in einen Zustand hinein und eine
Strong-Transition hinaus}

\begin{figure}[h]
\centering
\begin{minipage}{0.95\textwidth}
  \centering
  $\vcenter{\hbox{\includegraphics[height=1.3cm]
      {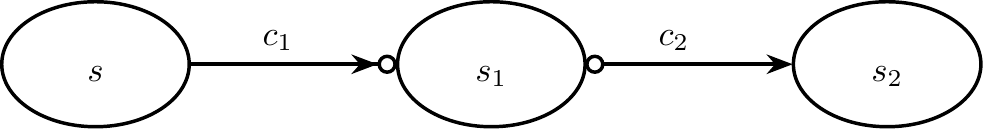}}}$

  \vspace{15pt}
  $\downarrow$
  \vspace{15pt}

  $\vcenter{\hbox{\includegraphics[height=2.3cm]
      {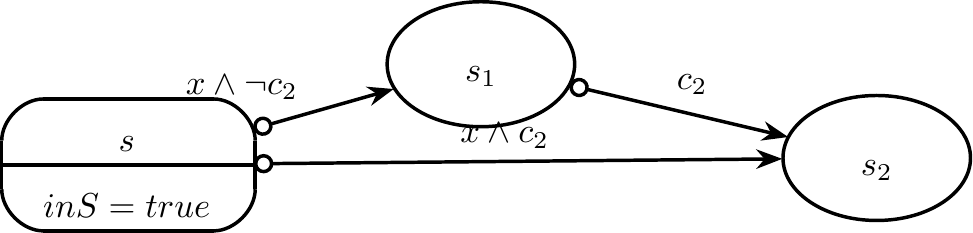}}}$
\end{minipage}
\caption{Übersetzung einer Weak- gefolgt von einer Strong-Transition}
\label{fig:scade_weak_strong}
\end{figure}

Hierfür sei der erste Automat in \figRef{scade_weak_strong} gegeben.

Wenn $s$ aktiv ist und $c_1$ gilt, dann wird $s_1$ ausgewählt.
Wenn nun aber im nächsten Takt $c_2$ gilt, wird $s_2$ als
aktiv angenommen. $s_1$ wird also zu diesem Zeitpunkt nicht aktiv sein.
Daher fügen
wir bei der Übersetzung eine zusätzliche Transition ein, die
das „Überspringen“ von $s_1$ umsetzt. Die Bedingung $x$ ist dabei
die gleiche wie oben.

\subsubsection{3. Eine Strong- führt in $s$ hinein und eine
Weak-Transition hinaus}

\begin{figure}[h]
\centering
\begin{minipage}{0.95\textwidth}
  \centering
  $\vcenter{\hbox{\includegraphics[height=2cm]
      {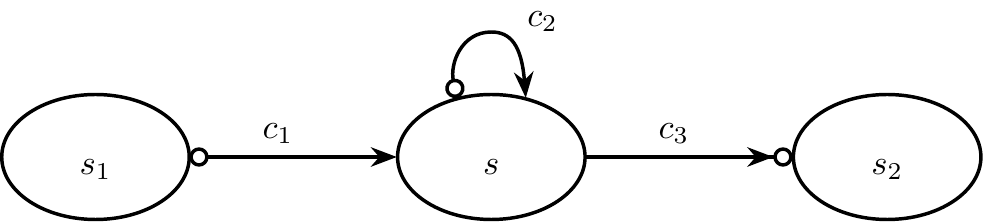}}}$

  \vspace{15pt}
  $\downarrow$
  \vspace{15pt}

  $\vcenter{\hbox{\includegraphics[height=2cm]
      {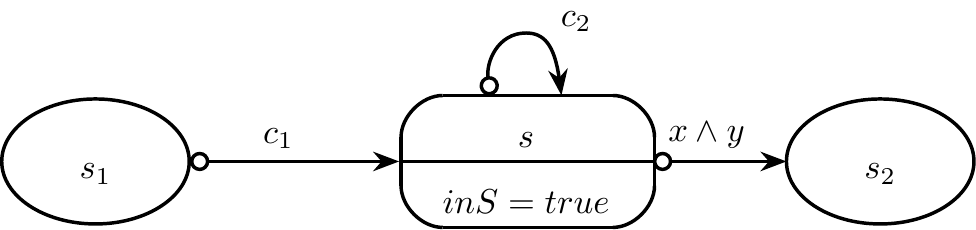}}}$
\end{minipage}
\caption{Übersetzung einer Weak-Transition,
wenn eine Strong-Transition in den Zustand führt}
\label{fig:scade_strong_weak}
\end{figure}

Im letzten Fall müssen wir die Bedingung $x$ noch ergänzen.
Wenn $s$ in einem Takt durch eine Strong-Transition betreten
worden ist, darf die Weak-Transition nicht mehr ausgeführt werden.
Wenn also $c_3$ in $x$ gespeichert wird, darf die Transition nur
ausgeführt werden, wenn vorher $s$ nicht mit $c_1$ oder $c_2$
aktiviert wurde. Dazu dient die Bedingung $y$:
\begin{definition}
  \begin{equation*}
    y(0) = false, \; y' =
      (c_1 \longrightarrow wasS) \wedge (c_2 \longrightarrow \neg wasS)
  \end{equation*}
  Wichtig ist, dass hier $wasS$ statt $inS$ benutzt wird. Damit
  wird in der Bedingung $inS$ aus \emph{zwei} Takten vorher verwendet.
\end{definition}

\begin{remark}
In allen Fällen können die Übersetzungen auf mehrere parallele
Strong-Transitions erweitert werden, indem die Disjunktion der
Transitionsbedingungen gebildet wird.
\end{remark}

\subsection{default-Deklarationen}
\label{sec:default_decls}

\Scade unterstützt wie \Lama (eigentlich umgekehrt)
Standardflüsse für Variablen. Diese werden bei der Deklaration
angegeben:
\begin{lstlisting}[language=scade]
  x : A default = M
\end{lstlisting}
Dabei ist $A$ ein Typ und $M$ ein Ausdruck. Wenn die Variable
$x$ einem Zustand eines Automaten keine Definition hat,
wird dieser Standardfluss verwendet. Es ergibt sich also folgende
Übersetzung:

\begin{translation}
  Sei $x$ wie oben in einem Knoten $N$ deklariert und $\widetilde{N}$
  der entsprechende \Lama-Knoten. Wenn

  \begin{enumerate}
  \item $x$ nirgendwo in $\widetilde{N}$ definiert wird,
    wird zu $\widetilde{N}$ ein Fluss $x = M$ hinzugefügt.

  \item $x$ in mindestens einem Zustand eines Automaten $A$
    definiert wird, wird $A$ eine Deklaration
    \lstLm!default x = M! hinzugefügt.

  \item $x$ global definiert ist, wird die Deklaration ignoriert.
  \end{enumerate}
\end{translation}

Gibt es für eine Variable $x$ keine
\lstSc!default!-Deklaration, wird
\begin{lstlisting}[language=scade]
  x : A default = last 'x
\end{lstlisting}
angenommen.
Dann muss aber eine Initialisierung mit \lstSc!last! deklariert
worden sein (s. \ref{sec:pre_last_in_states}), wenn $x$ nicht in allen
Zuständen definiert ist und damit dieser Ausdruck verwendet wird.

\subsection{pre/last in Zuständen}
\label{sec:pre_last_in_states}

Sowohl \lstSc!pre x! als auch \lstSc!last 'x! (wir werden
$\lastApp x$ schreiben) benötigen einen Speicher. Sie unterscheiden
sich darin, dass \lstSc!pre! für jeden Zustand eines Automaten $x$
einen eigenen
Speicher besitzt, während bei \lstSc!last! ein gemeinsamer Speicher
für alle Zustände genutzt wird. Dadurch kann \lstSc!last! aber auch
nur für Variablen verwendet werden (angedeutet durch \lstSc!'!,
was für den Namen selber und nicht einen Ausdruck stehen soll).
Darin liegt auch der Grund, warum \lstSc!pre!
einen getrennten Speicher für alle Zustände hat.

Der Wert von $\pre M$ in einem Zustand $A$ ist dabei der Wert von $M$ 
zu einem Zeitpunkt, an dem $A$ zuletzt aktiv gewesen ist.
Der Wert von $\lastApp x$ ist dabei der Wert von $x$
aus dem letzten Takt.

\begin{figure}[h]
\centering
\begin{minipage}{0.95\textwidth}
  \centering
  $\vcenter{\hbox{\includegraphics[width=0.45\textwidth]
      {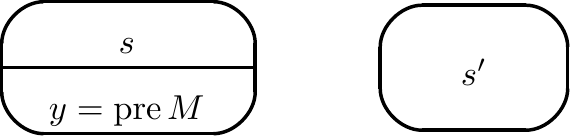}}}$
  $\leadsto$
  $\vcenter{\hbox{\includegraphics[width=0.45\textwidth]
      {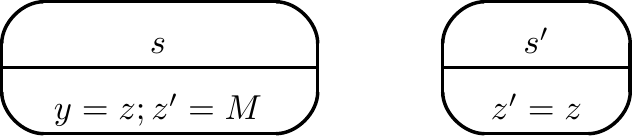}}}$
\end{minipage}
\caption{Übersetzung von $\pre$ in Zuständen}
\label{fig:scade_state_pre}
\end{figure}

Dies ergibt die folgende Übersetzung:

\begin{translation}
  Sei $s$ ein Zustand mit einem Datenfluss $y = M$ mit
  $\pre N$ bzw.
  $P \arr \pre N \in \mathrm{Sub}(M)$. Dann erzeugen
  wir einen \Lama-Datenfluss $y = \widetilde{M}; z' = M$ in $s$
  zusammen mit einem Fluss $z' = z$ und ggf. $z(0) = P$.
  Dabei ist
  $\widetilde{M} = M[z/\pre N] \text { bzw. } M[z/P \arr \pre N]$
  (s. \figRef{scade_state_pre}).
\end{translation}

\lstSc!last! kann auch eine Initialisierung durch
eine Deklaration erhalten:

\begin{lstlisting}[language=scade]
  x : A last = M
\end{lstlisting}

Wenn eine solche Deklaration existiert, müssen wir diese
bei der Transformation beachten.

\begin{figure}[h]
\centering
\begin{minipage}{0.95\textwidth}
  \centering
  $\vcenter{\hbox{\includegraphics[width=0.15\textwidth]
      {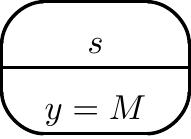}}}$
  $\leadsto$
  $\vcenter{\hbox{\includegraphics[width=0.5\textwidth]
      {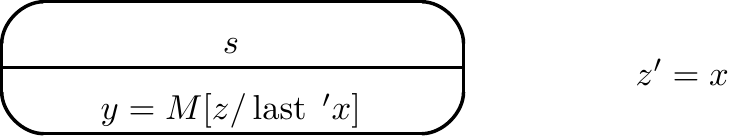}}}$
\end{minipage}
\caption{Übersetzung von $\last$ in Zuständen}
\label{fig:scade_state_last}
\end{figure}

\begin{translation}
  Sei $s$ ein Zustand mit einem Datenfluss $y = M$, wobei
  $\lastApp x \in \mathrm{Sub}(M)$. Dann ist
  dies äquivalent zu einem \Lama-Datenfluss $y = M[z/\lastApp x]$
  in $s$ zusammen mit einem globalen Standardfluss $z' = x$
  (s. \figRef{scade_state_last}). Wenn eine $\last$-Deklaration $P$
  existiert, setzen wir $z(0) = P$.
\end{translation}

\subsection{restart/resume-Transitionen}
\label{sec:restart_resume}

\Scade unterscheidet \lstSc!restart!- und
\lstSc!resume!-Transitionen. Die Unterscheidung
ist, dass beim Betreten eines Zustandes im ersten Fall alle
Initialisierungen ausgeführt werden. Der zweite Fall repräsentiert
also das normale Verhalten \Lama und bedarf daher keiner weiteren
Behandlung.

Der erste Fall betrifft zwei Konstrukte: $\arr$ und Automaten.
Dabei ist $\arr$ sowohl im Datenfluss des Zielzustandes,
als auch in Bedingungen an Transitionen betroffen. Wenn
ein Datenfluss $M \arr N$ zurückgesetzt wird, wird $M$
zurückgegeben. $\pre$ etc. sind nicht betroffen. Insbesondere
wird auch die $\last$-Deklaration nicht erneut ausgeführt.
Ein Automat wird beim Zurücksetzen wieder in seinen Initialzustand
versetzt. (\cite[S.78-79]{ScadeRef})

\begin{figure}[h]
\centering
\begin{minipage}{0.95\textwidth}
  \centering
  $\vcenter{\hbox{\includegraphics[width=0.5\textwidth]
      {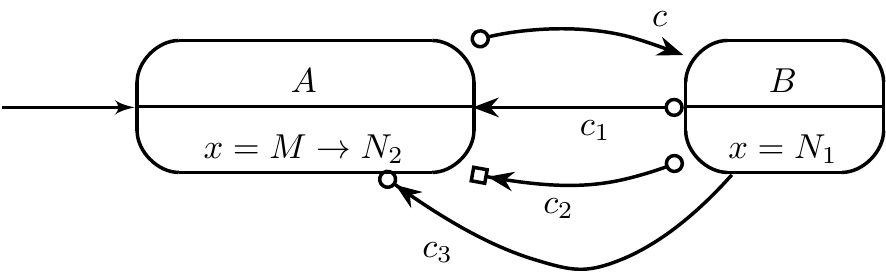}}}$

  \vspace{10pt}
  $\downarrow$
  \vspace{10pt}

  $\vcenter{\hbox{\includegraphics[width=0.9\textwidth]
      {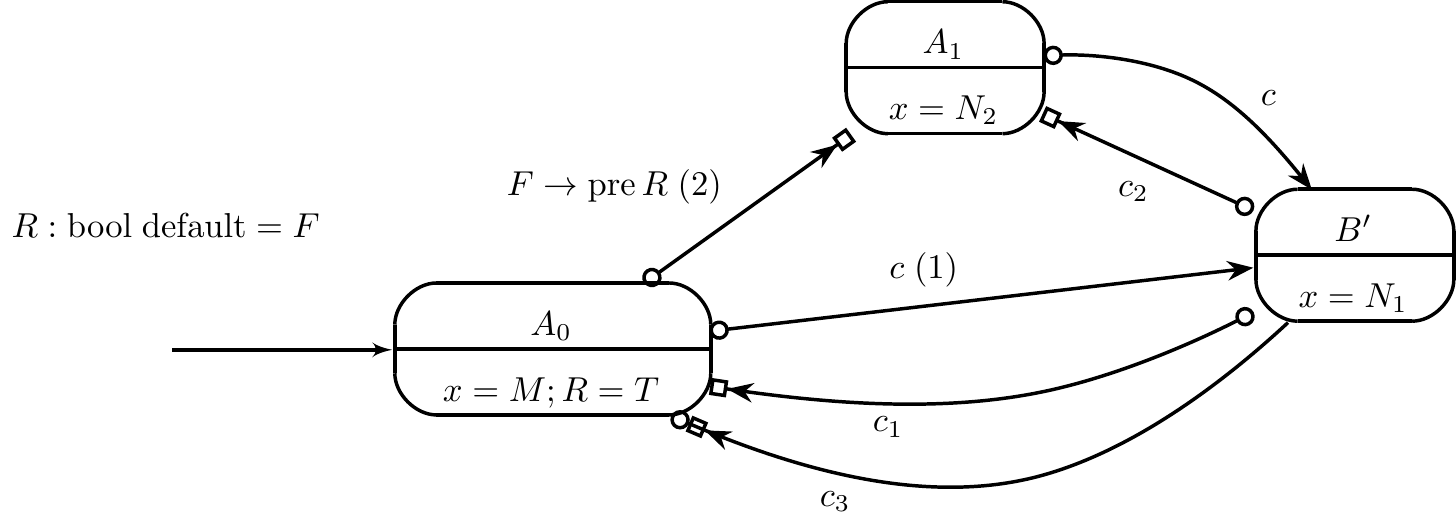}}}$
\end{minipage}
\caption{Übersetzung von restart-Transitionen}
\label{fig:scade_restart}
\end{figure}

Wir betrachten dazu \figRef{scade_restart}. Wir notieren hier aus
technischen Gründen \lstSc!resume!-Transitionen mit $\boxempty$ am
Ende der Transition. Um eingehende \lstSc!restart!- in
\lstSc!resume!-Transitionen zu transformieren, wird der Zielzustand
$A$ in einen Initialisierungszustand $A_0$ und einen „Arbeitszustand“
$A_1$ aufgetrennt. Dabei wird $A_0$ der Zielzustand aller in $A$
eingehenden \lstSc!restart!-Transitionen. \lstSc!resume!-Transitionen
erhalten dagegen $A_1$ als neuen Zielzustand. Außerdem fügen wir
eine Transition $A_0 \rightarrow A_1$ ein, um nach der Initialisierung
$N_2$ auszuführen.

\begin{remark}
  Die Transition unterscheidet sich
  $t_1 : A_0 \xrightarrow{F \arr \pre R} A_1$ von einer Weak-Transition
  $t_2 : A_0 \xrightarrow{T} A_1$. Zwar werden führen
  beide zu einer Transition von $A_0$ nach $A_1$ im nächsten Takt.
  Allerdings wird $t_2$ nicht ausgeführt, wenn $A_0$ mit einer
  Strong-Transition (hier z.B. $c_1$) betreten worden ist. Damit
  würde $x = M$ in zwei Takten gelten (wenn nicht zuvor $c$ wahr wird).
\end{remark}

\begin{remark}
  Diese Konstruktion ist eine Erweiterung von \transRef{lone_arrow}
  auf Datenfluss in Zuständen.
\end{remark}

\section{Abgeleitete Konstrukte}
\label{sec:derived}

In diesem Abschnitt werden die Konstrukte von \Scade beschrieben,
die durch ein anderes ausgedrückt werden können. Dabei werden
wir uns nur auf Konstrukte aus dem letzten Abschnitt stützen.

\subsection{Getaktete Blöcke}
\label{sec:clocked_blocks}

\Scade erlaubt neben Automaten noch weitere Konstrukte zur
Auswahl von Datenflüssen. Diese werden als
\term[Datenfluss!getakteter]{getaktet} bezeichnet, da die jeweiligen
Bedingungen als lokaler Takt interpretiert werden können, die einen
Datenfluss ein- oder ausschalten. Es gibt dabei \lstSc!if!-
und \lstSc!when-match!-Blöcke. Wir werden beide
auf Automaten zurückführen.

\begin{figure}[h]
\centering
\includegraphics[width=0.8\textwidth]{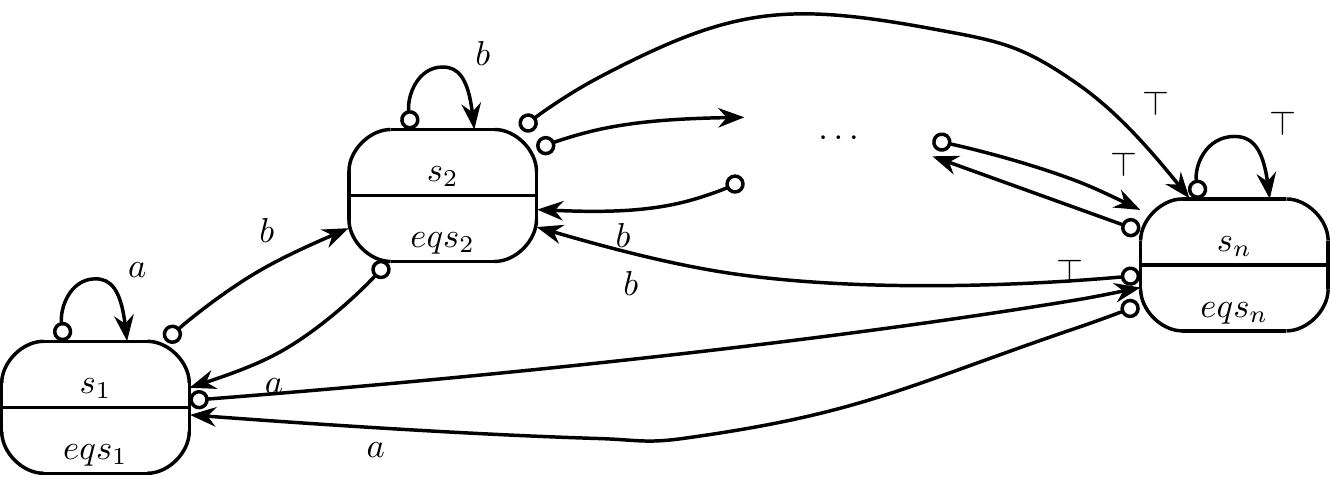}
\caption{Übersetzung von if-Blöcken}
\label{fig:scade_if_block_trans}
\end{figure}

\begin{translation}
\label{trans:clocked_if}
  Sei ein Block
  \begin{lstlisting}[language=scade,mathescape=true]
activate
  if a then $eqs_1$
  else if b then $eqs_2$
  $\vdots$
  else $eqs_n$
  \end{lstlisting}
  gegeben. Dabei ist $eqs_i$ jeweils ein Block von Gleichungen
  (Zuweisungen, Automaten, getaktete Blöcke). Wir erzeugen
  daraus einen Automaten wie in \figRef{scade_if_block_trans}.
  Ein Zustand repräsentiert hier also jeweils einen Block mit
  einer Strong-Transition, die die jeweilige Bedingung kodiert.
\end{translation}

\begin{translation}
  Es sei ein Block
  \begin{lstlisting}[language=scade,mathescape=true]
when M match
  | $P_1$ : $eqs_1$
  | $P_2$ : $eqs_2$
  $\vdots$
  | $P_n$ : $eqs_n$
  \end{lstlisting}
  gegeben. Wobei $P_i = E$ mit $E$ ein Enum oder $P_i = \_$.
  Wir kodieren zunächst die Berechnung des Zielzustandes in
  \begin{lstlisting}[language=scade,mathescape=true]
s = case M of
  | $P_1$ : $1$
  | $P_2$ : $2$
  $\vdots$
  | $P_n$ : $n$
  \end{lstlisting}
  und benutzen dies zur Konstruktion eines Automaten aus
  \begin{lstlisting}[language=scade,mathescape=true]
activate
  if s = 1 then $eqs_1$
  else if s = 2 then $eqs_2$
  $\vdots$
  else $eqs_n$
  \end{lstlisting}
  mit \transRef{clocked_if}.
\end{translation}

\subsection{Followed-By}
\label{sec:fby}

\lstSc!fby! ist eine kurze Schreibweise
für eine Kette von initialisierten
\lstSc!pre!-Ausdrücken.
\lstSc!fby(M;n;N)! ist dabei äquivalent
zu
\begin{equation*}
  \underbrace{M \arr \pre \:
    (M \arr \pre \: (\dotsm \: M \arr \pre }_{\text{n-mal}} N).
\end{equation*}

\subsection{times-Operator}
\label{sec:times_op}

Der \lstSc!times!-Operator ist kein
primitiver Operator. Eine mögliche Definition findet
sich in \cite[S. 98]{ScadeRef}:

\begin{translation}
  Ein Ausdruck

  \begin{lstlisting}[language=scade]
N times M
  \end{lstlisting}

  ist äquivalent zu

  \begin{lstlisting}[language=scade]
times_behaviour(N, M).
  \end{lstlisting}
  Dabei ist \lstSc!times_behaviour! durch den
  folgenden Knoten gegeben.

\begin{lstlisting}[language=scade]
node times_behavior (n : int; c : bool) returns (o : bool)
var
  v3, v4 : int;
let
  v4 = n -> pre (v3);
  v3 = if (v4 < 0)
    then v4
    else (if c then v4 - 1 else v4);
  o = c and (v3 = 0);
tel
\end{lstlisting}

\end{translation}





\section{Optimierungen}
\label{sec:scade_optimisations}

Wir werden hier nur eine kleine Auswahl möglicher Optimierungen
bei der Transformation beschreiben. Zunächst kann die Zahl der
Speichervariablen durch die Eliminierung von $\pre$ und die von
nicht initialisierten Variablen reduziert werden
(\ref{sec:optimise_pre}).

Auch ist die Zahl der Variablen, die
bei der automatischen Generierung von \Scade-Code aus dem
graphischen Modell entstehen recht hoch. Häufig ist an eine
solche Variable auch nur ein Teilausdruck gebunden, der nur
einmal verwendet wird. Diesem Umstand kann mit Inlining
(\ref{sec:scade_inlining}) begegnet werden.

\subsection{pre}
\label{sec:optimise_pre}

Bei dieser Optimierung wird versucht, $\pre$ in einem Ausdruck
möglichst weit an die Wurzel des Ausdrucks zu bringen. Dies hat
mehrere Gründe:

\begin{enumerate}
\item Ein $\pre$, dass tiefer in einem Ausdruck steckt, wird
  herausgezogen und an eine neue Variable zugewiesen
  (\seeR{sec:stream_functions}). Es kann also ggf. die Generierung
  zusätzlicher Variablen verhindert werden.
\item Ein $\pre$ ohne Initialisierung durch $\arr$ lässt den
  Wert der generierten Zustandsvariable im Takt 0 offen.
  Da aber ein Fluss immer initialisiert sein muss, existiert
  irgendwo ein passender $\arr$, der dann aber ein $ite$
  erzeugt (\seeR{sec:stream_functions}). Dies kann  verhindert werden,
  wenn $\arr$ und $\pre$ zusammengebracht werden.
\item Wenn ein Ausdruck $f(\pre M_1, \dotsc, \pre M_k)$ existiert,
  dann werden $k$ Zustandsvariablen erzeugt (s. oben), obwohl
  eine genügen würde. Dabei ist $f$ eine beliebige Basisoperation
  (\seeR{sec:basic_functions}).
\end{enumerate}

Im Wesentlichen nutzen wir hier die Distributivität von $\pre$ über
nicht-Stream Operatoren aus:
\begin{equation}
  f(\pre M_1, \dotsc, \pre M_k) \equiv \pre f(M_1, \dotsc, M_k).
  \label{eq:distr_pre}
\end{equation}

Wobei $f$ eine Basisoperation ist. Diese Termersetzung stoppt,
wenn $M \arr \pre N$ oder eine Zuweisung erreicht ist. Letzteres
kann zusätzlich unterstützt werden:

\begin{equation*}
  x = \pre M \leadsto (x = M, [x \mapsto \pre x])
\end{equation*}

Dies erzeugt eine Substitution, die im Gültigkeitsbereich
von $x$ angewendet werden muss.

\eqRef{distr_pre} kann auch auf initialisierte Ströme
verallgemeinert werden, wenn alle $M_i$ den gleichen Typ haben:
\begin{equation*}
  f(N \arr \pre M_1, \dotsc, N \arr \pre M_k)
    \equiv N \arr \pre f(M_1, \dotsc, M_k).
  \tag{\ref{eq:distr_pre}'}
\end{equation*}

\subsection{Inlining}
\label{sec:scade_inlining}

Wie bereits erwähnt erzeugt \Scade bei der Generierung von Code
aus dem graphischen Modell viele Zwischenvariablen. Dies geschieht
an jeder Kante in einem Datenfluss. Wir betrachten
dazu das Beispiel in \figRef{scade_chrono}. Der generierte Code
für den Zustand \lstSc!START! ist in
\lstRef{chrono_code_start} dargestellt.

\begin{listing}[h]
\begin{lstlisting}[language=scade,
    numbers=left,xleftmargin=25pt]
state START
  unless
    if StSt resume STOP;
  var
    d : int;
    _L2 : int;
    _L3 : int;
    _L6 : int;
    _L8 : bool;
  let
    d= _L2;
    _L2= 0 -> (pre d + 1) mod 100;
    s= _L3;
    _L3= if d < (0 -> pre d)
          then (last 's + 1) mod 60
          else last 's;
    _L6= if s < last 's
          then (last 'm + 1) mod 60
          else last 'm;
    m= _L6;
    run= _L8;
    _L8= true;
  tel
\end{lstlisting}

  \caption{Zustand START des Knotens Chrono}
  \label{lst:chrono_code_start}
\end{listing}

Hier lässt sich nun gut erkennen, dass jede Zuweisung zunächst
eine lokale Variable erhält (Zeile 10, 13, 20 und 21) und
der jeweilige Ausdruck an diese Variable zugewiesen wird.
In diesem Beispiel ist die Zahl der Variablen relativ niedrig,
da textuelle Ausdrücke verwendet worden sind. Wären die Ausdrücke
graphisch modelliert worden, wäre für jeden Teilausdruck eine
solche Variable erzeugt worden.

Wie bei jedem Optimierungsverfahren, muss man abwägen, wann eine
Variable durch ihren definierenden Ausdruck ersetzt werden soll.
Wir werden dazu zwei Parameter verwenden: Ersetzungtiefe $D$ („depth“)
und -häufigkeit $B$ („branching“). Weiterhin werden wir Ausdrücke
mit temporalen Operatoren an der Wurzel eines Ausdrucks nicht
ersetzen, da diese später ohnehin wieder abgerollt werden
(\seeR{sec:stream_functions}).

\begin{algorithm}[h]
\caption{Inlining in \Scade-Code}
\label{alg:scade_inlining}
\begin{algorithmic}[1]
  \REQUIRE \Scade-Knoten mit Gleichungen $E$,
    $O \subseteq E$ Gleichungen für Ausgabevariablen
  \ENSURE Gleichungen $E'$
  \STATE $W \colonequals (E \setminus O) \times \{0\}$
    \COMMENT {Die zweite Komponente ist die bisherige
      Einsetzungstiefe der Gleichung.}
  \STATE $E' \colonequals O$
  \WHILE {$e \colonequals (x = M, d_x) \in W$, so dass kein
    $x' \in FV(M)$ mit einer Gleichung
    $(x' = M', d_{x'}) \in W$ existiert} \label{inlining_for}
    \IF{$M$ hat keinen Streamoperator an der Wurzel, $d_x < D$
      und $x$ wird höchstens $B$-mal verwendet}
      \label{inlining_cond}
      \STATE Ersetze $x$ durch $M$ in allen Gleichungen
        in $(y = N, d_y) \in W$ mit $y \not= x$ und $d_y \leq D$
      \STATE Setze bei all diesen $d_y \colonequals d_y + d_x$
    \ELSE
      \STATE $E' \colonequals E' \cup \{x = M\}$
    \ENDIF
    \STATE $W \colonequals W \setminus \{e\}$
  \ENDWHILE
\end{algorithmic}
\end{algorithm}

Das Inlining ist in \algRef{scade_inlining} beschrieben.
$FV(M)$ bezeichnet dort die Menge der freien Variablen des Ausdrucks
$M$.\footnote{Da in \Scade-Ausdrücken keine Variablen gebunden werden
können, sind dies trivialerweise alle Variablen, die in $M$
verwendet werden.} Wir wollen kurz auf die Bedingungen der Schleife
und Verzweigungen eingehen. Die Auswahl der
nächsten Gleichung in Schritt \ref{inlining_for} sorgt dafür,
dass wir „bottom-up“ Ausdrücke einsetzen. Damit bekommt
$B$ die erwartete Bedeutung, dass jede Gleichung höchstens
$B$-mal eingesetzt wird.

Die Bedingung in Schritt \ref{inlining_cond} sorgt dafür, dass
wir keine Ausdrücke einsetzen, die später wieder heraus gezogen werden.
Außerdem wird $x$ nur wenn es noch nicht seine maximale Einsetztiefe
erreicht hat und eingesetzt werden darf, durch seine Definition
ersetzt. Die Ersetzung darf nicht in die eigene Definition erfolgen,
da dies dazu führen würde, dass wir eine nicht terminierende
Ersetzung vornehmen würden. Danach wird die Gleichung als abgearbeitet
angesehen (die Gleichung wird aus $W$ entfernt). Wenn die
Gleichung nicht eingesetzt werden konnte, muss sie erhalten bleiben
und wird in $E'$ gespeichert.

Um die Definitionen von Ausgaben zu erhalten, wird das Inlining
für diese Gleichungen komplett unterbunden.

Am Ende müssen alle Variablen, die keine Definition mehr in $E'$
haben, aus den Deklarationen des Gültigkeitsbereiches entfernt werden.

\begin{listing}[h]
\begin{lstlisting}[language=scade,
    numbers=left,xleftmargin=25pt]
state START
  unless
    if StSt resume STOP;
  var
    d : int;
  let
    d= 0 -> (pre d + 1) mod 100;
    s=  if d < (0 -> pre d)
          then (last 's + 1) mod 60
          else last 's;
    m=  if s < last 's
          then (last 'm + 1) mod 60
          else last 'm;
    run= true;
  tel
\end{lstlisting}

  \caption{Zustand START des Knotens Chrono nach dem Inlining}
  \label{lst:chrono_code_start_inlined}
\end{listing}

Das Ergebnis für das obige Listing mit $D \geq 1$ und $B = 1$
ist in \lstRef{chrono_code_start_inlined} dargestellt.
Die Variablen $s,m,run$ sind Ausgaben,
werden also nicht weiter ersetzt. $d$ kann nicht in der eigenen
Definition (Zeile 7) ersetzt werden und auch nicht in der
Definition von $s$, da $d$ insgesamt dreimal verwendet wird.

\chapter{Implementierung}
\label{chap:implementation}

In diesem Kapitel wird die Implementierung der zuvor beschriebenen
Transformationen vorgestellt. Die Implementierung erfolgte in
Haskell.

\section{Architektur}
\label{sec:architecture}

Die Implementierung setzt sich aus folgenden Teilen zusammen:
\begin{itemize}
\item eine Bibliothek für die Sprache \Lama (language-lama,
  \chapRef{lama})
\item ein Programm zur Transformation von \Scade nach
  \Lama (scade2lama, \chapRef{scade2lama})
\item ein Programm zur Verifikation von \Lama-Programmen
  mit SMT (lamasmt, \chapRef{lama_smt})
\item ein Interpreter für \Lama (lama-interpreter, \chapRef{lama})
\end{itemize}

Die Implementierung nutzt zwei von Henning Günther
am Institut entwickelte Bibliotheken:
\begin{itemize}
\item language-scade \cite{ScadeBib} -- \Scade-Syntax in Haskell
\item smtlib2 \cite{SMTLib2Haskell} -- SMTLib2-Umsetzung in Haskell
\end{itemize}

\begin{figure}[h]
  \centering
  \begin{tikzpicture}[anchor=base,scale=0.7, every node/.style={scale=0.7},baseline=(current bounding box.center)]
    \pgfsetcolor{black}
  \draw [->] (294.14bp,75.947bp) .. controls (278.27bp,65.748bp) and (256.5bp,51.748bp)  .. (229.73bp,34.542bp);
  \draw [->] (64.495bp,71.831bp) .. controls (64.281bp,64.131bp) and (64.027bp,54.974bp)  .. (63.511bp,36.413bp);
  \draw [->] (316bp,71.831bp) .. controls (316bp,64.131bp) and (316bp,54.974bp)  .. (316bp,36.413bp);
  \draw [->] (201.01bp,71.831bp) .. controls (201.44bp,64.131bp) and (201.95bp,54.974bp)  .. (202.98bp,36.413bp);
  \draw [->] (93.129bp,75.43bp) .. controls (113.63bp,64.812bp) and (141.67bp,50.284bp)  .. (173.5bp,33.798bp);
\begin{scope}
  \definecolor{strokecol}{rgb}{0.0,0.0,0.0};
  \pgfsetstrokecolor{strokecol}
  \draw (316bp,18bp) ellipse (30bp and 18bp);
  \draw (316bp,13bp) node {smtlib2};
\end{scope}
\begin{scope}
  \definecolor{strokecol}{rgb}{0.0,0.0,0.0};
  \pgfsetstrokecolor{strokecol}
  \draw (204bp,18bp) ellipse (60bp and 18bp);
  \draw (204bp,13bp) node {language-lama};
\end{scope}
\begin{scope}
  \definecolor{strokecol}{rgb}{0.0,0.0,0.0};
  \pgfsetstrokecolor{strokecol}
  \draw (65bp,90bp) ellipse (47bp and 18bp);
  \draw (65bp,85bp) node {scade2lama};
\end{scope}
\begin{scope}
  \definecolor{strokecol}{rgb}{0.0,0.0,0.0};
  \pgfsetstrokecolor{strokecol}
  \draw (316bp,90bp) ellipse (34bp and 18bp);
  \draw (316bp,85bp) node {lamasmt};
\end{scope}
\begin{scope}
  \definecolor{strokecol}{rgb}{0.0,0.0,0.0};
  \pgfsetstrokecolor{strokecol}
  \draw (200bp,90bp) ellipse (64bp and 18bp);
  \draw (200bp,85bp) node {lama-intepreter};
\end{scope}
\begin{scope}
  \definecolor{strokecol}{rgb}{0.0,0.0,0.0};
  \pgfsetstrokecolor{strokecol}
  \draw (63bp,18bp) ellipse (63bp and 18bp);
  \draw (63bp,13bp) node {language-scade};
\end{scope}
  \end{tikzpicture}
  \caption{Abhängigkeiten der Programme}
  \label{fig:prog_deps}
\end{figure}
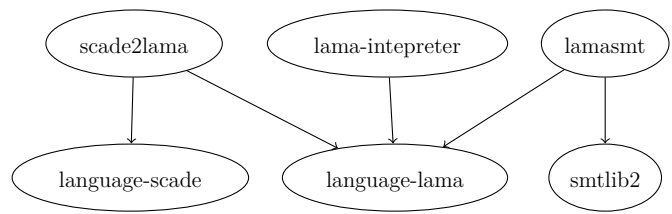

Die Abhängigkeiten sind in \figRef{prog_deps} dargestellt.

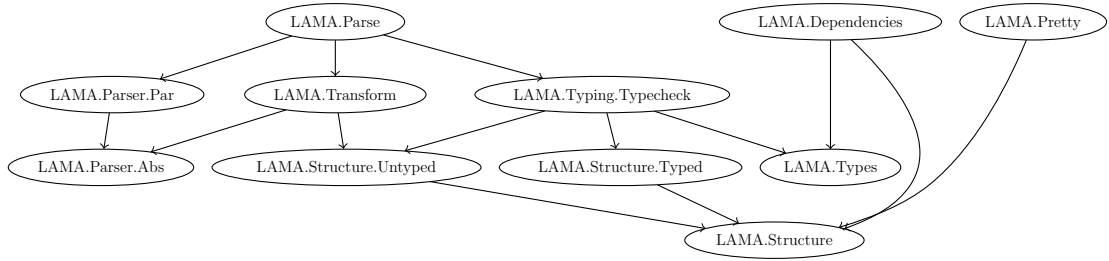
\begin{figure}[h]
  \centering
  \begin{tikzpicture}[anchor=base,scale=0.38, every node/.style={scale=0.5},baseline=(current bounding box.center)]
    \pgfsetcolor{black}
  \draw [->] (278.25bp,219.95bp) .. controls (244.09bp,208.71bp) and (195.9bp,192.87bp)  .. (148.81bp,177.39bp);
  \draw [->] (807bp,215.76bp) .. controls (807bp,191.2bp) and (807bp,147.25bp)  .. (807bp,108.09bp);
  \draw [->] (368.45bp,220.96bp) .. controls (409.39bp,209.71bp) and (469.01bp,193.32bp)  .. (524.92bp,177.96bp);
  \draw [->] (637.17bp,72.754bp) .. controls (658.39bp,62.634bp) and (685.37bp,49.771bp)  .. (717.17bp,34.609bp);
  \draw [->] (526.67bp,145.84bp) .. controls (488.75bp,134.96bp) and (438.57bp,120.57bp)  .. (389.01bp,106.35bp);
  \draw [->] (272.5bp,146.82bp) .. controls (236.69bp,135.61bp) and (187.77bp,120.29bp)  .. (139.83bp,105.28bp);
  \draw [->] (587.54bp,143.83bp) .. controls (589.49bp,136.05bp) and (591.81bp,126.77bp)  .. (596.4bp,108.41bp);
  \draw [->] (414.23bp,75.904bp) .. controls (489.61bp,62.981bp) and (600.43bp,43.984bp)  .. (684.51bp,29.57bp);
  \draw [->] (634.11bp,145.57bp) .. controls (669.75bp,134.11bp) and (717.22bp,118.86bp)  .. (763.28bp,104.05bp);
  \draw [->] (323.78bp,143.83bp) .. controls (324.95bp,136.13bp) and (326.35bp,126.97bp)  .. (329.19bp,108.41bp);
  \draw [->] (99.224bp,143.83bp) .. controls (98.048bp,136.13bp) and (96.649bp,126.97bp)  .. (93.813bp,108.41bp);
  \draw [->] (827.52bp,216.09bp) .. controls (860.35bp,185.23bp) and (917.77bp,120.68bp)  .. (885bp,72bp) .. controls (872.21bp,53.002bp) and (851.46bp,40.685bp)  .. (820.63bp,29.433bp);
  \draw [->] (321bp,215.83bp) .. controls (321bp,208.13bp) and (321bp,198.97bp)  .. (321bp,180.41bp);
  \draw [->] (999.18bp,215.92bp) .. controls (985.77bp,182.72bp) and (952.87bp,111.83bp)  .. (903bp,72bp) .. controls (880.43bp,53.972bp) and (851.35bp,41.687bp)  .. (815.07bp,30.597bp);
\begin{scope}
  \definecolor{strokecol}{rgb}{0.0,0.0,0.0};
  \pgfsetstrokecolor{strokecol}
  \draw (332bp,90bp) ellipse (132bp and 18bp);
  \draw (332bp,85bp) node {LAMA.Structure.Untyped};
\end{scope}
\begin{scope}
  \definecolor{strokecol}{rgb}{0.0,0.0,0.0};
  \pgfsetstrokecolor{strokecol}
  \draw (583bp,162bp) ellipse (125bp and 18bp);
  \draw (583bp,157bp) node {LAMA.Typing.Typecheck};
\end{scope}
\begin{scope}
  \definecolor{strokecol}{rgb}{0.0,0.0,0.0};
  \pgfsetstrokecolor{strokecol}
  \draw (102bp,162bp) ellipse (90bp and 18bp);
  \draw (102bp,157bp) node {LAMA.Parser.Par};
\end{scope}
\begin{scope}
  \definecolor{strokecol}{rgb}{0.0,0.0,0.0};
  \pgfsetstrokecolor{strokecol}
  \draw (321bp,162bp) ellipse (89bp and 18bp);
  \draw (321bp,157bp) node {LAMA.Transform};
\end{scope}
\begin{scope}
  \definecolor{strokecol}{rgb}{0.0,0.0,0.0};
  \pgfsetstrokecolor{strokecol}
  \draw (91bp,90bp) ellipse (91bp and 18bp);
  \draw (91bp,85bp) node {LAMA.Parser.Abs};
\end{scope}
\begin{scope}
  \definecolor{strokecol}{rgb}{0.0,0.0,0.0};
  \pgfsetstrokecolor{strokecol}
  \draw (321bp,234bp) ellipse (68bp and 18bp);
  \draw (321bp,229bp) node {LAMA.Parse};
\end{scope}
\begin{scope}
  \definecolor{strokecol}{rgb}{0.0,0.0,0.0};
  \pgfsetstrokecolor{strokecol}
  \draw (807bp,234bp) ellipse (109bp and 18bp);
  \draw (807bp,229bp) node {LAMA.Dependencies};
\end{scope}
\begin{scope}
  \definecolor{strokecol}{rgb}{0.0,0.0,0.0};
  \pgfsetstrokecolor{strokecol}
  \draw (752bp,18bp) ellipse (88bp and 18bp);
  \draw (752bp,13bp) node {LAMA.Structure};
\end{scope}
\begin{scope}
  \definecolor{strokecol}{rgb}{0.0,0.0,0.0};
  \pgfsetstrokecolor{strokecol}
  \draw (1006bp,234bp) ellipse (72bp and 18bp);
  \draw (1006bp,229bp) node {LAMA.Pretty};
\end{scope}
\begin{scope}
  \definecolor{strokecol}{rgb}{0.0,0.0,0.0};
  \pgfsetstrokecolor{strokecol}
  \draw (601bp,90bp) ellipse (119bp and 18bp);
  \draw (601bp,85bp) node {LAMA.Structure.Typed};
\end{scope}
\begin{scope}
  \definecolor{strokecol}{rgb}{0.0,0.0,0.0};
  \pgfsetstrokecolor{strokecol}
  \draw (807bp,90bp) ellipse (69bp and 18bp);
  \draw (807bp,85bp) node {LAMA.Types};
\end{scope}
  \end{tikzpicture}
  \caption{Abhängigkeiten von language-lama}
  \label{fig:lama_language_deps}
\end{figure}

In \figRef{lama_language_deps} sind die verfügbaren Module
mit ihren internen Abhängigkeiten dargestellt. Es gibt
ein Modul zum Einlesen von \Lama-Programmen
(\lstinline!LAMA.Parse!). Dies umfasst auch die Typprüfung.
Außerdem existiert ein Modul zur Abhängigkeitsberechnung
(\lstinline!LAMA.Dependencies!)
und eines zum Erzeugen von \Lama-Programmen
(\lstinline!LAMA.Pretty!).

Die Grammatik von \Lama wurde mit Hilfe von \lang{BNFC}
(\cite{BNFC}) implementiert. Damit können Grammatiken in
annotierter Backus-Naur-Form (BNF in der Label für jede
Alternative einer Produktion vergeben werden) beschrieben werden,
um anschließend daraus Dokumentation und Code generieren zu können.

Die Struktur eines \Lama-Programms
ist in dem Modul \lstinline!LAMA.Structure! abstrakt
umgesetzt und mit/ohne Typen in den Modulen
\lstinline!LAMA.Structure.Typed! bzw.
\lstinline!LAMA.Structure.Untyped! spezialisiert. Weiterhin
ist der Bezeichner abstrakt gehalten, um bspw. beim Einlesen
Informationen über die Position im Quelltext speichern zu können.

Der Interpreter setzt die beschriebene Semantik um. Dabei baut
dieser auf den berechneten Abhängigkeiten auf
(\lstinline!LAMA.Dependencies!).

scade2lama führt die Transformation analog zu der Beschreibung
in \chapRef{scade2lama} in drei Schritten durch:
\begin{enumerate}
\item Optimierungen in \Scade
\item Termersetzung in \Scade
\item Transformation nach \Lama
\end{enumerate}

Dabei wird jeder Ersetzungschritt auch in einem Modul umgesetzt
(\lstinline!Rewrite.*!). Dies ist zwar relativ ineffizient in
der Berechnung, aber dafür konzeptionell klarer.

Die Transformation ist in mehrere Teiltransformationen aufgeteilt:
\begin{itemize}
\item Ausdrücke
\item Einfache Gleichungen
\item Automaten
\item Pakete und Knoten
\end{itemize}

Die Verifikation von \Lama-Programmen mit lamasmt ist
in die Transformation in \lang{SMTLib2}, die Prüfung der
gewünschten Eigenschaft und das Auslesen von Modellen aufgeteilt.
Die Prüfung kann dabei mit verschiedenen Strategien
(BMC und k-Induktion) durchgeführt werden.

Zum Lösen des erzeugten SMT-Problems benutzen wir eine
\term[inkrementell]{inkrementelle} Prozedur und \term{Backtracking}
(\cite[S.843]{HBSat}). Beides wird von der Bibliothek
\cite{SMTLib2Haskell} unterstützt. Wie in \secRef{induction} erwähnt,
teilen sich bei BMC und auch bei der k-Induktion die jeweiligen
Schritte ein Großteil der Formeln. Daher kann ein lernender Solver
hier bessere Geschwindigkeiten erreichen. Um dies voll ausnutzen zu
können, muss der Solver auch Backtracking unterstützen.
In \lang{SMTLib2} wird dies als
\term{Assertion-Stack} realisiert. Hier kann eine Menge von Formeln
auf diesem Stack angelegt und zusammen mit allen anderen Formeln
auf Erfüllbarkeit geprüft werden. Danach kann diese Menge wieder vom
Stack genommen werden.

\section{Umgesetzte Transformationen}
\label{sec:implemented_transformations}

Von den beschriebenen Transformationen von \Scade nach \Lama konnten
noch nicht alle umgesetzt werden. Es fehlen noch
\begin{itemize}
\item die Transformation von $\pre$ und $\last$ in \Scade-Zuständen,
\item die Behandlung von in Zuständen deklarierten Variablen
  (dies wird derzeit teilweise über Inlining unterstüzt),
\item \lstSc!when .. match!,
\item \lstSc!restart!-Transitionen und
\item Inlining außerhalb von Zuständen.
\end{itemize}

In der Transformation von \Lama nach SMT werden $\ty{uint[n]}$ und
$\ty{sint[n]}$ derzeit nicht berücksichtigt. Dies kann z.B. auf
Basis von Bitvektoren oder auch mit Hilfe von $\ty{int}$ und
entsprechenden zusätzlichen Bedingungen geschehen.

\section{Korrektheit}
\label{sec:soundness}

Die Korrektheit der Implementierung wurde anhand
von passend gewählten Beispielen getestet. Dabei prüft jedes
Beispiel jeweils eine spezielle Transformation
oder ein Zusammenspiel von Transformationen. Die entsprechenden
Beispiele finden sich zum Teil in \secRef{source_examples}.

\section{Benutzung der Programme}
\label{sec:usage}
Um eine Eigenschaft eines \Scade-Programms zu verifizieren bzw.
zu debuggen, sind zwei Schritte nötig:
\begin{enumerate}
  \item Transformation nach \Lama
  \item Ausführung in SMT
\end{enumerate}

Beide Programme bieten die Option „-h“ an, mit der eine Hilfe
zur Benutzung ausgegeben werden kann.

Der erste Schritt erfolgt mit dem Programm scade2lama. Dabei muss
der Name des Knotens übergeben werden, dessen Eigenschaften
verifiziert werden sollen. Diese Eigenschaft kann ebenfalls
als Parameter angegeben werden. Das Programm generiert ein
\Lama-Programm, dass an lamasmt übergeben werden kann. Weiterhin
können Optimierungen parametrisiert werden.

lamasmt benötigt als Eingabe ein solches \Lama-Programm. Außerdem
kann die verwendete Strategie (BMC, k-Induktion) mit weiteren
Parametern (Abrolltiefe, Ausgaben) konfiguriert werden.
lamasmt hat die Möglichkeit im Falle eines Fehlers,
ein vollständiges Modell für alle Variablen auszugeben oder eine
\Scade-Szenariodatei zu generieren. Zuletzt können noch verschiedene
Kodierungen (\seeR{sec:comparison_implementation}) konfiguriert
werden.

\section{Performance}
\label{sec:performance}

Wir wollen hier einerseits mit dem Referenzsystem, dem
\Scade DV (\term{Design Verifier}), vergleichen. Andererseits
wollen wir uns ansehen, wie sich verschiedene Kodierungen für
den SMT-Solver auf die Berechnungsgeschwindigkeit auswirken.

Die hier angegebenen Zeiten für \Scade wurden \cite{Huhn12} entnommen,
da kein nativ laufendes Windows verfügbar gewesen ist. Die Zeiten
für \Lama wurden mit Z3 (\cite{Z3Internet,Z3DeMoura08}) in der
Version 4.0 unter Linux gemessen. Es wurde ein Laptop mit einem
Intel(R) Core(TM)2 Duo CPU P9600 @ 2.53GHz und 4GB Arbeitsspeicher
benutzt. Dabei ist zu beachten, dass Z3 unter Linux in dieser
Version noch kein Multi-Threading unterstützt und daher nur
ein Kern des Prozessors genutzt wurde. 

\subsection{Vergleich mit \Scade DV}
\label{sec:comparison_prover}

Die Performance des Systems wurde anhand eines Modells eines
lokal gesteuerten Bahnübergangs evaluiert. Dieses Beispiel ist auch
Grundlage von \cite{Huhn12}. Wir werden hier fünf der beschriebenen
Fehlerfälle und den fehlerfreien Fall nutzen. Die Terminologie
ist dabei dem obigen Papier zu entnehmen.

\begin{table}[h]
  \centering
  \begin{tabular}{c|c|c|c|c}
    & \multicolumn{2}{c|}{\Scade DV}
      & \multicolumn{2}{c}{Via \Lama} \\ \hline
    Art & Bewiesen & Zeit (in Sek.)
      & Bewiesen & Zeit (in Sek.) \\ \hline
    (1) & $\checkmark$ & 12 & x (bis Tiefe 46) & (27 Std.) \\
    (2) & x & - & - & - \\
    (3) & $\checkmark$ & 82914 & x (bis Tiefe 50) & (68 Std.)
  \end{tabular}
  \caption{Vergleich der Zeiten zur Beweisführung}
  \label{tab:comparison_proof}
\end{table}

Die \tabRef{comparison_proof} legt nahe, dass k-Induktion alleine
nicht stark genug ist, um die gewünschte Eigenschaft zu beweisen.
Da dies aber in anderen Fällen (s. \secRef{source_examples})
möglich ist, muss auf eine andere Art die Induktionsvoraussetzung
gestärkt werden. Darauf gehen wir in \secRef{future_research} ein.

\begin{table}[h]
  \centering
  \begin{tabular}[h]{c|c|c|c|c}
    & \multicolumn{2}{c|}{\Scade DV}
      & \multicolumn{2}{c}{Via \Lama} \\ \hline
    Fehler & Gefunden & Zeit (in Sek.)
      & Gefunden & Zeit (in Sek.) \\ \hline
    L1         & $\checkmark$ & 1 & $\checkmark$ & 422 \\
    BS13       & $\checkmark$ & 1 & $\checkmark$ & 491 \\
    B7 + BS11  & $\checkmark$ & 11 & $\checkmark$ & 418 \\
    B9         & x & - & x & - \\
    L3 + BS11  & x & - & x & -
  \end{tabular}
  \caption{Vergleich der Zeiten zum Finden von Fehlern}
  \label{tab:comparison_errors}
\end{table}

In \tabRef{comparison_errors} ist für die untersuchten Fehlerfälle
jeweils angegeben, ob sie gefunden worden sind und wie lange
dies jeweils gedauert hat. Die Prüfungen wurden jeweils die
„debug“-Strategie von \Scade bzw. BMC bei \Lama verwendet.
Für B9 bzw. L3 + BS11 konnte damit aber jeweils kein Fehler
gefunden werden. Daher wurde jeweils versucht, nachzuweisen, dass
das Modell fehlerfrei ist. Bei B9 wurde das BMC nach 24 Min. bei
einer Tiefe von 35 abgebrochen, bei L3 + BS11 nach 13 Min. bei
Tiefe 30. Für B9 wurde die k-Induktion nach 8 Std. bei Tiefe 30
abgebrochen.

\subsection{Vergleich verschiedener Implementierungstechniken}
\label{sec:comparison_implementation}

Es werden zwei Arten der Implementierung jeweils für die Kodierung
der natürlichen Zahlen $\N$ (als Indizes von Strömen)
und von Enumerations bereitgestellt. Zum einen können jeweils
\term[abstrakter Datentyp]{abstrakte Datentypen} genutzt werden.
Dies muss vom Solver unterstützt werden (wie z.B. von Z3). Dabei wird
$\N$ durch einen Typ
\begin{lstlisting}
  nat ::= zero | succ nat
\end{lstlisting}
repräsentiert und ein Enum $E = \{c_1, \dotsc, c_k\}$ durch
\begin{lstlisting}[mathescape=true]
  E ::= $c_1$ | $\dotsm$ | $c_k$.
\end{lstlisting}

Als alternative Kodierung für $\N$ bietet sich $\Z$ an. Dabei müssen
Variablen (z.B. die Induktionsvariable $n$) auf nicht-negative
Zahlen eingeschränkt werden.

Die zweite Kodierung für Enums erfolgt mit Bitvektoren.
Dabei werden die Konstruktoren als Bitvektor kodiert. Diese
Kodierung wird gegenüber z.B. $\Z$ vorgezogen, da die (endliche)
Zahl der verwendeten Elemente bei der Transformation bekannt ist.

Im Falle von Z3 hat sich gezeigt, dass die Kodierung von $\N$
als Datentyp vorteilhaft ist. Dagegen kann Z3 mit Enumerations als
Bitvektoren besser umgehen. In \tabRef{comparison_encodings_l1}
sind die Laufzeiten für das Finden des Fehlers L1 aus
\ref{sec:comparison_prover} zusammengestellt.
\begin{table}[H]
  \centering
  \begin{tabular}{c|c|c}
              & Datentyp & Ganze Zahlen \\ \hline
    Datentyp  & 812      & 820 \\ \hline
    Bitvektor & 422      & 673 \\
  \end{tabular}
  \caption{Verschiedene Kodierungen von Enums (Zeilen) und
    $\N$ (Spalten)}
  \label{tab:comparison_encodings_l1}
\end{table}


\chapter{Zusammenfassung}
\label{chap:conclusion}

Wir wollen hier kurz darauf eingehen, was noch an offenen Arbeiten
verbleibt. Hier sind einerseits fehlende Elemente der Sprache
\Scade zu nennen (\secRef{missing_scade}). Andererseits ermöglichen
die Transformationen dieser Arbeit weitere Untersuchungen
hinsichtlich von Implementierungstechniken (\secRef{future_research}).

Zuletzt fassen wir die Ergebnisse der Arbeit zusammen.

\section{Fehlende Sprachelemente \Scade}
\label{sec:missing_scade}

Wir wollen zunächst zusammenstellen, welche Sprachelemente von
\Scade in \chapRef{scade2lama} nicht berücksichtigt worden sind.
Die Struktur orientiert sich dabei an den Abschnitten
der Grammatik in \cite[A-1]{ScadeRef}. Wir werden jeweils angeben,
ob und wie sich die jeweiligen Sprachelemente im aktuellen System
umsetzen lassen.

\begin{itemize}
\item Packages
  \begin{itemize}
  \item Interfaceart (z.B. Sichtbarkeit) -- statische Analyse bei
    der Transformation. Kann ggf. aber auch ignoriert werden, da
    wir davon ausgehen, dass wir ein gültiges Modell als Eingabe
    erhalten.
  \item \lstSc!open! -- ebenfalls statische analysierbar.
  \end{itemize}
\item Typen
  \begin{itemize}
  \item \lstSc!char! -- kann bspw. als $\ty{uint[8]}$ interpretiert
    werden.
  \item Records -- als Produkt darstellbar.
  \item Typvariablen -- Termersetzung bei Instantiierung von
    Operatoren.
  \end{itemize}
\item \lstSc!group! -- entsprechen den Produkten in \Lama. Im Falle
  von Knoten wird dies bereits implizit umgesetzt.
\item Globale Deklarationen
  \begin{itemize}
  \item \lstSc!sensor! -- globale, nur lesbare Variable. Kann also
    als Eingabe realisiert werden, die zur Stelle der Verwendung
    durchgereicht wird.
  \end{itemize}
\item Variablendeklarationen
  \begin{itemize}
  \item \lstSc!clock! / \lstSc!when! -- s. bei getaktete Ausdrücken.
  \item \lstSc!probe! -- garantiert, dass eine Variable erhalten
    bleibt. Dies ist nur zum Debuggen in Tools relevant. Kann hier also
    ignoriert werden.
  \end{itemize}
\item Operatoren
  \begin{itemize}
  \item \lstSc!function! -- kann durch Knoten umgesetzt oder direkt
    ersetzt.
  \item Interfaceart (z.B. Sichtbarkeit) -- auch hier statische
    Analyse (s. bei Paketen).
  \item Statische Eingaben (Größenangaben) -- können statisch bei
    Benutzung von Knoten ersetzt werden.
  \item \lstSc!where ... numeric! -- dies erlaubt polymorphe
    Operatoren in den numerischen Typen
    $\{\ty{int}, \ty{integer}, \ty{real}, \ty{float}\}$
    (\cite[\secRefO{7.1}]{ScadeRef},
    \cite[\secRefO{8.2}]{ScadePrimer}). Dies kann ebenfalls statisch
    aufgelöst werden.
  \item \lstSc!specialize! -- dies erlaubt mit \lstSc!imported! und
    \lstSc!where ... numeric! zusammen die Angabe und Implementierung
    von polymorphen Schnittstellen
    (\cite[\secRefO{7.1}]{ScadeRef},
    \cite[\secRefO{8.3}]{ScadePrimer}). Dies kann also ebenfalls durch
    statische Analyse aufgelöst werden.
  \end{itemize}
\item Scope-Deklarationen
  \begin{itemize}
  \item Signale -- Dies sind eine besondere Form lokaler \ty{bool}
    Variablen (\cite[\secRefO{7.4, 8.1, 8.3.3, 9.1.1}]{ScadeRef}),
    können also auch so in \Lama dargestellt werden.
  \end{itemize}
\item Gleichungen
  \begin{itemize}
  \item \lstSc!guarantee! -- dies kann als zusätzliche Invariante
    angenommen werden (\cite[\secRefO{8.1}]{ScadeRef},
    \cite[\secRefO{6.1}]{ScadePrimer}).
  \item \lstSc!emit! -- s. bei den Signalen
  \item \lstSc!returns! wird derzeit ignoriert (im Moment nehmen wir
    „\lstSc!..!“ an).
  \end{itemize}
\item Automaten
  \begin{itemize}
  \item Datenfluss an Transitionen (\cite[S. 79]{ScadeRef})
    -- im Wesentlichen eine Transformation von Mealy- in
    Moore-Automaten.
  \item \lstSc!final! -- bekommt nur in Verbindung mit \lstSc!synchro!
    eine Bedeutung, s. dort.
  \item Verzweigende Kanten -- der entsprechende Hypergraph kann
    auf einen einfachen Graphen zurückgeführt werden
  \item \lstSc!synchro! -- eine solche Transition wird ausgelöst,
    wenn alle Subautomaten eines Zustandes, ihren Finalzustand
    erreicht haben (\cite[\secRefO{8.3.1}]{ScadeRef},
    \cite[\secRefO{3.5.2}]{ScadePrimer}). Dies kann also durch eine
    Konjunktion von Variablen simuliert werden, die wahr sind, wenn
    der jeweilige Automat im Finalzustand ist.
  \end{itemize}
\item Ausdrücke
  \begin{itemize}
  \item Getaktete Ausdrücke -- können z.B. durch zusätzlichen Fluss
    mit getakteten if-Blöcken umgesetzt werden
  \item Tupel (Listenausdrücke) -- können durch Produkte dargestellt
    werden (s. oben bei \lstSc!group!)
  \end{itemize}
\item Sequentielle Operatoren
  \begin{itemize}
  \item \lstSc!when!/\lstSc!merge! (s.o. bei getakteten Ausdrücken)
  \end{itemize}
\item Kombinatorische Operatoren
  \begin{itemize}
  \item Casts -- benötigen primitive Operatoren in \Lama. Wenn die
    SMT-Logik eine Funktion $real$ zur Verfügung stellt,
    kann das Abrunden durch \eqRef{is_floor} dargestellt werden.
    \begin{align}
      isFloor = \lambda x \; y.\; real(y) \leq x \wedge real(y) > x - 1
      \label{eq:is_floor} 
    \end{align}
  \item \lstSc!#(...)! -- xor mit mehreren Parametern
  \end{itemize}
\item Arrays und Structs
  \begin{itemize}
  \item \lstSc!reverse!,\lstSc!transpose! etc.
    -- können durch entsprechende Operationen auf Produkten
    dargestellt werden
  \item Konstruktion und Zugriff auf Structs -- durch entsprechende
    Produkt-Introduktionen/-Eliminationen
  \end{itemize}
\item \lstSc!case!-Ausdrücke -- im Falle von Enums können diese
  direkt übersetzt werden. Ansonsten müssen sie durch \lstLm!ite!
  ausgedrückt werden.
\item Operator-Applikationen höherer Ordnung
  und getaktete Applikationen -- können entsprechend
  \cite[9.5 (S.113ff.)]{ScadeRef} abgerollt bzw. auf
  lokale Takte zurückgeführt werden.
\end{itemize}

\section{Ausblick}
\label{sec:future_research}

Die in dieser Arbeit beschriebene Transformation ist ein erster
Grundstein zur Verifikation von \Scade-Modellen. Damit steht nun eine
Plattform zur Verfügung, auf der Optimierungen und verschiedene
Techniken in der Übersetzung ausprobiert werden können.

Zum einen können weitere Optimierungen bei der Transformation von
\Scade nach \Lama durchgeführt werden. Wobei die meisten Optimierungen
aber wahrscheinlich in \Lama vorgenommen werden könnten (s.u.). Damit
bleiben deren Implementierung unabhängig von \Scade.

Die Sprache \Lama selbst hat auch Verbesserungspotential.
So wird erwartet,
dass ein Knoten höchstens einmal benutzt wird. Dadurch muss bei
jeder weiteren Benutzung eine Kopie im Quelltext angelegt werden.
Dies führt aber zu vielen Wiederholungen oder auch Symmetrien.
Würde dies nicht verlangt werden, könnten sehr einfach Techniken
wie \term{Symmetry Reduction} angewendet werden.
Siehe dazu z.B.
\cite{Emerson97} oder \cite{Sistla04}. Um dies zu ermöglichen,
müssen die Gültigkeitsbereiche von Knoten in \Lama erweitert werden.
In der aktuellen Form ist die Generierung von SMT-Formeln sehr einfach.
Dies kann erhalten werden, indem die Benutzung eines Knotens vorab
deklariert werden muss. Eine mögliche Umsetzung ist in
\lstRef{lama_node_modified} dargestellt.

\lstdefinelanguage[refnode]{lama}[]{lama}{%
  morekeywords={reference,as,unique}}

\begin{listing}[h]
  \begin{lstlisting}[language={[refnode]lama}]
nodes
  node f ($\ldots$) returns ($\ldots$) let $\ldots$ tel
  node g unique ($\ldots$) returns ($\ldots$) let $\ldots$ tel
$\vdots$
nodes
  $\ldots$
  reference f as f_1, f_2;
definition
  x = (use f_1 $\ldots$);
  y = (use f_2 $\ldots$);
  -- z = (use g $\ldots$); -- not possible
  \end{lstlisting}  
  \caption{Modifizierte Benutzung von Knoten in \Lama}
  \label{lst:lama_node_modified}
\end{listing}

In dem \lstLm!nodes!-Teil eines Knotens muss vor Benutzung
ein Name deklariert werden
(\lstinline[language={[refnode]lama}]!reference .. as ..!),
unter dem der Knoten im aktuellen
Gültigkeitsbereich verwendet wird. Darf ein Knoten nicht
referenziert und damit höchstens einmal benutzt werden, muss
dies entsprechend gekennzeichnet werden
(\lstinline[language={[refnode]lama}]!unique!).
Die Verwendung solcher Knoten entspricht der in der jetzigen
\Lama-Variante.

Nicht zuletzt bietet die Sprache \Lama eine Ausgangsbasis für die
Untersuchung
von Optimierungs"~, Abstraktions- und Implementierungstechniken.
Durch ihren Aufbau können auch andere Datenflusssprachen
wie z.B. \lang{Lustre} nach \Lama konvertiert werden. Die andere
Richtung wäre auch denkbar. Dies würde einen Vergleich mit
bestehenden Tools wie z.B. \emph{Kind} oder \emph{NBAC}
ermöglichen.\footnote{Man beachte, dass ggf. nicht alle benötigten
Sprachfeatures von \lang{Lustre} unterstützt werden. Beispielsweise
wären lokale Takte nötig, um Automaten korrekt zu simulieren. Dann
muss auf die Analyse numerischer Fehler verzichtet werden.}

Insbesondere sind bereits genannte Techniken wie Pfadkompression
(\secRef{induction}) von Interesse um mehr Eigenschaften verifizieren
zu können (s. Probleme in \secRef{comparison_prover}). Außerdem wurde
\Lama gerade aus dem Grund entwickelt, dass Abstraktionen in
Modellen erhalten bleiben. Dies kann z.B. auf der in
\cite{HagenTinelli} beschriebenen „Structural Abstraction“ aufbauen.

Weitere mögliche Untersuchungen betreffen vor allem die Korrektheit der
Transformationen. Zunächst sind automatisierte Tests mit einer
größeren Menge von Beispielen nötig, um die Implementierung zu testen
und die Performance kontinuierlich vergleichen zu können.

Weiterhin wäre eine Formalisierung von \Scade interessant. Dies kann
evtl. durch eine Transformation in eine einfachere Teilsprache (wie
hier bereits teilweise geschehen) und Formalisierung dieser
umgesetzt werden. Den Automaten wurde auf diese Art in
\cite{SynchFlowState} eine Semantik gegeben.

Um die Korrektheit all unserer Transformationen von Automaten in
andere Automaten zu zeigen, kamen (nicht Teil dieser Arbeit) häufiger
Argumente auf, die \term[Bisimulation]{Bisimulationen} ähneln.
Dies korrekt zu
definieren (evtl. für Datenfluss im Allgemeinen) wäre ein weiterer
Schritt zu einem Korrektheitsbeweis der Transformationen.

\section{Fazit}
\label{sec:result}

Wir konnten zeigen, dass es möglich ist, \Scade in eine SMT-Logik
zu transformieren. Dabei wurden aufgrund des Umfangs nicht alle
Sprachkonstrukte behandelt. Die Transformation der verbleibenden
kann aber meist auf bereits vorhandene zurückgeführt werden
(Ausnahme: Casts).

Die Implementierung benötigt wie erwartet mehr Zeit als der
Design Verifier, findet dabei aber die gleichen Fehler. Leider ist die
einfache k-Induktion nicht ausreichend, um die Korrektheit gewisser
Modelle nachzuweisen. Da hiermit aber eine offene Plattform
zugänglich ist, können diese Probleme in einem weiteren Schritt
behandelt werden. Weiterhin ermöglicht dies weitere Untersuchungen
von Optimierungen, Abstraktions- und Implementierungstechniken.


\bibliography{ScadeSMT}
\printindex

\chapter{Anhang}
\label{chap:appendix}

\section{\Lama-Grammatik}
\label{sec:lama_grammar}

\ExecuteMetaData[content/Grammar.tex]{grammar}

\section{\Lama-Typregeln}
\label{sec:lama_type_rules}

\ExecuteMetaData[content/Types.tex]{typerules}


\section{\Scade/\Lama-Beispiele}
\label{sec:source_examples}

Hier wird eine Auswahl der verwendeten Beispiele zur Prüfung der
Korrektheit der Implementierung vorgestellt.

\begin{example}[Transformation von Weak-Transitions]
Dieses Beispiel soll die Korrektheit der Übersetzung von
Weak-Transitions (\secRef{weak_strong_transitions})
demonstrieren. Dabei werden die Automaten aus
\secRef{weak_strong_transitions} in einen kombiniert.
Das \Scade-Modell ist graphisch in \figRef{mixed_trans} dargestellt.
\begin{figure}[h]
  \centering
  \includegraphics[height=8cm]{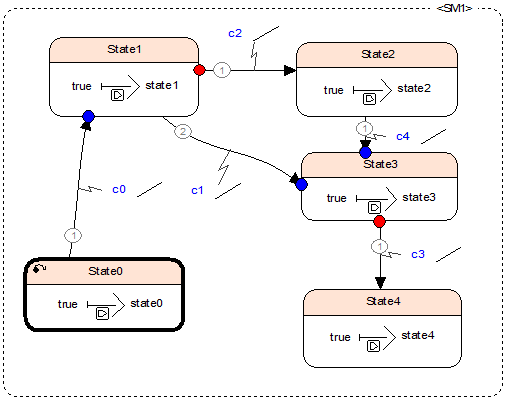}
  \caption{Automat mit gemischten Weak-/Strong-Transitions}
  \label{fig:mixed_trans}
\end{figure}

In \lstRef{mixed_trans} ist der Automat in Textform dargestellt. Um
die gewünschten Eigenschaften ausdrücken zu können, wird ein
zusätzlicher Knoten \lstSc!Prop! eingeführt
(\lstRef{mixed_trans_observe}). Dieser führt den Knoten
\lstSc!WeakStrong!, in dem der Automat implementiert
ist, aus und speichert einige Variablen, die für die
zu verifizierende Eigenschaft gebraucht werden.
\begin{listing}[h]
\begin{lstlisting}[language=scade,linerange={1-25}]
node WeakStrong(c0, c1, c2, c3, c4 : bool)
  returns (state0, state1 : bool default = false;
    state2, state3, state4 : bool default = false)
let
  automaton SM1
    initial state State0
      let state0 = true; tel
      until if c0 do restart State1;

    state State1
      unless if c2 restart State2;
      let state1 = true; tel
      until if c1 do restart State3;

    state State2
      let state2 = true; tel
      until if c4 do restart State3;

    state State3
      unless if c3 restart State4;
      let state3 = true; tel

    state State4 let state4 = true; tel
  returns ..;
tel
\end{lstlisting}
  \caption{\figRef{mixed_trans} in Textform}
  \label{lst:mixed_trans}
\end{listing}

\begin{listing}[h]
\begin{lstlisting}[language=scade,linerange={27-43}]
node Prop(c0, c1, c2, c3, c4 : bool)
  returns (state0, state1, state2, state3, state4 : bool;
    c0_1, c1_1, c1_2, c2_1, c4_1 : bool;
    state0_1, state1_1, state1_2 : bool)
let
  state0, state1, state2, state3, state4
    = WeakStrong(c0, c1, c2, c3, c4);

  c0_1 = false -> pre c0;
  c1_1 = false -> pre c1;
  c1_2 = false -> pre c1_1;
  c2_1 = false -> pre c2;
  c4_1 = false -> pre c4;
  state0_1 = true -> pre state0;
  state1_1 = false -> pre state1;
  state1_2 = false -> pre state1_1;
tel
\end{lstlisting}
  \caption{Beobachterknoten für WeakStrong}
  \label{lst:mixed_trans_observe}
\end{listing}

In \lstRef{mixed_trans_prop} findet sich die Invariante, die
von dem Modell erwartet wird. Diese umfasst einerseits die Tabelle
aus \ref{sec:weak_strong_par}. Andererseits werden die Kombinationen
von Weak- und Strong-Transitions geprüft.

\begin{listing}[h]
\begin{lstlisting}[language=lama,linerange={50-69}]
  (and (and (and (and (and (and
    -- FF
    (=> (and state1_1 (and (not c1_1) (not c2)))
      state1)
    -- FT
    (=> (and state1_1 (and (not c1_1) c2))
      state2))
    -- TF
    (=> (and state1_1 (and c1_1 (not c2)))
      (or state3 state4)))
    -- TT
    (=> (and state1_1 (and c1_1 c2))
      (or state3 state4)))
    -- Transitive transition 1 -> 4
    (=> (and state1_1 (and c1_1 c3)) state4))
    -- non Transitive transition 1 -> 3
    (=> (and state1_1 (and c1_1 (not c3))) state3))
    -- Don't activate weak after strong transition
    (=> (and (and state1_2 (not c1_2)) (and c2_1 c4_1))
      state2))
\end{lstlisting}
  \caption{Gewünschte Eigenschaften von WeakStrong}
  \label{lst:mixed_trans_prop}
\end{listing}

\end{example}

\begin{example}[Transformation von Subautomaten]
In \lstRef{subautomaton_ex} ist \lstSc!SM1! ein Subautomaten
in dem Zustand \lstSc!A!.

\begin{listing}[h]
\begin{lstlisting}[language=scade,linerange={1-26}]
node NestedStates(c1 : bool; c2 : bool)
  returns (inA, inA1, inA2, inB : bool default = false)
let

  automaton SM1
    initial state A
      unless if c2 restart B;
      let
        inA = true;

        automaton SM2
          initial state A1
            unless if c1 restart A2;
            let inA1 = true; tel

          state A2
            unless if c1 restart A1;
            let inA2 = true; tel
        returns ..;
      tel

    state B
      unless if c2 resume A;
      let inB = true; tel
  returns ..;
tel
\end{lstlisting}
  \caption{Subautomat in \Scade (Knoten NestedStates)}
  \label{lst:subautomaton_ex}
\end{listing}

\begin{listing}[h]
\begin{lstlisting}[language=scade,linerange={28-39}]
node Observer(c1 : bool; c2 : bool)
  returns (inA, inA1, inA2, inB : bool;
    inA_2, inB_1, inA1_2, inA2_2 : bool)
let
  inA, inA1, inA2, inB
    = NestedStates(c1, c2);

  inA_2 = fby(inA; 2; true);
  inB_1 = fby(inB; 1; false);
  inA1_2 = fby(inA1; 2; true);
  inA2_2 = fby(inA2; 2; false);
tel
\end{lstlisting}
  \caption{Beobachterknoten für NestedStates}
  \label{lst:subautomaton_observe}
\end{listing}

Die Invariante in \lstRef{subautomaton_prop} drückt aus,
dass der Zustand des Subautomaten erhalten bleibt, wenn zwei Takte
zuvor \lstSc!A! aktiv gewesen ist, danach \lstSc!B! und dann wieder
\lstSc!A!. In \lstRef{subautomaton_observe} ist der entsprechende
Beobachterknoten dargestellt.

\begin{listing}[h]
\begin{lstlisting}[language=lama,linerange={46-48}]
  (=> (and (and inA_2 inB_1) (and inA (not c1)))
    (and (= inA1 inA1_2)
      (= inA2 inA2_2)))
\end{lstlisting}
  \caption{Gewünschte Eigenschaften von WeakStrong}
  \label{lst:subautomaton_prop}
\end{listing}
\end{example}

\end{document}


\maketitle


\newcommand{\emptyP}{\mbox{$\epsilon$}}
\newcommand{\terminal}[1]{\mbox{{\textbf{\texttt {#1}}}}}
\newcommand{\nonterminal}[1]{\mbox{$\langle \mbox{\textsl{#1}} \rangle$}}
\newcommand{\arrow}{\coloncolonequals}
\newcommand{\delimit}{\mbox{$|$}}
\newcommand{\reserved}[1]{\mbox{{\texttt {#1}}}}
\newcommand{\literal}[1]{\mbox{{\texttt {#1}}}}
\newcommand{\symb}[1]{\mbox{{\texttt {#1}}}}



This document was automatically generated by the {\em BNF-Converter}. It was generated together with the lexer, the parser, and the abstract syntax module, which guarantees that the document matches with the implementation of the language (provided no hand-hacking has taken place).

\section*{The lexical structure of LAMA}

\subsection*{Literals}
Integer literals \nonterminal{Int}\ are nonempty sequences of digits.

Identifier literals are recognized by the regular expression
\(({\nonterminal{letter}} \mid \mbox{`\_'}) ({\nonterminal{letter}} \mid {\nonterminal{digit}} \mid \mbox{`\_'})*\)

StateId literals are recognized by the regular expression
\(({\nonterminal{letter}} \mid \mbox{`\_'}) ({\nonterminal{letter}} \mid {\nonterminal{digit}} \mid \mbox{`\_'})* \mbox{`''}\)

\subsection*{Reserved words and symbols}
The set of reserved words is the set of terminals appearing in the grammar. Those reserved words that consist of non-letter characters are called symbols, and they are treated in a different way from those that are similar to identifiers. The lexer follows rules familiar from languages like Haskell, C, and Java, including longest match and spacing conventions.

The reserved words used in LAMA are the following: \\

\begin{tabular}{lll}
{\reserved{and}} &{\reserved{assertion}} &{\reserved{automaton}} \\
{\reserved{bool}} &{\reserved{constants}} &{\reserved{default}} \\
{\reserved{definition}} &{\reserved{div}} &{\reserved{edge}} \\
{\reserved{enum}} &{\reserved{false}} &{\reserved{initial}} \\
{\reserved{input}} &{\reserved{int}} &{\reserved{invariant}} \\
{\reserved{ite}} &{\reserved{let}} &{\reserved{local}} \\
{\reserved{location}} &{\reserved{match}} &{\reserved{mod}} \\
{\reserved{node}} &{\reserved{nodes}} &{\reserved{not}} \\
{\reserved{or}} &{\reserved{project}} &{\reserved{real}} \\
{\reserved{returns}} &{\reserved{sint}} &{\reserved{state}} \\
{\reserved{tel}} &{\reserved{transition}} &{\reserved{true}} \\
{\reserved{typedef}} &{\reserved{uint}} &{\reserved{use}} \\
{\reserved{xor}} & & \\
\end{tabular}\\

The symbols used in LAMA are the following: \\

\begin{tabular}{lll}
{\symb{;}} &{\symb{{$=$}}} &{\symb{\{}} \\
{\symb{\}}} &{\symb{,}} &{\symb{\^}} \\
{\symb{(}} &{\symb{\#}} &{\symb{)}} \\
{\symb{[}} &{\symb{]}} &{\symb{{$-$}}} \\
{\symb{/}} &{\symb{:}} &{\symb{.}} \\
{\symb{\_}} &{\symb{{$=$}{$>$}}} &{\symb{{$<$}}} \\
{\symb{{$>$}}} &{\symb{{$<$}{$=$}}} &{\symb{{$>$}{$=$}}} \\
{\symb{{$+$}}} &{\symb{*}} & \\
\end{tabular}\\

\subsection*{Comments}
Single-line comments begin with {\symb{{$-$}{$-$}}}. \\There are no multiple-line comments in the grammar.

\section*{The syntactic structure of LAMA}
Non-terminals are enclosed between $\langle$ and $\rangle$. 
The symbols  {\arrow}  (production),  {\delimit}  (union) 
and {\emptyP} (empty rule) belong to the BNF notation. 
All other symbols are terminals.\\

\begin{tabular}{lll}
{\nonterminal{Program}} & {\arrow}   & {\nonterminal{TypeDefs}} {\nonterminal{ConstantDefs}} \\
                                & & {\nonterminal{Inputs}} {\nonterminal{Declarations}} {\nonterminal{Flow}} \\
                                & & {\nonterminal{Initial}} {\nonterminal{Assertion}} {\nonterminal{Invariant}}  \\
\end{tabular}\\

\begin{tabular}{lll}
{\nonterminal{TypeDefs}} & {\arrow}  &{\emptyP} \\
 & {\delimit}  &{\terminal{typedef}} {\nonterminal{ListTypeDef}}  \\
\end{tabular}\\

\begin{tabular}{lll}
{\nonterminal{ListTypeDef}} & {\arrow}  &{\nonterminal{TypeDef}} {\terminal{;}}  \\
 & {\delimit}  &{\nonterminal{TypeDef}} {\terminal{;}} {\nonterminal{ListTypeDef}}  \\
\end{tabular}\\

\begin{tabular}{lll}
{\nonterminal{TypeDef}} & {\arrow}  &{\terminal{enum}} {\nonterminal{Identifier}} {\terminal{{$=$}}} {\terminal{\{}} {\nonterminal{ListEnumConstr}} {\terminal{\}}}  \\
\end{tabular}\\

\begin{tabular}{lll}
{\nonterminal{EnumConstr}} & {\arrow}  &{\nonterminal{Identifier}}  \\
\end{tabular}\\

\begin{tabular}{lll}
{\nonterminal{ListEnumConstr}} & {\arrow}  &{\nonterminal{EnumConstr}}  \\
 & {\delimit}  &{\nonterminal{EnumConstr}} {\terminal{,}} {\nonterminal{ListEnumConstr}}  \\
\end{tabular}\\


\begin{tabular}{lll}
{\nonterminal{Type}} & {\arrow}  &{\nonterminal{BaseType}}  \\
 & {\delimit}  &{\nonterminal{Identifier}}  \\
 & {\delimit}  &{\nonterminal{BaseType}} {\terminal{\^}} {\nonterminal{Natural}}  \\
 & {\delimit}  &{\terminal{(}} {\terminal{\#}} {\nonterminal{ListType}} {\terminal{)}}  \\
\end{tabular}\\

\begin{tabular}{lll}
{\nonterminal{ListType}} & {\arrow}  &{\nonterminal{Type}}  \\
 & {\delimit}  &{\nonterminal{Type}} {\nonterminal{ListType}}  \\
\end{tabular}\\

\begin{tabular}{lll}
{\nonterminal{BaseType}} & {\arrow}  &{\terminal{bool}}  \\
 & {\delimit}  &{\terminal{int}}  \\
 & {\delimit}  &{\terminal{real}}  \\
 & {\delimit}  &{\terminal{sint}} {\terminal{[}} {\nonterminal{Natural}} {\terminal{]}}  \\
 & {\delimit}  &{\terminal{uint}} {\terminal{[}} {\nonterminal{Natural}} {\terminal{]}}  \\
\end{tabular}\\


\begin{tabular}{lll}
{\nonterminal{ConstantDefs}} & {\arrow}  &{\emptyP} \\
 & {\delimit}  &{\terminal{constants}} {\nonterminal{ListConstantDef}}  \\
\end{tabular}\\

\begin{tabular}{lll}
{\nonterminal{ListConstantDef}} & {\arrow}  &{\nonterminal{ConstantDef}} {\terminal{;}}  \\
 & {\delimit}  &{\nonterminal{ConstantDef}} {\terminal{;}} {\nonterminal{ListConstantDef}}  \\
\end{tabular}\\

\begin{tabular}{lll}
{\nonterminal{ConstantDef}} & {\arrow}  &{\nonterminal{Identifier}} {\terminal{{$=$}}} {\nonterminal{Constant}}  \\
\end{tabular}\\

\begin{tabular}{lll}
{\nonterminal{Natural}} & {\arrow}  &{\nonterminal{Integer}}  \\
\end{tabular}\\

\begin{tabular}{lll}
{\nonterminal{IntegerConst}} & {\arrow}  &{\nonterminal{Integer}}  \\
 & {\delimit}  &{\terminal{(}} {\terminal{{$-$}}} {\nonterminal{Integer}} {\terminal{)}}  \\
\end{tabular}\\

\begin{tabular}{lll}
{\nonterminal{Constant}} & {\arrow}  &{\nonterminal{BoolV}}  \\
 & {\delimit}  &{\nonterminal{IntegerConst}}  \\
 & {\delimit}  &{\nonterminal{IntegerConst}} {\terminal{/}} {\nonterminal{IntegerConst}}  \\
 & {\delimit}  &{\terminal{sint}} {\terminal{[}} {\nonterminal{Natural}} {\terminal{]}} {\terminal{(}} {\nonterminal{IntegerConst}} {\terminal{)}}  \\
 & {\delimit}  &{\terminal{uint}} {\terminal{[}} {\nonterminal{Natural}} {\terminal{]}} {\terminal{(}} {\nonterminal{Natural}} {\terminal{)}}  \\
\end{tabular}\\

\begin{tabular}{lll}
{\nonterminal{BoolV}} & {\arrow}  &{\terminal{true}}  \\
 & {\delimit}  &{\terminal{false}}  \\
\end{tabular}\\

\begin{tabular}{lll}
{\nonterminal{Inputs}} & {\arrow}  &{\emptyP} \\
 & {\delimit}  &{\terminal{input}} {\nonterminal{VarDecls}}  \\
\end{tabular}\\

\begin{tabular}{lll}
{\nonterminal{Initial}} & {\arrow}  &{\emptyP} \\
 & {\delimit}  &{\terminal{initial}} {\nonterminal{ListStateInit}} {\terminal{;}}  \\
\end{tabular}\\

\begin{tabular}{lll}
{\nonterminal{Assertion}} & {\arrow}  &{\emptyP} \\
 & {\delimit}  &{\terminal{assertion}} {\nonterminal{Expr}} {\terminal{;}}  \\
\end{tabular}\\

\begin{tabular}{lll}
{\nonterminal{Invariant}} & {\arrow}  &{\emptyP} \\
 & {\delimit}  &{\terminal{invariant}} {\nonterminal{Expr}} {\terminal{;}}  \\
\end{tabular}\\

\begin{tabular}{lll}
{\nonterminal{ListStateInit}} & {\arrow}  &{\nonterminal{StateInit}}  \\
 & {\delimit}  &{\nonterminal{StateInit}} {\terminal{,}} {\nonterminal{ListStateInit}}  \\
\end{tabular}\\

\begin{tabular}{lll}
{\nonterminal{StateInit}} & {\arrow}  &{\nonterminal{Identifier}} {\terminal{{$=$}}} {\nonterminal{ConstExpr}}  \\
\end{tabular}\\

\begin{tabular}{lll}
{\nonterminal{ConstExpr}} & {\arrow}  &{\nonterminal{Expr}}  \\
\end{tabular}\\

\begin{tabular}{lll}
{\nonterminal{TypedVar}} & {\arrow}  &{\nonterminal{Identifier}} {\terminal{:}} {\nonterminal{Type}}  \\
\end{tabular}\\

\begin{tabular}{lll}
{\nonterminal{ListTypedVar}} & {\arrow}  &{\nonterminal{TypedVar}}  \\
 & {\delimit}  &{\nonterminal{TypedVar}} {\terminal{,}} {\nonterminal{ListTypedVar}}  \\
\end{tabular}\\

\begin{tabular}{lll}
{\nonterminal{MaybeTypedVars}} & {\arrow}  &{\emptyP} \\
 & {\delimit}  &{\nonterminal{ListTypedVar}}  \\
\end{tabular}\\

\begin{tabular}{lll}
{\nonterminal{Node}} & {\arrow}   & {\terminal{node}} {\nonterminal{Identifier}} {\terminal{(}} {\nonterminal{MaybeTypedVars}} {\terminal{)}} \\
                                & & {\terminal{returns}} {\terminal{(}} {\nonterminal{ListTypedVars}} {\terminal{)}} {\terminal{let}} \\
                                & & \hspace{5pt} {\nonterminal{Declarations}} \\
                                & & \hspace{5pt} {\nonterminal{Flow}} \\
                                & & \hspace{5pt} {\nonterminal{ControlStructure}} \\
                                & & \hspace{5pt} {\nonterminal{Initial}} \\
                                & & \hspace{5pt} {\nonterminal{Assertion}} \\
                                & & {\terminal{tel}}  \\
\end{tabular}\\

\begin{tabular}{lll}
{\nonterminal{ListNode}} & {\arrow}  &{\nonterminal{Node}}  \\
 & {\delimit}  &{\nonterminal{Node}} {\nonterminal{ListNode}}  \\
\end{tabular}\\

\begin{tabular}{lll}
{\nonterminal{Declarations}} & {\arrow}  &{\nonterminal{NodeDecls}} {\nonterminal{LocalDecls}} {\nonterminal{StateDecls}}  \\
\end{tabular}\\

\begin{tabular}{lll}
{\nonterminal{VarDecls}} & {\arrow}  &{\nonterminal{TypedVar}} {\terminal{;}}  \\
 & {\delimit}  &{\nonterminal{TypedVar}} {\terminal{;}} {\nonterminal{VarDecls}}  \\
\end{tabular}\\

\begin{tabular}{lll}
{\nonterminal{NodeDecls}} & {\arrow}  &{\emptyP} \\
 & {\delimit}  &{\terminal{nodes}} {\nonterminal{ListNode}}  \\
\end{tabular}\\

\begin{tabular}{lll}
{\nonterminal{LocalDecls}} & {\arrow}  &{\emptyP} \\
 & {\delimit}  &{\terminal{local}} {\nonterminal{VarDecls}}  \\
\end{tabular}\\

\begin{tabular}{lll}
{\nonterminal{StateDecls}} & {\arrow}  &{\emptyP} \\
 & {\delimit}  &{\terminal{state}} {\nonterminal{VarDecls}}  \\
\end{tabular}\\

\begin{tabular}{lll}
{\nonterminal{Flow}} & {\arrow}  &{\nonterminal{LocalDefinitions}} {\nonterminal{Transitions}}  \\
\end{tabular}\\

\begin{tabular}{lll}
{\nonterminal{LocalDefinitions}} & {\arrow}  &{\emptyP} \\
 & {\delimit}  &{\terminal{definition}} {\nonterminal{ListInstantDefinition}}  \\
\end{tabular}\\

\begin{tabular}{lll}
{\nonterminal{Transitions}} & {\arrow}  &{\emptyP} \\
 & {\delimit}  &{\terminal{transition}} {\nonterminal{ListTransition}}  \\
\end{tabular}\\

\begin{tabular}{lll}
{\nonterminal{ListInstantDefinition}} & {\arrow}  &{\nonterminal{InstantDefinition}} {\terminal{;}}  \\
 & {\delimit}  &{\nonterminal{InstantDefinition}} {\terminal{;}} {\nonterminal{ListInstantDefinition}}  \\
\end{tabular}\\

\begin{tabular}{lll}
{\nonterminal{ListTransition}} & {\arrow}  &{\nonterminal{Transition}} {\terminal{;}}  \\
 & {\delimit}  &{\nonterminal{Transition}} {\terminal{;}} {\nonterminal{ListTransition}}  \\
\end{tabular}\\

\begin{tabular}{lll}
{\nonterminal{InstantDefinition}} & {\arrow}  &{\nonterminal{Identifier}} {\terminal{{$=$}}} {\nonterminal{Expr}}  \\
 & {\delimit}  &{\nonterminal{Identifier}} {\terminal{{$=$}}} {\terminal{(}} {\terminal{use}} {\nonterminal{Identifier}} {\nonterminal{ListExpr}} {\terminal{)}}  \\
\end{tabular}\\

\begin{tabular}{lll}
{\nonterminal{Transition}} & {\arrow}  &{\nonterminal{StateId}} {\terminal{{$=$}}} {\nonterminal{Expr}}  \\
\end{tabular}\\

\begin{tabular}{lll}
{\nonterminal{ControlStructure}} & {\arrow}  &{\nonterminal{ListAutomaton}}  \\
\end{tabular}\\

\begin{tabular}{lll}
{\nonterminal{Automaton}} & {\arrow}  & {\terminal{automaton}} {\terminal{let}} \\
                                    & & \hspace{5pt} {\nonterminal{ListLocation}} \\
                                    & & \hspace{5pt} {\nonterminal{InitialLocation}} \\
                                    & & \hspace{5pt} {\nonterminal{ListEdge}} \\
                                    & & \hspace{5pt} {\nonterminal{Defaults}} \\
                                    & & {\terminal{tel}}  \\
\end{tabular}\\

\begin{tabular}{lll}
{\nonterminal{Location}} & {\arrow}  &{\terminal{location}} {\nonterminal{Identifier}} {\terminal{let}} {\nonterminal{Flow}} {\terminal{tel}}  \\
\end{tabular}\\

\begin{tabular}{lll}
{\nonterminal{InitialLocation}} & {\arrow}  &{\terminal{initial}} {\nonterminal{Identifier}} {\terminal{;}}  \\
\end{tabular}\\

\begin{tabular}{lll}
{\nonterminal{Edge}} & {\arrow}  &{\terminal{edge}} {\terminal{(}} {\nonterminal{Identifier}} {\terminal{,}} {\nonterminal{Identifier}} {\terminal{)}} {\terminal{:}} {\nonterminal{Expr}} {\terminal{;}}  \\
\end{tabular}\\

\begin{tabular}{lll}
{\nonterminal{ListLocation}} & {\arrow}  &{\nonterminal{Location}}  \\
 & {\delimit}  &{\nonterminal{Location}} {\nonterminal{ListLocation}}  \\
\end{tabular}\\

\begin{tabular}{lll}
{\nonterminal{ListEdge}} & {\arrow}  &{\nonterminal{Edge}}  \\
 & {\delimit}  &{\nonterminal{Edge}} {\nonterminal{ListEdge}}  \\
\end{tabular}\\

\begin{tabular}{lll}
{\nonterminal{ListAutomaton}} & {\arrow}  &{\emptyP} \\
 & {\delimit}  &{\nonterminal{Automaton}} {\nonterminal{ListAutomaton}}  \\
\end{tabular}\\

\begin{tabular}{lll}
{\nonterminal{Defaults}} & {\arrow}  &{\emptyP} \\
 & {\delimit}  &{\terminal{default}} {\nonterminal{ListDefault}} {\terminal{;}}  \\
\end{tabular}\\

\begin{tabular}{lll}
{\nonterminal{ListDefault}} & {\arrow}  &{\nonterminal{Default}}  \\
 & {\delimit}  &{\nonterminal{Default}} {\terminal{,}} {\nonterminal{ListDefault}}  \\
\end{tabular}\\

\begin{tabular}{lll}
{\nonterminal{Default}} & {\arrow}  &{\nonterminal{Identifier}} {\terminal{{$=$}}} {\nonterminal{Expr}}  \\
\end{tabular}\\

\begin{tabular}{lll}
{\nonterminal{Atom}} & {\arrow}  &{\nonterminal{Constant}}  \\
 & {\delimit}  &{\nonterminal{Identifier}}  \\
\end{tabular}\\

\begin{tabular}{lll}
{\nonterminal{Expr}} & {\arrow}  &{\nonterminal{Atom}}  \\
 & {\delimit}  &{\terminal{(}} {\nonterminal{UnOp}} {\nonterminal{Expr}} {\terminal{)}}  \\
 & {\delimit}  &{\terminal{(}} {\nonterminal{BinOp}} {\nonterminal{Expr}} {\nonterminal{Expr}} {\terminal{)}}  \\
 & {\delimit}  &{\terminal{(}} {\nonterminal{TernOp}} {\nonterminal{Expr}} {\nonterminal{Expr}} {\nonterminal{Expr}} {\terminal{)}}  \\
 & {\delimit}  &{\terminal{(}} {\terminal{\#}} {\nonterminal{ListExpr}} {\terminal{)}}  \\
 & {\delimit}  &{\terminal{(}} {\terminal{project}} {\nonterminal{Identifier}} {\nonterminal{Natural}} {\terminal{)}}  \\
 & {\delimit}  &{\terminal{(}} {\terminal{match}} {\nonterminal{Expr}} {\terminal{\{}} {\nonterminal{ListPattern}} {\terminal{\}}} {\terminal{)}}  \\
\end{tabular}\\

\begin{tabular}{lll}
{\nonterminal{ListExpr}} & {\arrow}  &{\emptyP} \\
 & {\delimit}  &{\nonterminal{Expr}} {\nonterminal{ListExpr}}  \\
\end{tabular}\\

\begin{tabular}{lll}
{\nonterminal{ListPattern}} & {\arrow}  &{\nonterminal{Pattern}}  \\
 & {\delimit}  &{\nonterminal{Pattern}} {\terminal{,}} {\nonterminal{ListPattern}}  \\
\end{tabular}\\

\begin{tabular}{lll}
{\nonterminal{Pattern}} & {\arrow}  &{\nonterminal{PatHead}} {\terminal{.}} {\nonterminal{Expr}}  \\
\end{tabular}\\

\begin{tabular}{lll}
{\nonterminal{PatHead}} & {\arrow}  &{\nonterminal{EnumConstr}}  \\
 & {\delimit}  &{\terminal{\_}}  \\
\end{tabular}\\

\begin{tabular}{lll}
{\nonterminal{List2Id}} & {\arrow}  &{\nonterminal{Identifier}} {\nonterminal{Identifier}}  \\
 & {\delimit}  &{\nonterminal{Identifier}} {\nonterminal{List2Id}}  \\
\end{tabular}\\

\begin{tabular}{lll}
{\nonterminal{UnOp}} & {\arrow}  &{\terminal{not}}  \\
\end{tabular}\\

\begin{tabular}{lll}
{\nonterminal{BinOp}} & {\arrow}  &{\terminal{or}}  \\
 & {\delimit}  &{\terminal{and}}  \\
 & {\delimit}  &{\terminal{xor}}  \\
 & {\delimit}  &{\terminal{{$=$}{$>$}}}  \\
 & {\delimit}  &{\terminal{{$=$}}}  \\
 & {\delimit}  &{\terminal{{$<$}}}  \\
 & {\delimit}  &{\terminal{{$>$}}}  \\
 & {\delimit}  &{\terminal{{$<$}{$=$}}}  \\
 & {\delimit}  &{\terminal{{$>$}{$=$}}}  \\
 & {\delimit}  &{\terminal{{$+$}}}  \\
 & {\delimit}  &{\terminal{{$-$}}}  \\
 & {\delimit}  &{\terminal{*}}  \\
 & {\delimit}  &{\terminal{/}}  \\
 & {\delimit}  &{\terminal{div}}  \\
 & {\delimit}  &{\terminal{mod}}  \\
\end{tabular}\\

\begin{tabular}{lll}
{\nonterminal{TernOp}} & {\arrow}  &{\terminal{ite}}  \\
\end{tabular}\\
